\documentclass[12pt]{article}
\usepackage{graphicx}
\usepackage{amsmath}

\usepackage{amscd,amssymb}
\usepackage[cmtip,arrow,
matrix]{xy}

\usepackage{pstricks}
\usepackage{curves}

\newcommand{\cE}{\mathcal{E}}

\DeclareMathOperator{\rank}{rank}

\def\cR{{\mathcal R}}
\def\cS{{\mathcal S}}
\def\cJ{{\mathcal J}}

\def\cN{{\mathcal N}}
\def\cA{{\mathcal A}}

\def\cC{{\mathcal C}}

\def\cD{{\mathcal D}}

\def\cH{{\mathcal H}}
\def\cP{{\mathcal P}}

\def\cE{{\mathcal E}}

\def\cO{{\mathcal O}}
\def\cQ{{\mathcal Q}}
\def\cU{{\mathcal U}}

\newtheorem{theorem}{Theorem}[section]

\newtheorem{definition}{Definition}[section]

\newtheorem{remark}{Remark}[section]

\newtheorem{comment}{Comment}[section]

\begin{document}


\begin{center}
{\large\bf String Partition
Functions, Hilbert Schemes, and \\ Affine
Lie Algebra Representations on Homology Groups}

\vspace{0.2in}

Loriano Bonora$^{a,}$\footnote{E-mail: bonora@sissa.it}, Andrey Bytsenko$^{b,}$\footnote{E-mail: abyts@uel.br}  and  Emilio Elizalde$^{c,}$\footnote{E-mail: elizalde@ieec.uab.es}

\vspace{0.2in}
 {\small        $^a$  International School for Advanced Studies (SISSA/ISAS) \\ Via Bonomea 265, 34136 Trieste\,
                                and INFN, Sezione di Trieste, Italy
\\
\vspace{0.1in}
         $^b$  Universidade Estadual de Londrina,  Caixa Postal 6001 Londrina-PR,  Brazil\\


\vspace{0.1in}
         $^c$  Consejo Superior de Investigaciones Cient\'{\i}ficas, ICE/CSIC and  IEEC \\ Campus UAB, Facultat de Ci\`encies, Torre C5-Par-2a, 08193 Bellaterra (Barcelona) Spain\\   }
$\,$
\\
\end{center}

\begin{abstract}

This review paper contains a concise introduction to highest weight
representations of infinite dimensional Lie algebras, vertex
operator algebras and Hilbert schemes of points, together with their
physical applications to elliptic genera of superconformal quantum
mechanics and superstring models. The common link of all these concepts
and of the many examples considered in the paper
is to be found in a very important feature of the theory of infinite
dimensional Lie algebras: the modular properties of the
characters (generating functions) of certain representations.
The characters of the highest weight modules
represent the holomorphic parts of the partition functions on the
torus for the corresponding conformal field theories. We discuss
the role of the unimodular (and modular) groups and the
(Selberg-type) Ruelle spectral functions of hyperbolic geometry in
the calculation of elliptic genera and associated $q$-series. For
mathematicians, elliptic genera are commonly associated to new
mathematical invariants for spaces, while for physicists elliptic
genera are one-loop string partition function (therefore they are
applicable, for instance, to topological Casimir effect calculations).
We show that elliptic genera can be conveniently transformed into
product expressions which can then inherit the homology properties
of appropriate polygraded Lie algebras.
\end{abstract}

\begin{flushleft}
\\
\vspace{0.3in}
June 2012
\end{flushleft}

\newpage

\tableofcontents


\section{Introduction}

There is a known (but sometimes inadverted) connection between representation theory of infinite dimensional Lie algebras and vertex operator algebras and their homologies, on one side, and partition functions (in particular elliptic genera) of physical models, on the other. This connection manifests itself in the fact that the just mentioned quantities can be expressed in a quite universal way in terms of Ruelle functions. Still this connection has not yet been
carefully studied in the literature. Here we would like to call the reader's attention on this link by reviewing first the mathematical aspects associated with it and then a certain number of physical applications.

To start with we would like to recall that application of Grassmannian algebras to differentiable manifolds led to the algebra of exterior differential forms and exterior operators. Subsequently Clifford algebras and spinor representations, applied to Riemannian spin manifolds, led to the theory of Dirac operators. The role of Lie algebras in conformal field theories, considered in this paper, is somehow in the same spirit. Highest weight modules of affine Lie algebras (the Virasoro algebra, for example) underlie conformal field theories. Vertex operator algebras can be constructed from highest weight representations of infinite dimensional Lie algebras. The characters of (integrable) highest weight modules have been identified with the holomorphic parts of the partition functions on the torus for the corresponding field theories. All these structures arise naturally, but not exclusively, in string theory, and are particularly clear and treatable when supersymmetry is involved.
In the course of the paper we will consider various applications of the previous ideas to physical models: elliptic genera of supersymmetric sigma models, chiral primary states, partition functions of brane in CY spaces, Hilbert schemes of points, and elliptic genera of symmetric products and orbifolds. All those examples are intended to accumulate evidence on the existence of the link mentioned at the beginning, although, of course, this does not make a proof.

The organization of the paper is as follows.
As already anticipated, the first part is a review of some mathematical preliminaries.
We begin with elements of representation theory of Lie algebras in Sect. \ref{Represent}. We study highest weight representations of Lie algebras with their connection to vertex algebra, and then in Sect. \ref{Super} we move on to vertex operator algebra bundles and modules. Sect. \ref{Lefschetz} is also of preparatory character:
it introduces the Lefschetz fixed-point formula. Sect. \ref{SymSpace} up to  Sect. \ref{Symmetric} are devoted to collecting examples of partition functions and elliptic genera of physical models with the aim to show that
all of them, although very different in nature, can be expressed in a generic way in terms of Ruelle functions. This common feature
encodes in our opinion the connection with infinite dimensional Lie algebras and their homologies. This suggestion is explicitly spelt out (although not proved) in Sect. \ref{Identities}.

Having given the scheme let us now explain, for the benefit of the reader, the connection between the specific contents of the various sections in more detail. We find this necessary given the length of this work and the many different concepts considered in it.
A relevant concept in the paper is that of elliptic genus. Elliptic genera are natural topological invariants, which generalize classical genera. They appear when one considers supersymmetric indices of the superconformal vertex algebras (SCVA). For mathematicians, elliptic genera (and the respective elliptic cohomology) may be associated to new mathematical invariants for spaces, while for physicists elliptic genera are one-loop string partition functions and, therefore, they are
applicable to topological Casimir effect calculations \cite{cas1}-\cite{cas3}. They have also been proven to be applicable to black hole entropy computations \cite{Strominger}. This explains why we devote here several pages to the construction and properties of SCVA, Sect. \ref{Super}. In Sect. \ref{N=1} we consider $N=1$ SCVA's from complex vector spaces, Riemannian manifolds, and characters and $K$-theories of vector algebra bundles. In Sect. \ref{N=2} we study $N=2$ SCVA from any complex manifold. There are two ways to twist an $N = 2$ SCFT to obtain topological vertex algebras. The BRST cohomology groups of these two vertex algebras correspond to the algebras of the primary chiral and anti-chiral fields, respectively, of the original $N = 2$ SCFT. In many applications, when the space $X$ is a Calabi-Yau manifold, there is an $N = 2$ SCFT associated to $X$, with the two twists leading to the so-called $A$-theory and $B$-theory, respectively. This has been discussed over the years in an impressive number of papers. The mathematical formalization of these concepts also exists since long
time ago. A sheaf of topological vertex algebras for any Calabi-Yau manifold has been constructed in \cite{Malikov}, where the theory involved is $A$-theory. A different approach based on standard techniques in differential geometry has been developed in \cite{Zhou00}, where holomorphic vector bundles of $N = 2$ SCVA on a complex manifold $X$ and the $\overline{\partial}$ operator on such bundles have been used.

In Sect. \ref{Lefschetz} we consider the Lefschetz fixed-point formula (named after Solomon Lefschetz, who first stated it in 1926), which counts the number of fixed points of a continuous mapping from a compact topological space $X$ to itself by means of traces of the induced mappings on the homology groups of $X$. One can use this formula to compute Lefschetz numbers of automorphisms of elliptic complexes in terms of the index of associated elliptic symbol classes on fixed-point sets. In particular, we discuss in this section a cohomological form of the Lefschetz fixed-point formula for elliptic complexes.

Sect. \ref{SymSpace} is devoted to Laplacians on forms and the Ruelle spectral function, Sect. \ref{Norm2}, and to the spectral functions of hyperbolic geometry, Sect. \ref{Dirac}, where the Lefschetz formula has been used for stable characteristic Pontryagin and Chern classes. This is one of the central issues here, since it provides the spectral function method, which we repeatedly use in the course of the paper. The simpler case of hyperbolic three-geometry and associated (Selberg-type) Ruelle spectral functions is considered in Sect. \ref{Three-geometry}. Our interest in this example stems from the $AdS_3/CFT_2$ correspondence. The geometric structure of three-dimensional gravity (and black holes) allows for exact computations, since its Euclidean counterpart is locally isomorphic to the constant curvature hyperbolic space. There is an agreement between spectral functions related to Euclidean $AdS_3$ and modular-like functions (Poincar\'e series). This occurs when the arguments of spectral functions take values on a Riemann surface, viewed as the conformal boundary of $AdS_3$ \cite{Bonora11}. According to the holographic principle, a strong correspondence exists between certain field theory quantities in the bulk of an $AdS_3$ manifold and related quantities on its boundary at infinity. To be more precise, the classes of Euclidean $AdS_3$ spaces are quotients of the real hyperbolic space by a discrete group (a Schottky group). The boundary of these spaces can be compact oriented surfaces with conformal structure (compact complex algebraic curves).

More physical applications along this same line, namely the elliptic genera of nonlinear sigma models, are considered in Sect. \ref{Elliptic}. Elliptic genera of (2,2) supersymmetric Landau-Ginzburg models over vector spaces were first computed in \cite{Witten94}; they
provide effective ways to compute elliptic genera of corresponding nonlinear sigma models. The Landau-Ginzburg elliptic genus has a natural meaning in the equivariant elliptic cohomology
\cite{Grojnowski,Greenlees}: it is the equivariant genus, or Euler class, of a virtual representation of $U(1)$ associated to the sigma orientation \cite{Ando01,Ando03,Ando07}. Most of those Landau-Ginzburg models do not live over smooth spaces, or orbifolds thereof, but over more complicated spaces, forming what are sometimes called {\it hybrid Landau-Ginzburg models} \cite{Sharpe09}.
We mainly concentrate on the Witten genus and elliptic genera of Landau-Ginzburg models over vector spaces, Sect. \ref{Landau}. For these examples we show that elliptic genera can be written in terms of spectral functions of the hyperbolic geometry associated with $q$-series.

In Sect. \ref{Chiral} we consider applications of the previous mathematical tools to the case (very important in physics) of generating functions in the black hole geometry. Namely, we consider the issue that, in the $AdS_{2}\times S^2\times CY_3$ attractor geometry of a Calabi-Yau black hole with $D4$ brane charges, superconformal quantum mechanics describes $D0$ branes. This model contains
a large degeneracy of chiral primary bound states, which arise from $D0$ branes in the lowest Landau levels that tile the $CY_3\times S^2$ horizon. It is well known that the primary chiral fields of an $N=2$ superconformal field theory form an algebra. The proof of this relevant fact in the physics literature \cite{Lerche,Warner93} shares many common features with the Hodge theory of K\"{a}hler manifolds in mathematics since, apparently, such a multi-$D0$ brane superconformal field theory is holographically dual to IIA string theory on $AdS_{2}\times S^2\times CY_3$. We evaluate the chiral primary generating function in terms of Patterson-Selberg spectral functions.

In Sect. \ref{HH} we survey Hilbert schemes of points. They allow
us to construct a representation of products of Heisenberg and
Clifford algebras on the direct sum of homology groups of all
components associated with schemes. Hilbert schemes of points of
surfaces are discussed in Sect. \ref{Surface}; we rewrite there the
character formulas and G\"{o}ttsche's formula, Sect.
\ref{Gottsche}, in terms of Ruelle spectral functions.

Elliptic genera of symmetric products are discussed in Sect. \ref{Symmetric}. The relationship with certain equivariant genera of instanton moduli spaces can be exploited to study the string partition functions of some local Calabi-Yau geometries and, in particular, the Gopakumar-Vafa conjecture for them \cite{Gopakumar}. The Gromov-Witten invariants are in general rational numbers. However, as conjectured in \cite{Gopakumar}, the generating series of Gromov-Witten invariants in all degrees and all genera has a very peculiar form, determined by some integers. Various proposals for the proof of this conjecture have appeared \cite{Katz, Hosono, Maulik}. A geometric approach was proposed in \cite{Li-Liu04}. Some progress toward the calculation of Gromov-Witten invariants is present in  physical approaches \cite{Aganagic02,Iqbal02,Aganagic03} and in  mathematical treatments \cite{Zhou03,Li04}. Our special attention here is devoted to the spectral function reformulation of the Gopakumar-Vafa conjecture, Sect. \ref{GV}, and orbifold elliptic genera, Sect. \ref{Orbifold}.

The common link of all these examples is to be found, in our
opinion, in one important feature of the theory of infinite
dimensional Lie algebras, that is, the modular properties of the
characters (generating functions) of certain representations. The
highest weight modules of the affine Lie algebras underlie
conformal field theories. The characters of the highest weight
modules can be interpreted as the holomorphic parts of the
partition functions (elliptic genera) on the torus for the
corresponding conformal field theories. Finally, elliptic genera
can be converted into product expressions which can inherit the
homology properties of appropriate polygraded Lie algebras, what is discussed in Sect.
\ref{Identities}. In many physical applications, quantum generating
functions can be reproduced in terms of spectral functions of Selberg type
(see \cite{booksz1} for related applications).
Therefore, the role of the unimodular group $SL(2;{\mathbb C})$ and
of the modular group $SL(2; {\mathbb Z})$ constitute a clear manifestation
of the remarkable link that exists between all the above and hyperbolic three-geometry.

\section{Elements of representation theory of Lie algebras}
\label{Represent}

Vertex operators and vertex operator algebras were introduced in the physical literature at the inception of string theory. In this section we will present a more mathematical approach to this subject, due in particular
to \cite{Borcherds86,Frenkel89,Frenkel93,Zhou96,Dong97,Dong04}.
Heisenberg vertex operator algebras and vertex operator algebras associated to the highest weight representations of affine Kac-Moody algebras will also be briefly introduced. They have played an important role in the study of elliptic genus and the Witten genus \cite{Witten88}.

Many vertex algebras can be constructed from highest weight representations of infinite dimensional Lie algebras. In the sequel we will consider some of them. Let $\mathfrak g$ be a complex Lie algebra with symmetric bilinear form $(\,\cdot\,|\,\cdot)$, invariant under the adjoint action:
$
([a, b]|c) + (b|[a, c]) = 0, \,\,\,\,\, a,b,c \in {\mathfrak g}\,.
$
The affinization of $(\mathfrak g, (\,\cdot\,|\,\cdot))$ is the Lie algebra
$\hat{\mathfrak g}$ with nonzero commutation relations:
\begin{equation}
\hat{\mathfrak g} = {\mathbb C}[t, t^{-1}]\otimes_{\mathbb C}
{\mathfrak g}\oplus {\mathbb C}K,\,\,\,\,\,\,\,
[a_m, b_n] = [a, b]_{m+n} + m (a|b)\delta_{m, -n}K\,,
\label{affine}
\end{equation}
where for each $a\in {\mathfrak g}$ and $n\in {\mathbb Z}$ $a_n$ stands for $t^n\otimes a$, ${\mathbb C}[t, t^{-1}]$ denotes the Laurent polynomial algebra in the variable $t$. This is referred to as a {\it current algebra}.

{\bf Examples:} Suppose that $\mathfrak g$ is a simple Lie algebra with Killing form, then $\hat{\mathfrak g}$ is the affine Kac-Moody algebra
(see also Sect. \ref{Super}).
If $\mathfrak g$ is the one-dimensional Lie algebra with a nondegenerate bilinear form, then $\hat{\mathfrak g}$ is the oscillator algebra.

Consider the decomposition
\begin{equation}
\hat{\mathfrak g} = \hat{\mathfrak g}^+\oplus \hat{\mathfrak g}^0\oplus \hat{\mathfrak g}^-\,,
\label{compose}
\end{equation}
where (see Eq. (\ref{compose}))
\begin{equation}
\hat{{\mathfrak g}}^+ = t{\mathfrak g}[t], \,\,\,\,\,\,\,
\hat{\mathfrak g}^- = t^{-1}{\mathfrak g}[t^{-1}],
\,\,\,\,\,\,\, \hat{\mathfrak g}^0 = {\mathfrak g}\oplus {\mathbb C}K\,.
\end{equation}
Then $\hat{\mathfrak g}^{\geq 0} = {\mathfrak g}[t]\oplus {\mathbb C}K$ is a Lie subalgebra of $\hat{\mathfrak g}$. Suppose $\pi: \, \hat{\mathfrak g}^{\geq 0}\rightarrow {\rm End}\,W$ is a representation. One can define the {\it induced $\hat{\mathfrak g}$-module} by
$
{\rm Ind}_{\hat{\mathfrak g}^{\geq 0}}^{\hat{\mathfrak g}}
\cong U(\hat{\mathfrak g})\otimes_{U(\hat{\mathfrak g}^{\geq 0})} W.
$\footnote{
$U({\hat{\mathfrak g}})$ is the quotient algebra of the tensor
algebra
$
\bigoplus_{n=0}^{\infty}
(\overbrace{{\hat{\mathfrak g}}\otimes ...
\otimes {\hat{\mathfrak g}}}^{n-\mbox{times}})
$
by the ideal generated by elements of the form
$
[{\hat{\mathfrak g}}_1, {\hat{\mathfrak g}}_2] - ({\hat{\mathfrak g}}_1\otimes{\hat{\mathfrak g}}_2 -
{\hat{\mathfrak g}}_2 \otimes {\hat{\mathfrak g}}_1)
$.
$
{\rm CoInd}_{\widehat{\mathfrak g}^{\geq 0}}^{\widehat{\mathfrak g}}
\cong {\rm Hom}_{U(\widehat{\mathfrak g}^{\geq 0})}
(U(\hat{\mathfrak g}), W).
$
}
In this formula, $U(\hat{\mathfrak g})$ is viewed as a right $U(\hat{\mathfrak g}^{\geq 0})$-module. Then,
$
{\rm Ind}_{\hat{\mathfrak g}^{\geq 0}}^{\hat{\mathfrak g}}
\cong {S}\,(\hat{\mathfrak g}^{-})W
$
as vector spaces, ${S}\,(\hat{\mathfrak g}^{-})$ being the symmetric algebra. Each $a_n$ acts on ${\rm Ind}_{\hat{\mathfrak g}^{\geq 0}}^{\hat{\mathfrak g}}$. 
One defines the normal ordered product by
\begin{equation}
: a_kb_\ell : \, \, =
\left\{ \begin{array}{ll}
a_kb_\ell, $ \,\,\,\,\,\,\,\,\,\,\,\,\,\,\,\,\,\,\,\,\,\,
for $\ell\geq 0,
\\
\\
(-1)^{|a||b|}b_\ell a_k,\,\,\,\, {\rm for}\,\,\,\, \ell< 0.
\end{array} \right.
\end{equation}
\begin{definition}
Let us define a {\it field} on a vector space $\bf V$ as a formal power series $a(z) = \sum_{n \in {\mathbb Z}} a_n z^{-n-1}$,\, $a_n \in {\rm End}\,{\bf V}$, such that for any $v\in {\bf V}$, $a_nv=0$, for $n$ sufficiently large. Suppose $a(z)$ and $b(w)$ are two fields on a vector space $\bf V$. Then, an equality of the form
\begin{equation}
a(z)b(w) = \sum_{k=0}^{N-1}\frac{c^k(w)}{(z-w)^{k+1}}
+ : a(z)b(w) :\,
\end{equation}
{\rm (}or, simply, $a(z)b(w)\sim \sum_{k=0}^{N-1}c^k(w)/(z-w)^{k+1}$, where
$c^k(w)$ are local fields{\rm )}, is called {\it operator product
expansion} {\rm (}OPE{\rm )} of the fields $a(z)$, $b(w)$.
Let $\lim_{z\rightarrow w}\, :a(z)b(w):\,\,\, = \,\,:a(w)b(w):$, then
$: a(w)b(w) :$ is called {\it the regular part} of the OPE, while the rest of the OPE is termed as {\it the singular part}.
\end{definition}
The normal ordered product can be equivalently defined as:
$
: a(z)b(w): \, = a(z)_+b(w) + (-1)^{|a||b|}b(w)a(z)_{-},
$
where $|a|$ is the order of $a$,
\begin{equation}
a(z)_+ \, := \sum_{n\in {\mathbb Z}_{-}}a_n z^{-n-1},\,\,\,\,\,\,
a(z)_{-} \, := \sum_{n\in {\mathbb Z}_{+}\,\cup\,\{0\} }a_n z^{-n-1}\,.
\end{equation}
The formal power series
$
a(z)b(w) = \sum_{k \in {\mathbb Z}}\sum_{\ell \in {\mathbb Z}}
b_\ell a_k z^{-k-1}w^{-\ell -1}
$
always makes sense, but the limit $z\rightarrow w$ of that expression may not exist.

Let us come now to the definition of {\it vertex algebra}.
\begin{definition}
A vertex algebra consists of the following data:
\begin{itemize}
\item{}
A $\mathbb Z$-graded vector space {\rm (}the state space{\rm )} ${\bf V}=\bigoplus_{n\in{\mathbb Z}}{\bf V}_n$;
\item{}
A vector $|0\rangle$ {\rm (}is called the vacuum vector{\rm )};
\item{}
A map $Y: {\bf V} \rightarrow {\rm End}({\bf V})[[z, z^{-1}]]$, whose image lies in the set of fields. The latter is called the state-field correspondence.
\end{itemize}
\end{definition}
Let $Y(a, z) = \sum_{n \in {\mathbb Z}}a_nz^{-n-1}, a \in {\bf V}$. Define $Ta = a_{-2}|0\rangle$. Then the following axioms are required:
\begin{enumerate}
\item{}
$[T, Y(a, z)]= \partial Y(a, z)$\, (translation coinvariance);
\item{}
$Y(|0\rangle, z)= Id_{\bf V}, \, Y(a, z)|0\rangle|_{z=0}=a$\, (vacuum);
\item{}
$Y(a, z)$ and $Y(b, z)$ are local, $a, b \in {\bf V}$\, (locality).
\end{enumerate}

{\bf Heisenberg vertex operator algebra.}
A well-known vertex operator algebra is the Heisenberg vertex operator algebra $M(1)$ of dimension $d$.
Let ${{\mathfrak h}}$ be a complex vector space of dimension
$d$ with a nondegenerate symmetric bilinear form $(\,,\,).$ Let us view
${{\mathfrak h}}$ as an abelian Lie algebra and, as before (Eq. (\ref{affine})), consider the corresponding affine Lie algebra ${\hat{{\mathfrak h}}}={{\mathfrak h}}\otimes {\mathbb C}[t,t^{-1}]\oplus {\mathbb C} {K}$, with bracket
\begin{equation}
[x\otimes t^m,y\otimes t^n]=\delta_{m+n,0}(x,y)K,\,\,\,\,\,\,\,
[K,{\hat{\mathfrak h}}]=0,
\end{equation}
where $x_m =x\otimes t^m$ for $x\in \mathfrak h$ and $m\in {\mathbb Z}.$ Let us form the induced module
\begin{equation}
{M}(1)=U({\hat{{\mathfrak h}}})\otimes_{U({\mathfrak h}\otimes {\mathbb C}[t]\oplus {\mathbb C} K)}{\mathbb C},
\end{equation}
where ${\mathfrak h}\otimes {\mathbb C}[t]$ acts trivially on $\mathbb C$ and $K$ acts as 1. Let $\{\alpha_1,...,\alpha_d\}$ be an orthonormal basis of ${{\mathfrak h}}.$ Then ${M}(1)$ is isomorphic linearly to the symmetric algebra
\begin{equation}
{S}\,({\mathfrak h}\otimes t^{-1}{\mathbb C}[t^{-1}])={\mathbb C}[\alpha_i(-n)|i=1,...,d,\, n>0].
\end{equation}
Set now ${\bf 1}=1$ and $\omega= (1/2)\sum_{i=1}^d\alpha_i(-1)^2.$
Then, $M(1)$ is the Heisenberg vertex operator algebra with vacuum
${\bf 1},$ and Virasoro element $\omega$ (cf. \cite{Frenkel89}, Chapter 8). Let $O({\mathfrak h})$ be the orthogonal group of ${\mathfrak h}$.
The automorphism group of ${M}(1)$ is exactly $O({\mathfrak h})$ \cite{ZhouLectures}.

{\bf Highest weight modules for affine Kac-Moody algebras.}
Let us analyze the vertex operator algebra associated to the highest weight modules of affine Kac-Moody algebras (cf. \cite{Dong93,Frenkel92,Li96}). Let ${\mathfrak g}={\mathfrak h}+\oplus_{\alpha\in \Delta}{\mathfrak g}_{\alpha}$ be a finite dimensional simple Lie algebra with a Cartan subalgebra ${\mathfrak h}$ and the corresponding root system $\Delta.$ We fix the positive roots $\Delta_+$ and assume that $\theta$ is the longest positive root. Denote $P_+$ the set of dominant weights. Let $(\cdot ,\cdot)$ be a nondegenerate symmetric invariant bilinear form on $\mathfrak g$ such that $(\theta,\theta)=2.$
The affine Kac-Moody algebra is
$
{\hat{\mathfrak g}}={\mathfrak g}\otimes{\mathbb C}[t,t^{-1}]\oplus {\mathbb C} {K},
$
with bracket Eq.~(\ref{affine}),
where $a_m=a\otimes t^m$ for $a\in {\mathfrak g}$ and $m\in {\mathbb Z}.$ If $M$ is an irreducible ${\hat{\mathfrak g}}$-module then the center $K$ acts as a constant $k$. The latter is called the {\it level} of the module. Let $M$ be a $\mathfrak g$-module and $k$ be a complex number. The generalized Verma module of level
$k$ associated to $M$ is
\begin{equation}
{\hat M}_k=U({\hat{\mathfrak g}})\otimes_{U({\mathfrak g}\otimes{\mathbb C}[t]\oplus {\mathbb C} K)}M\,,
\end{equation}
where ${\mathfrak g}\otimes t{\mathbb C}[t] \cdot M=0,$ ${\mathfrak g}\otimes t^0$ acts on $M$ as ${\mathfrak g}$ and $K = k$ on $M$.
Let $L(\lambda)$ be the  highest weight module for $\mathfrak g$ with
highest weight $\lambda\in{\mathfrak h}$; consider set
$V(k,\lambda) = \widehat{L(\lambda)}_k$ and denote $L(k,\lambda)$ the unique irreducible quotient. Then $L(k,\lambda)$ is integrable or unitary if and only if $k$ is a nonnegative integer, $\lambda\in P_{+},$ and $(\lambda,\theta)\leq k.$ Denote the dual Coxeter number of $\mathfrak g$ by $h^{\vee}.$ Then $h^{\vee}$ can be defined as  $\sum_{\alpha\in\Delta}(1/2)(\alpha,\alpha)=d h^{\vee}$,
where $d$ is the rank of $\mathfrak g.$

Any ${\hat{\mathfrak g}}$-quotient module of $V(k,0)$ is a vertex operator algebra, if $k+h^{\vee}\ne 0.$ $L(k,0)$ is rational if and only if $k$ is a nonnegative
integer. In this case, the irreducible $L(k,0)$-modules are
exactly the level $k$ unitary highest weight modules. In
particular, if $\mathfrak g$ is the Lie algebra of type $D_d,$ then $\hat{\mathfrak g}$ is the affine Lie algebra $D_{d}^{(1)}$ which has exactly four level 1 unitary highest weight modules. These modules are used in the
construction of elliptic genera. The automorphism groups of $V(k,0)$ and $L(k,0)$  are exactly the automorphism group of the Lie algebra $\mathfrak g$ (cf. \cite{Dong98}).

\section{Superconformal vertex algebras}
\label{Super}

Superconformal vertex algebras are familiar objects in superstring theories. We will present them here in a more mathematical way than the one familiar in the physical literature.

{\bf Wedging operators.}
Let us first recall some formulas on semi-infinite wedge products.
Let set $e_S := e_{s_1}\wedge e_{s_2}\wedge \ldots$, denote by $\Lambda ^{\infty/2}(\cH)$ the vector space with $\{e_S\}$ as an orthonormal basis. The vector $e_{-1/2}\wedge e_{-3/2}\wedge e_{-5/2}\wedge \cdots$
is called {\it the vacuum vector}. Take $\ell \in {\mathbb Z}+ 1/2$ and
denote by $\psi_{\ell}$ the wedge product by $e_\ell$ on $\Lambda^{\infty/2}(\cH)$, and $\psi_{\ell}^*$ the wedge product by its adjoint. One can verify the following relations:
\begin{equation}
[\psi_\ell, \, \psi_s] = [\psi_\ell^*, \, \psi_s^*] = 0,\,\,\,\,\,\,\,\,
[\psi_\ell, \, \psi_s^*] = \delta_{\ell, s}\,.
\end{equation}
For $\ell > 0$ their action on the vacuum vector is
\begin{equation}
\psi_\ell \mid 0\rangle \,= \,e_\ell \wedge e_{-1/2}\wedge e_{-3/2}\wedge \cdots,
\,\,\,\,\,\,\,
\psi_{-\ell}^* \mid 0\rangle \,= \, (-1)^{\ell-1/2} e_{-1/2}\wedge
e_{-3/2}\wedge\cdots \wedge \widehat{e}_{-\ell}\wedge \cdots
\end{equation}
In the physical terminology both $\psi_\ell$ and $\psi_{-\ell}^*$ for $\ell > 0$ are {\it creators}: $\psi_\ell$ {\it creates} an excitation of energy $\ell$ and $\psi_{-\ell}^*$ {\it creates} a anti-excitation of energy $\ell$. (For $\ell > 0$ dually, both $\psi_\ell^*$ and $\psi_{-\ell}$ are {\it annihilators}.) The generating series of the above operators can be written as follows:
\begin{equation}
b(z)\,:=\, \sum_{\ell \in {\mathbb Z}+1/2}b_\ell z^{-\ell+1/2},\,\,\,\,\,\,\,\,
c(z)\,:=\, \sum_{\ell \in {\mathbb Z}+1/2}c_\ell z^{-\ell+1/2}\,,
\label{series}
\end{equation}
$b_\ell = \psi_\ell^*,\, c_\ell = \psi_{-\ell}$.
In Eq.~(\ref{series}) the series are formally regarded as meromorphic fields of operators on ${\mathbb C}^*={\mathbb C}/\{0\}$, acting on $\Lambda^{\infty/2}(\cH)$.

Let $F$ be the Grassmannian algebra generated by $\{b_\ell,\,c_\ell\}_{\ell< 0}$. There is a natural isomorphism $F\rightarrow \Lambda^{\infty/2}(\cH)$ defined by
$b_{\ell_1}\cdots b_{\ell_n}c_{s_1}\cdots c_{s_n} \mapsto
b_{\ell_1}\cdots b_{\ell_n}c_{s_1}\cdots c_{s_n}\mid 0\rangle$, \,
$\ell_1, \ldots, \ell_m$, $s_1, \ldots, s_n \in {\mathbb Z}_{-}+1/2$.
Thus, under this isomorphism, we can also regard $b(z)$ and $c(z)$ as fields on $F$.

{\bf Example: Virasoro fields.} The energy operator can be
expressed in terms of normally ordered products as follows:
\begin{equation}
H = \sum_{r \in {\mathbb Z}_+-1/2}rc_{-r}b_r -
\sum_{r \in {\mathbb Z}_- + 1/2}rb_rc_{-r} =
\sum_{r \in {\mathbb Z}+1/2}r : c_{-r}b_r :\,.
\end{equation}
It can be generalized as follows. Introduce the field
\begin{equation}
L(z) : = (1/2): \partial_zb(z)c(z) : \, + \,\, (1/2) : \partial_zc(z)b(z) :\,.
\end{equation}
If $L(z) = \sum_{n \in {\mathbb Z}} L_n z^{-n-2}$, then it can be
shown that $L_n = (1/2)\sum_{r+s = n}(r-s): c_sb_r :$, and in
particular $H= L_0$.
\begin{definition}
A field $L(z)$ is called a Virasoro field
{\rm (}energy momentum tensor{\rm )} if it satisfies the following OPE:
\begin{equation}
L(z)L(w) \sim \frac{L'(\omega)}{z-\omega} + \frac{2L(\omega)}{(z-\omega)^2} + \frac{c/2}{(z-\omega)^4}\,,
\end{equation}
where the constant $c$ is called the central charge of the Virasoro field. If a field $a(z)$ satisfies
$
L(z)a(\omega) \sim ha(\omega)/(z-\omega)+ \partial_{\omega}a(\omega)/(z-\omega)^2 + O((z-\omega)^{-3})
$
then $a$ has a conformal weight $h$. If
$
L(z)a(\omega) \sim ha(\omega)/(z-\omega)+ \partial_{\omega}a(\omega)/(z-\omega)^2
$
then $a$ is primary of conformal weight $h$.
\end{definition}
Suppose $L(z)$ is a Virasoro field of central charge $c$ and, as before,
$L(z) = \sum_{n \in {\mathbb Z}} L_n z^{-n-2}$, then
\begin{equation}
[L_m , L_n] = (m-n)L_{m+n} + \frac{m^3-m}{12}\delta_{m, -n}c\,.
\end{equation}
If there is a Virasoro field of central charge $c$ on $V$ , then V is a representation of the Virasoro algebra (of central charge c).

We recall connections betwen the theory of higher-weight modules over the Virasoro algebra and the partition function of conformal field theory. In Sects. \ref{Elliptic} - \ref{Symmetric}, and specially in Sect. \ref{Identities}, we shall discuss these connections in details.
We start with very well known Lie algebra ${\mathfrak g}{\mathfrak l}(n, {\bf k})$
\footnote{
The symbol $\bf k$ denotes the field of real numbers $\mathbb R$
or the field $\mathbb C$ of complex numbers. In particular,
${\mathfrak g}{\mathfrak l}(n, {\mathbb C})$ is the Lie algebra of all complex $n\times n$ matrices with the operation $A, B \mapsto [A, B]
= AB - BA$.
}
.
Results for ${\mathfrak g}{\mathfrak l}(n, {\bf k})$
survive the passage to the limit $n\rightarrow \infty$, if one assumes that ${\mathfrak g}{\mathfrak l}(\infty, {\bf k})$ is the Lie algebra of infinite finitary matrices, which means
$\bigcup_n {\mathfrak g}{\mathfrak l}(n, {\bf k})$. In this remark we deal with the Lie algebra ${\mathfrak g}{\mathfrak l}_{\widehat{\cJ}} (\bf k)$ of generalized Jacobian matrices\footnote{
The bilateral matrix $\Vert a_{ij}\Vert_{i,j\in {\mathbb Z}}$ is called a generalized Jacobian matrix if it has a finite number of nonzero diagonals (that is, if there exists a positive $N$ such that $a_{ij}=0$ for $\vert j-i\vert> N$). It is clear that the set of generalized Jacobian matrices constitutes a Lie algebra, with respect to the usual commutation rule.
}.
The algebra ${\mathfrak g}{\mathfrak l}_{\widehat{\cJ}} (\bf k)$ can be considered as a nontrivial one-dimensional central extension of the Lie algebra ${\mathfrak g}{\mathfrak l}_{\cJ}({\bf k})$ (for details, see \cite{Fuks}). It is obvious that ${\mathfrak g}{\mathfrak l}_{\cJ}({\bf k})\supset {\mathfrak g}{\mathfrak l}_{\cJ}(\infty, {\bf k})$.
The importance of this Lie algebra stems from the following facts:
\begin{itemize}
\item{}
Many of the classical constructions of the theory of representations
of the Lie algebra ${\mathfrak g}{\mathfrak l}_{\cJ}({\bf k})$ can be also applied to the algebra ${\mathfrak g}{\mathfrak l}_{\widehat{\cJ}}({\bf k})$. This creates a sizable supply of
${\mathfrak g}{\mathfrak l}_{\widehat{\cJ}}({\bf k})$-modules.
\item{}
Important infinite-dimensional Lie algebras can be embedded in ${\mathfrak g}{\mathfrak l}_{\widehat{\cJ}}({\bf k})$. Thus, the already mentioned representations of ${\mathfrak g}{\mathfrak l}_{\widehat{\cJ}}({\bf k})$  become representations of these algebras.
\item{}
{\it Examples}. The subalgebra of ${\mathfrak g}{\mathfrak l}_{\cJ} ({\bf k})$ composed of $n$-periodic matrices, $\Vert a_{ij}\Vert$ with $a_{i+n, j+n}=a_{ij}$, is isomorphic to the algebra of currents \cite{Fuks}\footnote{
Recall that the space of smooth maps $X \rightarrow {\mathfrak g}$,
where $X$ is a smooth manifold and ${\mathfrak g}$ is a finite-dimensional Lie algebra, with the $\cC^\infty$-topology and the commutator $[f, g](x) = [f(x), g(x)]$, is the (topological) {\it current} Lie algebra and is denoted by ${\mathfrak g}^X$. Together with the algebra ${\mathfrak g}^{S^1}$ \,($X=S^1$) one can consider its subalgebra
$({\mathfrak g}^{S^1})^{\rm pol}$, consisting of maps described by trigonometric polynomials. For any commutative associative algebra $A$, the tensor product ${\mathfrak g}\otimes A$ is a Lie algebra with respect to the commutators $[g_1\otimes a_1, g_2\otimes a_2] = [g_1, g_2]\otimes a_1 a_2$; also $({\mathfrak g}^{S^1})^{\rm pol} = {\mathfrak g}\otimes
{\mathbb C}[t, t^{-1}]$.
}.
A non-trivial central extension of
${\mathfrak g}^X$ -- a Kac-Moody algebra -- is embedded in ${\mathfrak g}{\mathfrak l}_{\widehat{\cJ}} ({\bf k})$. The Lie algebra
${L}^{\rm pol} = {\mathbb C}({\rm Vect}\,S^1)^{\rm pol}$ of complex polynomial vector fields on the circle can be embedded in
${\mathfrak g}{\mathfrak l}_{\cJ}({\bf k}= {\mathbb C})$. Recall that
${L}^{\rm pol}$ has a basis ${\bf e}_i$ and commutators of the form
\begin{equation}
[{\bf e}_i, {\bf e}_j]  =  (i-j) {\bf e}_{i+j}\,\,\,\,\,
(j\in {\mathbb Z}),
\,\,\,\,\,
{\bf e}_j  =   -z^{j+1}d/dz \,\,\,\,\,
{\rm on}\,\,\,\,\, {\mathbb C}\setminus \{0\}\,.
\label{basis}
\end{equation}
(The cohomologies of the algebra ${L}^{\rm pol}$ are known; in particular, $H^2({L}^{\rm pol})= {\mathbb C}$.)
The Virasoro algebra is a Lie algebra over $\mathbb C$
with basis $L_n$ ($n\in {\mathbb Z}$), $c$. Because of
Eq.~(\ref{basis}), the Lie Virasoro algebra is a (universal) central extension of the Lie algebra of holomorphic vector fields on the punctured complex plane having finite Laurent series. It is for this reason that the Virasoro algebra plays a key role in conformal field theory.
\end{itemize}
A remarkable link between the theory of highest-weight modules
over the Virasoro algebra, conformal field theory and statistical
mechanics was discovered in \cite{Belavin1,Belavin2,Dowker}. Here
we briefly note  some elements of the representation theory of the
Virasoro algebra which are, in fact, very similar to those for
Kac-Moody algebras. Let us consider the highest representation of
the Virasoro algebra. Let $M(c, h)\, (c, h \in {\mathbb C})$ be
the Verma module over the Virasoro algebra (see, for example,
\cite{Kac}). The {\it conformal central charge} $c$ acts on $M(c,
h)$ as $cI$. As $[{\bf e}_0, {\bf e}_{-j}] = n {\bf e}_{-j}$,
${\bf e}_0$ is diagonalizable on $M(c, h)$, with spectrum $h+
{\mathbb Z}_{+}$ and eigenspace decomposition given by: $ M(c, h)
=\bigoplus_{j\in {\mathbb Z}_{+}} M(c, h)_{h+j}, $ where $M(c,
h)_{h+j}$ is spanned by elements of the basis $\{{\bf
e}_{-j_k}\}_{k=1}^n$ of $M(c, h)$. The number $ Z_j = {\rm dim}\,
M(c, h)_{h+j}, $ is the {\it classical partition function}. This
means that the Konstant partition function for the Virasoro
algebra is the classical partition function. On the other hand,
the partition functions can be rewritten in the form
\begin{equation}
{\rm Tr}_{M(c, h)}\, q^{{\bf e}_0} := \sum_{\lambda}{\rm dim}\,
M(c, h)_{\lambda}\,q^{\lambda} = q^h\prod_{j=1}^\infty (1-q^j)^{-1}\,.
\label{ch}
\end{equation}
The series ${\rm Tr}_{\mathcal V}\,q^{{\bf e}_0}$ is called the formal character of the Virasoro-module ${\mathcal V}$.

{\bf State-field correspondence: free fermion space.}
Let $I= 0\leq i_1\leq \cdots \leq i_m$ and
$J= 0\leq j_1\leq \cdots \leq j_m$, and let us set
$v_{IJ} = b_{-i_1-1/2}\cdots b_{-i_m-1/2}c_{-j_1-1/2}\cdots
c_{-j_n-1/2}$ (they form a basis of $F$). Set also
$v_{IJ}(z) = \, : \partial^{i_1}b(z)\cdots \partial^{i_m}b(z)
\partial^{j_1}c(z)\cdots \partial^{j_n}c(z) :$.  One can check that
$v_{IJ}(z)\mid 0\rangle \!\mid_{z=0} = v_{IJ}$.

{\bf State-field correspondence: bosonic space.}
Let $I = i_1\geq \cdots \geq i_n \geq 0$, and define
$\alpha_I = \alpha_{-i_1-1}\cdots \alpha_{-i_n-1}$. Set also
$\alpha_I(z) = \,:\partial^{i_1}\alpha (z)\cdots \partial^{i_n}\alpha (z) :$. As before, one can check that $\alpha_I(z)\mid 0\rangle \!\mid_{z=0}= \alpha_I$.

Recall that the oscillator algebra is spanned by $\{\alpha_{m}\}_{m\in {\mathbb Z}}$ and the central element $h$ satisfying: $[\alpha_{m}, \alpha_{n}] = m \delta_{m, -n}h$.
The oscillator algebra acts on $U = {\mathbb C}[\alpha_{-1}, \alpha_{-2}, \ldots]$ as follows. The central element $h$ acts as multiplication by a constant $\hbar$, $\alpha_{0}$ acts as $0$.
If $n > 0$, then $\alpha_{-n}$ acts as multiplication by $\alpha_{-n}$,
while $\alpha_{n}$ acts as $n\hbar \partial/\partial\alpha_{-n}$.
Let $\alpha (z) := \sum_{n\in {\mathbb Z}}\alpha_{n}z^{-n-1}$, then $\alpha$ is a field on $U$, and $\alpha (z)\alpha (w)\sim (z-w)^{-2}.$

The Virasoro field on the bosonic space is defined as
$
L(z) :=\, (1/2) :\, \alpha(z)\alpha(z)\,:
$. By Wick's theorem it is straightforward to see that
$
L(z)L(\omega) \sim L'(\omega)/(z-\omega) + 2L(\omega)/(z-\omega)^2
+ (z-\omega)^{-4}.
$
This means that $L(z)$ is a Virasoro field of central charge 2 on the bosonic space.
The following theorem enables one to construct vertex algebras.
\begin{theorem} {\rm (J. Zhou \cite{ZhouLectures}
\footnote{
This theorem is due to V. Kac and is known as the Kac Reconstruction Theorem. It first appeared in \cite{Kac95} and is also presented in the book \cite{Kac96}.
}
, Theorem 9.1)} Let $\bf V$ be a graded vector space, $|0\rangle \in {\bf V}_0$ and let $T$ be an endomorphism of ${\bf V}$ of degree $0$. Suppose that $\{a^\alpha(z)\}_{\alpha \in I}$ is a collection of fields on ${\bf V}$ such that:
$[T, a^\alpha(z)] = \partial_z a^\alpha (z)\, {\rm (}\alpha \in
I{\rm )}$; $T|0\rangle = 0,\, a^\alpha(z)|0\rangle|_{z=0} = a^\alpha \,
{\rm (}\alpha \in I{\rm )}$, where $a^\alpha$ are linear independent
; $a^\alpha (z)$ and $a^\beta (z)$ are mutually local\,\, ${\rm (}\alpha, \beta \in I{\rm )}$. The vectors $a_{-j_1-1}^{\alpha_1} \cdots a_{-j_n-1}^{\alpha_n}|0\rangle$ with $j_k\geq 0$ span ${\bf V}$.
Then the formula
\begin{equation}
Y(a_{-j_1-1}^{\alpha_1} \cdots a_{-j_n-1}^{\alpha_n}|0\rangle, z)
= \,\,\, :\, \partial^{(j_1)}a^{\alpha_1}(z) \cdots  \partial^{(j_n)}a^{\alpha_n}(z) :
\end{equation}
defines a unique structure of a vertex algebra on ${\bf V}$ such that
$|0\rangle$ is the vacuum vector, $T$ is the infinitesimal translation operator, $Y(a^\alpha, z) = a^\alpha(z), \, \alpha\in I$, and
$\partial^{(j)}\equiv \partial^{j}/j!$.
\end{theorem}
Going back to highest weight representations, one can consider the following special induced module. Let $\pi$ be the one-dimensional representation such that $\mathfrak g$ acts as zero operator and $h$ acts as multiplication by a constant $k$. Denote by ${\mathfrak V}^k({\mathfrak g})$ the induced representation. Then, as vector space,
${\mathfrak V}^k({\mathfrak g})$ is spanned by elements of the form
$
a_I^J = a_{-i_1-1}^{j_1} \cdots a_{-i_n-1}^{j_n}\,,
$
where $a^{j_1}, \ldots , a^{j_n}\in {\mathfrak g}, \, i_1, \ldots, i_n\geq 0$. Then ${V}^k({\mathfrak g})$ is a vertex algebra, with
\begin{equation}
Y(a_I^J, z) = \,\,\,: \partial^{(i_1)}a^{j_1}(z) \cdots  \partial^{(i_n)}a^{j_n}(z):\,.
\end{equation}

{\bf Cohomological aspects.} An {\it ideal} $J$ of a vertex algebra $V$ is a $T$-invariant subspace not containing $|0\rangle$, such that
$
a_{(n)}J\subset J, \,\,\, \forall a \in V
$. Note that because of the quasi-symmetry property
$Y (a; z)v = (-1)^{|a| |b|}e^{zT}Y(v, -z)a$, we have
$a_{(n)}V \subset J,\,\,\, \forall a \in J$. As a consequence,
the quotient space $V/J$ has an induced structure of a vertex algebra.
A derivation of degree k on a vertex algebra $V$ is a linear map $\delta : V \rightarrow V$ of degree $k$, such that
\begin{equation}
\delta (a_{(n)}b) = (\delta a)_{(n)}b + (-1)^{k|a|}a_{(n)}\delta b\,,
\end{equation}
for all $a, b \in V,\,  n \in {\mathbb Z}$. A derivation $\delta$ of degree 1 is a {\it differential} if $\delta^2 = 0$. The standard
cohomology of a differential is (by definition)
$H(V, \delta) = {\rm Ker}\, \delta / {\rm Im} \,\delta$.
The following statement holds: If $\delta$ is a differential on a vertex algebra $V$, then ${\rm Ker}\, \delta$ is a
subalgebra of $V$, and ${\rm Im}\, \delta$ is an ideal ${\rm Ker}\, \delta$. Hence $H(V, \delta)$ has an induced structure of a vertex algebra.

\subsection{Bosons and fermions from vector spaces with inner products}

Let us consider the construction of vertex algebras from a vector space with an inner product $g(\cdot,\cdot)$. Define the Lie algebras \begin{equation}
{\cA}(X, g) = X[t; t^{-1}]\oplus {\mathbb C}K\,,
\end{equation} with commutation relations
\begin{equation}
[a_m,\, b_n] = mg(a,\, b)\delta_{m, -n}K,\,\,\,\,\,\,\,\,
[K, {\cA}(X, g)] = 0\,,\,\,\,\,\,\,\, a, b \in X\,,
\end{equation}
where $a_m$ stands for $at^m$. Suppose that ${\mathfrak V}$ is a representation of the Lie algebra $\cA(X, g)$, such that $a(z) = \sum_{n\in {\mathbb Z}}a_nz^{-n-1}$ is a field for any $a\in X$ (this representation is called {\it a field representation}). $\{a(z)\}$ is a collection of mutually local fields with OPE: $a(z)b(w)\sim kg(a, b)/(z-w)^2$. A well-known field representation of $\cA(X, g)$ is the one on
\begin{equation}
B(X, g) =  S(\bigotimes_{n\in {\mathbb Z}_{-}}t^nX) =
\bigotimes_{n \in {\mathbb Z}_{+}} S(t^{-n}X)\,.
\end{equation}
Recall that $K$ acts as multiplication by $k$; for any element $a\in X$,\, $a_n$ acts as symmetric product by $a_n$ if $n\in {\mathbb Z}_{-}$, and as $kn$ times the contraction by $a_{-n}$ if $n \in{{\mathbb Z}_+\cup \{0\}}$.
Denote the element 1 in $B(X, g)$, then the bosonic space is spanned by elements of the form
$
a^1_{-j_1-1} \cdots a^1_{-j_m-1}|0\rangle,
$
where $a^1, \cdots, a^m\in X$ and $j_1, \cdots, j_m \geq 0$. The space $B(X, g)$ is called the ({\it bosonic}) {\it Fock space}, and $a_n$
is called a {\it creation operator} if $n <0$, and an {\it annihilation operator} if $n\geq 0$.

{\bf Clifford algebras.} Let $(X, g)$ be a finite dimensional
complex vector space with an inner product $g(\cdot \,,\, \cdot)$.
Denote by $\widehat{\cA}_{NS}(X,g)$ and $\widehat{\cA}_{R}(X,g)$
the Lie superalgebras with even parts ${\mathbb C}K$, and odd
parts $\bigoplus_{r\in {\mathbb Z}+ 1/2}{\mathbb C}\varphi_r$ and
$\bigoplus_{r \in {\mathbb Z}}{\mathbb C}\varphi_r$, respectively,
which satisfy the commutation relations $[\varphi_r,\,\psi_s] =
g(\varphi, \psi)\delta_{r, -s}K$,\, $[K, \varphi_r] = 0$. Here
$\varphi, \psi \in X$,\, $r, s \in {\mathbb Z}+ 1/2$ for
$\widehat{\cA}_{NS}(X,g)$, and $r, s \in {\mathbb Z}$ for
$\widehat{\cA}_{R}(X,g)$ respectively. Then, a field representation
of $\widehat{\cA}_{NS}(X,g)$ is a representation $V$ such that for
all $\varphi \in X$ ,\, $\varphi(z)= \sum_{r\in {\mathbb
Z}+1/2}\varphi_rz^{-r-1/2}$. Thus, we have a collection of mutually
local field with the following OPE: $\varphi(z)\psi(w)\sim
kg(\varphi, \psi)/(z-w)$.

{\bf Free fermions.}
For $n \in {\mathbb Z}$ there is a field representation of $\widehat{\cA}_{NS}(X,g)$ on
\begin{equation}
F(X, g) = \Lambda(\bigotimes_{n \in {\mathbb Z}_+}t^{-n+1/2}X)
= \bigotimes_{n \in {\mathbb Z}_+}\Lambda(t^{-n+1/2}X)\,.
\end{equation}
Note that $K$ acts as $k Id$, and for any element $\varphi \in X$,
$\varphi_r$ acts as exterior product by $\varphi_r$ if $r \in {\mathbb Z}_{-}$, and as $k$ times the contraction by $\varphi_{-r}$ if $r \in {\mathbb Z}_{+}$. Let us denote the unit element in $F(X, g)$ as
$|0\rangle$\, then $F(X, g)$ is spanned by elements of
the form:
$
\varphi^1_{-j_1-1/2}\cdots \varphi^m_{-j_m-1/2} =
\varphi^1_{-j_1-1/2}\cdots \varphi^m_{-j_m-1/2}\mid 0\rangle,
$
where $\varphi^1, \cdots, \varphi^m \in X$ and $j_1, \cdots, j_m \geq 0$. The space $F(X, g)$ is called the {\it fermionic space}, moreover
$\{\varphi_{-r}\mid \varphi \in X, r\in {\mathbb Z}+1/2,\, r>0\}$
are the creation operators, and
$\{\varphi_{r}\mid \varphi \in X, r\in {\mathbb Z}+1/2,\, r>0\}$
are the annihilation operators.
\begin{itemize}
\item{}
There is a structure of conformal vertex algebra on
$B(X, g)$ defined by
\begin{equation}
Y(a^1_{-j_1-1} \cdots a^m_{-j_m-1}, z)
= \,\,\, :\partial^{(j_1)}a^1(z) \cdots \partial^{(j_m)}a^m(z):
\end{equation}
and $j_1, \cdots, j_m \geq 0$, with conformal vector
$\nu_B = (1/2)\sum_i a^i_{-1}a^i_{-1}$
of central charge $c= \dim X$, where $\{a^\ell\}$ is an orthonormal basis of $X$.
\item{}
There is a structure of a conformal vertex algebra on $F(X, g)$
defined by
\begin{equation}
Y(\varphi^1_{-j_1-1/2} \cdots \varphi^m_{-j_m- 1/2}, z)
= \,\,\, :\partial^{(j_1)}\varphi^1(z) \cdots \partial^{(j_n)}\varphi^m(z):
\end{equation}
$\{\varphi^\ell\}$ is an orthonormal basis of $X$, and integers $j_1, \cdots, j_m \geq 0$, with conformal vector
$\nu_F = (1/2)\sum_i \varphi^i_{-3/2}\varphi^i_{- 1/2}$
of central charge $c = (1/2) \dim X$.
\end{itemize}
Similar results can be obtained for a field representation of $\widehat{\cA}_{R}(X,g)$ (see Sect. \ref{Norm}, Eq.~(\ref{Pol1})).
Any commutative associative algebra $V$ over $\mathbb C$
is a vertex operator algebra with ${\bf 1} =1,$ $\omega=0$ and
$Y(u,z)v=uv$ for $u,v\in V.$ In particular, $\mathbb C$ itself is a
vertex operator algebra.

\subsection{Vertex operator algebra bundles and modules}
\label{Norm}

For a holomorphic vector bundle $\cE$ on $X$ and a formal variable $z$ we use the following identities
\begin{eqnarray}
S_q \left( z{\cE} \right) & = & 1 \: \oplus \: z q {\cE} \: \oplus \:
z^2 q^2 \mbox{Sym}^2 {\cE} \: \oplus \:
z^3 q^3 \mbox{Sym}^3 {\cE} \: \oplus \: \cdots = S_{zq} {\mathcal E}\,,
\\
\Lambda_q \left( z {\cE} \right) & = & 1 \: \oplus \: z q {\cE}
\: \oplus \:
z^2 q^2 \mbox{Alt}^2 {\cE} \: \oplus \:
z^3 q^3 \mbox{Alt}^3 {\cE} \: \oplus \: \cdots =  \Lambda_{zq} {\cE}\,.
\end{eqnarray}
Similarly,
\begin{equation}
S_q \left( z {\cE} \right)^{ {\mathbb C} } =
S_q \left( z {\cE}\right) \otimes S_q \left(
\overline{z} \overline{ {\cE} } \right),\,\,\,\,\,\,\,
\Lambda_q \left( z {\cE}\right)^{ {\mathbb C} } =
\Lambda_q \left( z{\cE} \right) \otimes \Lambda_q
\left( \overline{z} \overline{ {\cE} } \right)\,.
\end{equation}
These identities have good multiplicative properties and it elements should be understood as elements of the $K$-theory of the underlying space.
\begin{eqnarray}
S_q \left( {\mathcal E} \oplus {\mathcal F} \right)  & = &
\left( S_q {\mathcal E} \right) \otimes \left( S_q {\mathcal F} \right), \,\,\,\,\,\,\,
S_q \left( {\mathcal E} \ominus {\mathcal F} \right) =
\left( S_q {\mathcal E} \right) \otimes \left(
S_q {\mathcal F} \right)^{-1}\,,
\label{ident1}
\\
\Lambda_q \left( {\mathcal E} \oplus {\mathcal F} \right) & = &
\left( \Lambda_q {\mathcal E} \right) \otimes
\left( \Lambda_q {\mathcal F} \right), \,\,\,\,\,\,\,
\Lambda_q \left( {\mathcal E} \ominus {\mathcal F} \right) =
\left( \Lambda_q {\mathcal E} \right) \otimes
\left( \Lambda_q {\mathcal F} \right)^{-1}\,.
\label{ident2}
\end{eqnarray}
In Eqs. (\ref{ident1}), (\ref{ident2}) we have used the facts that
\begin{eqnarray}
\mbox{Sym}^n ({\mathcal E} \oplus {\mathcal F}) & = &
\bigoplus_{i=0}^n \, \mbox{Sym}^i({\mathcal E}) \otimes
\mbox{Sym}^{n-i}({\mathcal F})\,, \\
\mbox{Alt}^n ({\mathcal E} \oplus {\mathcal F}) & = &
\bigoplus_{i=0}^n \, \mbox{Alt}^i({\mathcal E}) \otimes
\mbox{Alt}^{n-i}({\mathcal F})\,.
\end{eqnarray}
In the case of a line bundle ${\mathcal L}$, we have
$
S_q {\mathcal L} = 1 \bigoplus_{n\in {\mathbb Z}_+} q^n {\mathcal L}^n
= (1 \ominus q {\mathcal L})^{-1} = ( \Lambda_{-q} {\mathcal L})^{-1},
$
 and therefore
$\left( S_q {\cE} \right)^{-1} = \Lambda_{-q} {\cE}$
for any vector bundle ${\cE}$, and similarly
$\left( \Lambda_q {\cE} \right)^{-1} = S_{-q} {\cE}$.

{\bf Examples.} Let us review some well known examples of vertex operator algebra bundles which have been used in the literature to study the elliptic genus and the Witten genus.
If $X$ is a Riemannian manifold, then the transition functions of the complex tangent bundle $T_{\mathbb C}X$ lie in the special orthogonal group $SO(d)$, where $d$ is the dimension of $X$. Then
$\bigotimes_{n\in {\mathbb Z}_+}S_{q^n}(T_{\mathbb C} X)$ is a $M(1)^{SO(d)}$-bundle.
As before, $M(1)$ is the Heisenberg vertex operator algebra of dimension $d$, with $SO(d)$ as a subgroup of Aut$(M(1))$, and $M(1)^{SO(d)}$ is the set of $SO(d)$-invariants of $M(1)$, which is a vertex operator subalgebra of $M(1)$. Similarly, $\bigotimes_{n\in {\mathbb Z}_+\cup \{0\}} \Lambda_{q^{n+1/2}}(T_{\mathbb C} X)$ is an $L(1,0)^{SO(d)}$-bundle where $L(1,0)$ is the level one module for the affine algebra
$D_{d/2}^{(1)}$. In this case we assume that $d$ is even.
If $X$ is further assumed to be a spin manifold, we denote the spin bundle by $\cS$. Then
$\cS\otimes\bigotimes_{n\in {\mathbb Z}_+}\Lambda_{q^{n}}(T_{\mathbb C} X)$ is also a $L(1,0)^{SO(d)}$-bundle.

{\bf Polarization.} Suppose the pair $(X, g)$ to be a finite dimensional real (complex) vector space, with an inner product $g: X \otimes X \to {\mathbb R} \, ({\mathbb C})$. This means that
$g(a; b) = g(b; a)$ for $a, b \in X$, and if $a \neq 0$; then
$g(a,\,\cdot) : X \rightarrow {\mathbb R}\, ({\mathbb C})$ is a nontrivial linear function on $X$. Let us call {\it polarization} of a vector space $X$ with an inner product $g$ a decomposition
$X = X^{'}\oplus X^{''}$, such that
$g(a_1,\, a_2) = g(b_1,\, b_2) = 0$ for $a_1, a_2 \in X^{'},\, b_1, b_2 \in  X^{''}$. It is clear that $g$ induces an isomorphism
$X^{''} \cong (X^{'})^*$. Let $X = X_R\otimes {\mathbb C}$ and
$g= g_R\otimes {\mathbb C}$. Regard $X$ as a complex vector space $X_c$, then $X^{'} \cong X_c$ and $X^{''} \cong {\overline{X_c}}$.
For a polarization $X = X^{'}\oplus X^{''}$, define the spaces
\begin{eqnarray}
\!\!\!\!\!\!\!\!
{F}_{R}(X, g)\!& = & \!
\Lambda(\!\!\!\!\bigoplus_{n \in {\mathbb Z}_{+}\cup \{0\}}\!\!\!t^{-n}X^{'})\otimes
\Lambda(\bigoplus_{n \in {\mathbb Z}_{+}} t^{-n}X^{''})
=
\Lambda(X^{'})\bigotimes_{n\in {\mathbb Z}_+}\Lambda(t^{-n}X^{'})
\bigotimes_{n\in {\mathbb Z}_+}\Lambda(t^{-n}X^{''}),
\label{Pol1}
\\
\!\!\!\!\!\!\!\!
F_{NS}(X, g)\! & = &\!
\bigotimes_{n \in {\mathbb Z}_{+}} \Lambda(t^{-n+ 1/2}X^{'})
\bigotimes_{n \in {\mathbb Z}_{+}} \Lambda(t^{-n+ 1/2}X^{''})\,.
\label{Pol2}
\end{eqnarray}

{\bf Characters and $K$-theories of vertex algebra bundles.}
Let $V$ be a vertex operator algebra such that the eigenvalues of $L_0$ form a countable set $\{h_1, h_2, \ldots \}$ on $\mathbb C$, and all eigenspaces are finite dimensional. Let
$
V = \bigoplus_{n\in {\mathbb Z}_+} V^{h_n}
$
be the eigenspace decomposition of $L_0$ on $V$. By definition, the character of $V$ is
\begin{equation}
{\rm ch}\,(V; q) = q^{- c/24}\,{\rm STr}\,q^{L_0} =
q^{- c/24} \sum_{n}{\rm STr}(Id \,|\, V^{h_n})
q^{h_n}\,,
\end{equation}
where ${\rm STr}$ is the supertrace which is just the ordinary trace in the even subspace, and the negative of the ordinary trace in the odd subspace.
The following auxiliary notation will be used:
\begin{equation}
G(V; q) = q^{-c/24}\sum_{n}V^{h_n}d^{h_n}.
\end{equation}
Therefore, ${\rm ch}(V , q)$ can be obtained by taking the supertrace of the identity map on $G(V , q)$ term by term.
Suppose V is a vertex operator algebra with a $U(1)$ current, and that the eigenvalues of $L_0$ and $J_0$ are two countable subsets of ${\mathbb C}$, $\{h_1, h_2, \ldots \}$ and $\{j_1, j_2, \ldots \}$, respectively.
Let $V = \bigoplus_{n, m \in {\mathbb Z}_+} V^{h_n, j_m}$
be the decomposition of $V$ into common eigenspaces of $L_0$, and
$J_0$. Suppose that each $V^{h, j_m}$ is finite dimensional. Then the character with $U(1)$ charge of a conformal vertex algebra $V$ with $U(1)$ current is defined by \cite{ZhouLectures}
\begin{eqnarray}
G(V; q, y) & = & \sum_{n, m}V^{h_n, j_m} q^{h_n-c/24}(-y)^{j_m}\,,
\\
{\rm ch}\,(V; q, y) & = & {\rm Tr}\, q^{L_0- c/24}(- y)^{J_0}
= \sum_{n, m}{\rm Tr}\,(Id \,|\, V^{h_n, j_m}) q^{h_n-c/24}(-y)^{j_m}\,.
\end{eqnarray}
If the $\mathbb Z$-grading given by the eigenspace decomposition of $J_0$ coincides with the $\mathbb Z$-grading on V, one has
${\rm ch}\,(V; q, 1) = {\rm ch}\,(V; q)$.
Finally note that the charged character for the fermionic Fock space is
\cite{ZhouLectures}:
\begin{equation}
{\rm ch}\,(V; q, y) = q^{-c_\lambda/24}\prod_{j\in {\mathbb Z}_+}
(1- yq^{j+1-\lambda})(1- y^{-1}q^{j+\lambda}),
\end{equation}
with respect to the conformal vector $\nu_\lambda$.

Denote by Aut$(V)$ the automorphism group of a vertex algebra $V$. Let $X$ be a smooth topological space. A vertex algebra bundle with fiber $V$ over $X$ is a vector bundle $\pi: \, E\rightarrow X$ with fiber $V$
such that the transition functions lie in  Aut$(V)$. (One can similarly define conformal and superconformal vertex algebra bundles.)
The isomorphism classes of vertex algebra bundles
on a topological space $X$ form an abelian monoid under the direct sum.
Denote by $K^{G}(X)$ the Grothendieck group of this monoid.
The tensor product induces a structure of a ring on this group.
We can similarly define the $K$ theory of conformal vertex algebra bundles, $K^{G}_c(X)$, charged conformal vertex algebra bundles, $K^{G}_{cc}(X)$, and $N = n$-superconformal vertex algebra bundles, $K^{G}_{n-sc}(X)$ \cite{Zhou00}. In the next sections we restrict our attentions to conformal vertex algebra bundles, $K^{G}_c(X)$,
such that in each fiber $L_0$ (see the notation in the next section) is diagonalizable with finite dimensional eigenspaces.
For such bundles, we define
\begin{eqnarray}
G_q(E) & := & \sum_n E_{c_n, h_n}q^{h_n - \frac{c_n}{24}}\,,
\label{bundleE1}
\\
G_{q, y}(E) & := & \sum_n E_{c_n, h_n, j_n}q^{h_n - \frac{c_n}{24}}y^{j_n}\,,
\label{bundleE2}
\end{eqnarray}
where in Eq.~(\ref{bundleE1}) $E_{c_n, h_n}$ is the subbundle of elements of central charge $c_n$ and conformal weight $h_n$. For an $N=2$ SCVA (see Section \ref{N=2} below) with bundle $E$, Eq.~ (\ref{bundleE2}) defines the subbundle $E_{c_n, h_n, j_n}$ of elements of central charge $c_n$ with conformal weight $h_n$ and $U(1)$ charge $j_n$. These extend to homomorphisms of the corresponding $K$-theories.

\subsection{$N=1$ SCVA}
\label{N=1}

Let us consider $N = 1$ SCVA from a vector space with inner product.
Suppose that $(X, g)$ is a finite dimensional complex vector space.
There is a natural structure of an $N = 1$ SCVA (see, for details, \cite{ZhouLectures}) on
\begin{equation}
V_{NS}(X, g) = B_R(X, g)F_{NS}(X, g) =
\bigotimes_{n\in {\mathbb Z}_+}S(t^{-n}X)
\bigotimes_{n\in {\mathbb Z}_+}\Lambda(t^{-n+1/2}X),
\end{equation}
with central charge $(3/2)\,{\rm dim}\,X$. Also, for an $N=1$ SCVA structure,
\begin{equation}
G_q(V_{NS}(X, g)) = q^{-\frac{{\rm dim}\,X}{16}}
\bigotimes_{n\in {\mathbb Z}_+}S_{q^n}(X)
\bigotimes_{n\in {\mathbb Z}_+}\Lambda_{q^{n-1/2}}(X)\,.
\end{equation}
A related formal power series,
$
q^{-\frac{{\rm dim}\,X}{16}}
\bigotimes_{n\in {\mathbb Z}_+}S_{q^n}(X)
\bigotimes_{n\in {\mathbb Z}_+}\Lambda_{-q^{n-1/2}}(X)
$,
can be obtained by introducing an operator $(-1)^F$. (Recall that, by definition, $(-1)^F: \Omega^*(X)\rightarrow \Omega^*(X),\, (-1)^F\alpha = (-1)^p\alpha$, where $\alpha\in \Omega^p(X)$ and $\Omega^p(X)$ is a vector space of $p$-forms on $X$.)

{\bf $N=1$ SCVA bundles over Riemannian manifolds.}
For any Riemannian manifold $(X, g)$, let us consider the principal bundle $\cE(X, g)$ of orthonormal frames. The structure group of $\cE(T, g)$ is $\cE(T_xX, g_x)$\, ($x \in X$) which acts on
$V(T_xX \otimes {\mathbb C}, g \otimes {\mathbb C})$ by automorphisms.
The following assertion holds \cite{ZhouLectures}:
for a Riemannian manifold $(X, g)$ an $N=1$ SCVA bundle is
\begin{eqnarray}
V_{NS}(TX\otimes {\mathbb C},\,g \otimes {\mathbb C})
& = & B_R(TX\otimes {\mathbb C},\, g)F(TX\otimes {\mathbb C}, \,g)
\nonumber \\
& = &
\bigotimes_{n\in {\mathbb Z}_+} S(t^{-n}TX\otimes {\mathbb C})
\bigotimes_{n\in {\mathbb Z}_+} \Lambda(t^{-n+\frac{1}{2}}TX \otimes {\mathbb C})\,.
\end{eqnarray}
Note that the bundle $V_{NS}(TX\otimes {\mathbb C},\, g\otimes {\mathbb C})$ has appeared in the theory of elliptic genera. Also,
\begin{equation}
G_q(V_{NS}(TX\otimes {\mathbb C},\, g\otimes {\mathbb C})
= q^{-\frac{{\rm dim}\,TX}{16}} \bigotimes_{n\in {\mathbb Z}_+} S_{q^n} (TX \otimes {\mathbb C})
\bigotimes_{n\in {\mathbb Z}_+} \Lambda_{q^{n -1/2}}(TX \otimes {\mathbb C})
\end{equation}
(cf. \cite{Witten88}). As before, a related formal power series
$
\bigotimes_{n\in {\mathbb Z}_+} S_{q^n} (TX \otimes {\mathbb C}) \bigotimes_{n\in {\mathbb Z}_+} \Lambda_{-q^{n -1/2}}(TX \otimes {\mathbb C})
$
(cf. \cite{Liu96}) can be obtained by introducing an operator $(-1)^F$.
\begin{remark}
The local holonomy groups of a nonsymmetric Riemannian manifold
can only be $O(n)$, $SO(n)$, $U(n/2)$, $SU(n/2)$, $Sp(n)$, $Sp(n)Sp(1)$,
$G_2$ and $Spin(7)$ {\rm (}a theorem of M. Berger {\rm \cite{Berger}}, see also {\rm \cite{Simons})}. The $N=1$ SCVA can be extended for Riemannian manifolds of special holonomy groups {\rm (}see, for example, {\rm \cite{Shatashvili}}{\rm )}. The OPE's for the
case of quaternionic K\"{a}hler manifolds can be found in {\rm \cite{ZhouLectures}}.
\end{remark}

\subsection{$N=2$ SCVA}
\label{N=2}

{\bf $N=2$ SCVA from a vector space with inner product and polarization properties.} Let, as before, $(X, g)$ admit a polarization
$X = X^{'} \oplus X^{''}$ and let $\{\varphi^i\}$ be a basis of $X^{'}$,  $\{\psi^i\}$ a basis of $X^{''}$, such that $\omega(\varphi^i, \psi^j) = \delta_{ij}$. Then $F(X, g)$ has a basis consists of elements of the form:
$\varphi^{i_1}_{-k_1} \cdots \varphi^{i_m}_{-k_m}
\psi^{j_1}_{-\ell_1-1}\cdots\psi^{j_n}_{-\ell_n-1}$,
where $k_1, \cdots, k_m, \ell_1, \cdots, \ell_n \geq 0$.

In the NS case, set
\begin{equation}
\varphi^i(z) = \sum_{r \in {\mathbb Z}_++1/2}
\varphi^i_{(r)} z^{-r - \frac{1}{2}}\,,\,\,\,\,\,\,\,\,\,
\psi^i(z)  = \sum_{r \in {\mathbb Z}_++1/2} \psi^i_{(r)} z^{-r - \frac{1}{2}}\,,
\end{equation}
where, for $r < 0$, $\varphi^i_{(r)}$ and $\psi^i_{(r)}$ are exterior products by $\varphi^i_r$ and $\psi^i_r$, respectively, and
for $ r > 0$, $\varphi^i_{(r)}$ and $\psi^i_{(r)}$ they are contractions by
$\varphi^i_{-r}$ and $\psi^i_{-r}$, respectively.

In the R case, set
\begin{equation}
\varphi^i(z)  = \sum_{n \in {\mathbb Z}_+} \varphi^i_n z^{-n}\,,
\,\,\,\,\,\,\,\,\,
\psi^i(z)  = \sum_{n \in {\mathbb Z}_+} \psi^i_n z^{-n-1}\,.
\end{equation}
The following assertion holds \cite{ZhouLectures}:
there is a structure of a vertex algebra on
$F_{NS}(X, g)$ defined by
\begin{eqnarray}
Y(\varphi^{i_1}_{-k_1 - 1/2} \cdots \varphi^{i_m}_{-k_m- 1/2}
\psi^{j_1}_{-l_1 - 1/2} \cdots \psi^{j_n}_{-l_n- 1/2})
= & : & \!\!\partial_z^{(k_1)}\varphi^{i_1}(z) \cdots \partial_z^{(k_m)}\varphi^{i_m}(z)
\nonumber \\
& \cdot &\!\!
\partial_z^{(l_1)}\psi^{j_1}(z)\cdots\partial_z^{(l_m)}\psi^{j_m}(z):
\end{eqnarray}
Using the polarization, we get
\begin{eqnarray}
\!\!\!\!
V_{R}(X, g) & = &
\bigotimes_{n\in {\mathbb Z}_+}S(t^{-n}X^{'})
\bigotimes_{n\in {\mathbb Z}_+}S(t^{-n}X^{''})
\bigotimes_{n\in {\mathbb Z}_+}\Lambda(t^{-n}X^{'})
\bigotimes_{n\in {\mathbb Z}_+}\Lambda(t^{-n}X^{''})\,,
\\
\!\!\!\!
V_{NS}(X, g) & = &
\bigotimes_{n\in {\mathbb Z}_+}S(t^{-n}X^{'})
\bigotimes_{n\in {\mathbb Z}_+}S(t^{-n}X^{''})
\bigotimes_{n\in {\mathbb Z}_+}\Lambda(t^{-n+ \frac{1}{2}}X^{'})
\bigotimes_{n\in {\mathbb Z}_+}\Lambda(t^{-n+ \frac{1}{2}}X^{''})\,,
\end{eqnarray}
\begin{eqnarray}
\!\!
G_{q, y}(V_R(X, g)) =
q^{-\frac{{\dim X}}{16}}\!
\bigotimes_{n\in {\mathbb Z}_+} S_{q^n}(X^{'}) \bigotimes_{n\in {\mathbb Z}_+} S_{q^n}(X^{''})
\bigotimes_{n\in {\mathbb Z}_+} \Lambda_{y^{-1}q^{n-\frac{1}{2}}}(X^{'})
\bigotimes_{n\in {\mathbb Z}_+} \Lambda_{yq^{n-\frac{1}{2}}}(X^{''})\,.
\label{Gqy}
\end{eqnarray}
In the so-called $A$ twist and $B$ twist cases, the BRST cohomology of the topological vertex algebras are isomorphic to $\Lambda (X^{''})$ and $\Lambda (X^{'})$ as graded commutative algebras, respectively (for details, see \cite{ZhouLectures}).

{\bf $N=2$ bundles from complex manifolds.}
Denote by $T_{\mathbb C}X$ the holomorphic tangent bundle over a complex manifold $X$. The fiberwise pairing between $T_{\mathbb C}X$ and $T_{\mathbb C}^*X$
induces a canonical complex inner product $\eta$ on
the holomorphic vector bundle $T_{\mathbb C}X \oplus T_{\mathbb C}^*X$
with a manifest polarization $X' = T_{\mathbb C}X$, $X''=T_{\mathbb C}^*X$. One can obtain an $N=2$ SCVA bundle
$V_R((TX)^{\mathbb C} \equiv T_{\mathbb C}X
\oplus T_{\mathbb C}^*X, \,\eta)$.
This bundle is holomorphic and we can consider the $\bar{\partial}$ operator on it:
\begin{equation}
\bar{\partial}: \Omega^{0, *}(V_R((TX)^{\mathbb C},\,\eta))
\longrightarrow \Omega^{0, *+1}(V_R((TX)^{\mathbb C},\,\eta))\,.
\end{equation}
For any complex manifold $X$, \,$\Omega^{0, *}(V_R((TX)^{\mathbb C},\,\eta))$ has a natural structure of an $N=2$ SCVA such that
$\bar{\partial}$ is a differential. Thus, the Dolbeault cohomology
$H^*(X,\, V_R((TX)^{\mathbb C}))$ is an $N=2$ SCVA, the BRST cohomology of its associated topological vertex algebras is isomorphic to $H^*(X, \,\Lambda(T_{\mathbb C}X))$\,\,
(or $H^*(X,\, \Lambda(T_{\mathbb C}^*X))$) \cite{ZhouLectures}.
A similar result one can get for $V_{NS}((TX)^{\mathbb C},\,\eta))$.
As in $N=1$ vertex algebra bundle in the Riemannian case
(cf. \cite{Liu96}), $V_R((TX)^{\mathbb C},\,\eta)$ is related to the elliptic genera.
By (\ref{Gqy}) we have
\begin{equation}
G_{q, y}(V_R((TX)_{\mathbb C}))
=
\bigotimes_{n\in {\mathbb Z}_+} S_{q^n}(T_{\mathbb C}X) \bigotimes_{n\in {\mathbb Z}_+} S_{q^n}(T_{\mathbb C}^*X)
\bigotimes_{n\in {\mathbb Z}_+} \Lambda_{y^{-1}q^{n-\frac{1}{2}}}(T_{\mathbb C}X)
\bigotimes_{n\in {\mathbb Z}_+}
\Lambda_{yq^{n-\frac{1}{2}}}(T_{\mathbb C}^*X)
\end{equation}
(cf. \cite{Hirzebruch}) and \cite{Dijkgraaf} for the $A$ and $B$ twists, respectively).
\begin{remark}
The reader can find $N=2$ SCVA bundles from K\"{a}hler manifolds and Calabi-Yau manifolds, $N=4$ SCVA's from hyperk\"{a}hler manifolds with its relations to loop spaces,  in the very good review papers
{\rm \cite{Zhou00,ZhouLectures}}.
\end{remark}

\section{Lefschetz formula}
\label{Lefschetz}

The purpose of this section is to establish the Lefschetz fixed point formula
which counts the number of fixed points of a continuous mapping from a compact
topological space to itself.
Let $X$ be a compact complex manifold and let $G$ be a Lie group acting on $X$ by biholomorphic maps. Let $g$ be a generator of $G$.
Furthermore, let $\pi: E \rightarrow X$ be a holomorphic vector bundle which admits a $G$-action compatible with the $G$-action on $X$.
Let $G$ be a compact Lie group, then a characteristic class $G$ can be defined as a functor which assigns to every principal $G$-bundle $P$
a cohomology class of $X = P/G$. The set of all characteristic classes forms a ring $H^*_G(A)$ when the coefficient ring $A$ has to be specify. Let $G= T$ be a torus, and $T^*$ a character group (or Pontryagin dual) of $T$. Thus $\{x_i\}_{i=1}^n$ are basis for $T^*$, and $H^*_G(A) = A[[x_1, \cdots x_n]]$ is the ring for formal power series in $x_1, \cdots, x_n$.

A crucial ingredient of the Lefschetz fixed point
formula is the Todd class. Let us define the {\it Todd class} ${\rm Td}({\mathbb E})$ and the {\it dual Todd class} ${\rm Td}^*({\mathbb E})$ associated with a complex vector bundle $\mathbb E$ over $X$. Formally, we can write these classes  ${\rm Td} = \prod_{j=1}^n(x_j/(1-e^{-x_j}))$,
\,${\rm Td}^* = \prod_{j=1}^n(-x_j/(1-e^{x_j}))$.
If ${\mathbb E}^*$ is dual to ${\mathbb E}$ then ${\rm Td}({\mathbb E})$ = ${\rm Td}({\mathbb E}^*)$. In particular, for the complexification of a real bundle,
${\mathbb E} = E\bigotimes_{\mathbb R}{\mathbb C}$,
${\mathbb E}\cong {\mathbb E}^*$, and thus
${\rm Td}({\mathbb E}) = {\rm Td}^*({\mathbb E})$.
The functor $E \mapsto {\rm Td}(E\bigotimes_{\mathbb R}{\mathbb C})$
defines a characteristic class of $O(n)$, and the image of $\mathfrak T$
in the homomorphism $H^*_{U(n)}({\mathbb Q})\rightarrow H^*_{O(n)}({\mathbb Q})$. This class is called {\it the index class}.
If $y_1, \cdots, y_m$ are the basic characters for the maximal torus of $O(n)$ ($m = [n/2]$), then
$
{\rm Td}(E) = {\rm Td}(E\bigotimes_{\mathbb R}{\mathbb C})
$, where
$
{\rm Td} = \prod (-y_j/(1- e^{y_j}))\prod (y_j/(1-e^{-y_j}).
$

We also introduce the sequences of polynomials ${\mathfrak R}(p_1, \cdots, p_r), {\mathfrak S}(c_1, \cdots, c_r)$ related to the Pontryagin and Chern classes \cite{Atiyah}:
\begin{eqnarray}
\prod_{j}\left[\left(\frac{1+e^{x_j}}{2}\right)\left(\frac{1+ e^{-x_j}}{2}\right)\right]^{-1} & = & \sum {\mathfrak R}_r(p_1, \cdots , p_r)
\\
\prod_j\left[\left(\frac{1-e^{y_j+ i\theta}}{1- e^{i\theta}}\right)\left(\frac{1- e^{-y_j-i\theta}}{1- e^{-i\theta}}\right)\right]^{-1} & = & \sum {\mathfrak S}_r^{\theta}(c_1, \cdots , c_r)\,.
\end{eqnarray}
Here the $p_\ell$ are symmetric functions of the $x_j$, and the $c_\ell$ are symmetric functions of the $y_j$. In order to define the $r$-th term of the sum, we take $H_j$ to be a product of $r$ terms:
$H_{j=1}^{N} = \sum {\mathfrak R}_{r, N}$ (for $r \leq N$, ${\mathfrak R}_{r, N}$ and ${\mathfrak S}_{r}^{\theta}$ is independent of $N$).

Let ${\mathfrak R} := \sum {\mathfrak R}_r$, ${\mathfrak S}^{\theta} := \sum {\mathfrak S}_r^{\theta}$.
\begin{eqnarray}
{\mathfrak R}(E) & = &  {\mathfrak R}(E) (p_1(E), \cdots , p_n(E))
\in H^*(X; {\mathbb C})\,\,\,\,({\rm real}\,\,\,\,{\rm vector}\,\,\,\,{\rm bundle}\,\,\,\, E)\,,
\\
{\mathfrak S}^{\theta}(F) & = &  {\mathfrak S}^{\theta}({\mathbb E}) (c_1({\mathbb E}), \cdots , c_n({\mathbb E}))
\in H^*(X; {\mathbb C})\,\,\,\,({\rm complex}\,\,\,\,{\rm vector}\,\,\,\,{\rm bundle}\,\,\,\, {\mathbb E})\,.
\end{eqnarray}
$\mathfrak R$ and ${\mathfrak S}^{\theta}$ depend only on the stable classes of $E, {\mathbb E}$, and not on the augmentation. For trivial line-bundles
$1_{\mathbb R}$ and $1_{\mathbb C}$ (real and complex respectively), we have
$
{\mathfrak R}(E\oplus 1_{\mathbb R}) = {\mathfrak R}(E)
$
and
$
{\mathfrak S}^{\theta}({\mathbb E}\oplus 1_{\mathbb C}) = {\mathfrak S}^{\theta}({\mathbb E}).
$
One can see that the value of ${\rm det}\,(1-g | N^g)$ is just a factor $2^{s(-1)}\prod_{\theta}(1-e^{i\theta})(1- e^{-i\theta})$.
It is constant on each component of $X^g$, and can be regarded as an element of $H^0(X^g; {\mathbb C})$ \cite{Atiyah}. Also, one can get
\begin{equation}
[ {\rm ch}\,\Lambda_{-1}(N^g \bigotimes_{\mathbb R}{\mathbb C}(g))
]^{-1} = \frac{{\mathfrak R}(N^g(-1))\prod_{\theta}{\mathfrak S}^{\theta}(N^g(\theta))}{{\rm det}\,(1- g | N^{g})}\,.
\end{equation}
The following theorem states a Lefschetz fixed-point formula, which computes
Lefschetz numbers of elliptic complexes in terms of the index of the associated
elliptic symbol on fixed-point sets.
\begin{theorem} {\rm (M. Atiyah and I.M. Singer \cite{Atiyah}, Lefschetz Theorem)}
Let $X^g$ denote the fixed-point set of $g$, $N^g$ the normal bundle of $X^g$ in $X$, and $N^g = N^g(-1)\oplus\sum_{0<\theta<\pi}N^g(\theta)$ the decomposition of $N^g$ determined by the action of $G$ {\rm (}compare with Eq.~{\rm (\ref{Northogonal}))}. Let $u\in K_G(TX)$ be the symbol class of $E$, $i^*u \in K_G(TX^g)$ its restriction to $X^g$.
Let ${\rm Td}\in H^*(X; {\mathbb Q})$ denote the index class of
$X$, and let ${\mathfrak R},\, {\mathfrak S}$ be the characteristic classes of the orthogonal and unitary groups. Then the Lefschetz number $L(X, E)(g)$ is given by
\begin{equation}
L(X, E)(g) = (-1)^n\left\{
\frac{{\rm ch}\,i^*u(g){\mathfrak R}(N^g(-1))\prod_{0<\theta<\pi}
{\mathfrak S}^{\theta}(N^g(\theta))\,{\rm Td}(X^g)}
{{\rm det}\,(1-g | N^g)}\right\}[TX^g]\,.
\label{LF}
\end{equation}
\end{theorem}
\begin{comment}
Eq.~{\rm (\ref{LF})} involves the following cohomological invariants of
$(X, G)$: Pontryagin classes of $X^g$; Pontryagin classes of $N^g(- 1)$;
Chern classes of all the $N^g(\theta)$.
For oriented $X^g$ one can replace the evaluation on $[TX^g]$, by an evaluation on $[X^g]$:
\begin{equation}
\iota :\,\, H^*(X^g, {\mathbb C}) \longrightarrow
H^*(TX^g; {\mathbb C})\,.
\end{equation}
For a complex manifold $X$ there is the Thom isomorphism in the
$K$-theory {\rm (}that is, there is a single elliptic symbol which, in a
sense, generates everything{\rm )}. Thus, it is natural to replace the
index homomorphism $K(TX)\rightarrow {\mathbb Z}$ of the general
theory by the homomorphism $K(X)\rightarrow {\mathbb Z}$ {\rm (see,
for example, \cite{BonoraBytsenko})}.
\end{comment}
Suppose that $V$ is a holomorphic vector bundle over $X$,\, $\cO(V)$
the sheaf of germs of holomorphic sections of $V$.
Let $G$ be a finite group of automorphisms of the pair $(X, V)$
(the holomorphic case, for example). The Lefschetz theorem
combined with the Riemann-Roch theorem gives
\begin{equation}
\sum (-1)^p {\rm Tr}\,(g \,|\, H^p(X; \cO(V)) =
\left\{\frac{{\rm ch}\,(V | X^g)(g){\rm Td}(X^g)}{{\rm ch}\,\Lambda_{-1}((N^g)^*)(g)}\right\}[X^g]\,.
\label{R-R}
\end{equation}

{\bf The case of an holomorphic vector bundle.}
Recall that the complex vector bundle $N^g$ has a decomposition
$N^g = \sum N^g(\theta)$, where $N^g(\theta)$ is the sub-bundle on which $g$ acts as $\exp(i\theta)$. Also,
\begin{equation}
{\rm ch}\,\Lambda_{-1}(N^g(\theta))^* = \prod_j (1-e^{-x_j-i\theta})
(N^g(\theta))\,,
\end{equation}
where $\prod_j (1- e^{-x_j-i\theta}) \in H^*_{U(m)}({\mathbb C})$,
$m={\rm dim}\,N^g(\theta)$. For $0 <\theta< 2\pi$ define the stable characteristic class
\begin{equation}
{\cU}^{\theta} = \sum {\cU}^{\theta}_r =
\prod_j\left[\frac{1- e^{-x_j-i\theta}}{1- e^{-i\theta}}\right]^{-1}\,.
\end{equation}
Thus each $\cU^{\theta}_r$ is a polynomial with complex coefficients in the Chern classes, and
\begin{equation}
[{\rm ch}\,\Lambda_{-1}(N^g(\theta))^*]^{-1} =
\frac{\cU^{\theta}(N^g(\theta))}{(1- e^{-i\theta})^m}\,.
\end{equation}
Taking the product over all $\theta$, we get
\begin{equation}
[{\rm ch}\,\Lambda_{-1}(N^g)^*]^{-1} =
\frac{\prod\,\cU^{\theta}(N^g(\theta))}{{\rm det}\,(1- g(N^g)^*)}\,,
\end{equation}
where as before ${\rm det}_{\mathbb C}(1 - g | (N^g)^*)\in H^0(X^g; {\mathbb C})$
assigns to the component of $x\in X^g$ the value ${\rm det}_{\mathbb C}(1 - g | (N^g_x)^*)$.

Finally, the following statement holds \cite{Atiyah}:
Let $X$ be a compact complex manifold and $V$ a holomorphic vector bundle over X. Suppose that $G$ is a finite group of automorphisms of the pair $(X, V)$. For any $g\in G$, let $X^g$ denote the fixed point set of $g$, and let, as before, $N^g = \sum N^g(\theta)$ denote the (complex) normal bundle of $X^g$ decomposed according to the eigenvalues $\exp (i\theta)$ of $g$. Let $\cU^{\theta}$ denote the characteristic class. Then,
\begin{equation}
\sum (-1)^p{\rm Tr}\,(g \,|\, H^p(X; \cO(V))) =
\left\{\frac{{\rm ch}\,(V | X^g)(g)\, \prod_{\theta}\cU^{\theta}(N^g(\theta)){\rm Td}(X^g)}
{{\rm det}\,(1- g | (N^g)^*)}\right\}[X^g]\,.
\end{equation}

\section{Bundles over locally symmetric spaces and spectral functions}
\label{SymSpace}

\subsection{Laplacian on forms and the Ruelle function}
\label{Norm2}

Let us consider an $N$-dimensional compact real hyperbolic
space $X$
with universal covering $\widetilde{X}$ and fundamental group
$\Gamma$. We can represent $\widetilde{X}$ as the symmetric space $G/{\mathcal K}$.
We regard $\Gamma$ as a discrete subgroup of $G$ acting isometrically on $\widetilde{X}$, and we take
$X$ to be the quotient space by that action:
$X=\Gamma\backslash \widetilde{X} = \Gamma\backslash G/{\mathcal K}$.
Let $\tau$ be an irreducible representation of
${\mathcal K}$ on a complex vector space $V_\tau$, and form the induced
homogeneous vector bundle $G\times_{\mathcal K} V_\tau$ (the fiber product of $G$ with $V_\tau$
over ${\mathcal K}$) $\rightarrow \widetilde{X}$ over
$\widetilde{X}$. Restricting the $G$ action to
$\Gamma$, we obtain the quotient bundle $E_\tau=\Gamma\backslash
(G\times_{\mathcal K}V_\tau)\rightarrow X =
\Gamma\backslash \widetilde{X}$ over $X$. The
natural Riemannian structure on $\widetilde{X}$
(therefore on $X$) induced by
the Killing form $(\;,\;)$ of $G$ gives rise to a connection Laplacian
$L$ on $E_\tau$. If $\Omega_{\mathcal K}$ denotes
the Casimir operator of ${\mathcal K}$, that is
$
\Omega_{\mathcal K}=-\sum y_j^2,
$
for a basis $\{y_j\}$ of the Lie algebra ${\mathfrak k}_0$ of ${\mathcal K}$, where
$(y_j\;,y_\ell)=-\delta_{j\ell}$, then
$\tau(\Omega_{\mathcal K})=\lambda_\tau{\bf 1}$
for a suitable scalar $\lambda_\tau$. Moreover for the Casimir operator
$\Omega$ of $G$, with $\Omega$ operating on smooth sections
$\Gamma^\infty E_\tau$ of $E_\tau$, one has
$
L=\Omega-\lambda_\tau{\bf 1}\;;
$
see Lemma 3.1 of \cite{Wallach}. For $\lambda\geq 0$, let
\begin{equation}
\Gamma^\infty\left(X\;,E_\tau\right)_\lambda=
\left\{s\in\Gamma^\infty E_\tau\left|-Ls=\lambda s\right.
\right\}
\end{equation}
be the space of eigensections of $L$ corresponding
to $\lambda$. Here we note that since $X$ is compact we can
order the spectrum of $-L$ by taking $ 0=\lambda_0<\lambda_1<\lambda_2<\cdots$;
$\lim_{j\rightarrow\infty}\lambda_j=\infty$.
We shall specialize $\tau$ to be the representation
$\tau^{(p)}$ of ${\mathcal K} =SO(N)$
on $\Lambda^p {\mathbb C}^{N}$. It will be convenient, moreover, to work
with the normalized Laplacian $L_p=-c(N)L$,
where $c(N)=2(N-1)$. $L_p$ has spectrum
$\left\{c(N)\lambda_j\;,m_j\right\}_{j=0}^\infty$, where the
multiplicity $m_j$ of the eigenvalue $c(N)\lambda_j$ is given by
$
m_j={\rm
dim}\;\Gamma^\infty\left(X\;,E_{\tau^{(p)}}\right)_{\lambda_j}\;.
$

If $L_p$ is a self-adjoint Laplacian on $p$-forms then
the following results hold. There exist $\varepsilon,\delta >0$
such that for $0<t<\delta$ the heat kernel expansion for
Laplace operators on a compact manifold $X$ is given by
\begin{equation}
{\rm Tr}\left(e^{-tL_p}\right)=
\sum_{0\leq \ell\leq \ell_0} a_\ell
(L_p)t^{-\ell}+ {O}(t^\varepsilon)
\mbox{.}
\end{equation}
The coefficients $a_\ell(L_p)$ are called Hadamard-Minakshisundaram-De Witt-Seeley coefficients (or, sometimes, heat kernel, or just heat coefficients).

Let $\chi$ be an orthogonal representation of $\pi_1(X)$. Using
the Hodge decomposition, the vector space $H(X;\chi)$ of twisted
cohomology classes can be embedded into $\Omega(X;\chi)$ as the
space of harmonic forms. This embedding induces a norm
$|\cdot|^{RS}$ on the determinant line ${\rm det}H(M;\chi)$. The
Ray-Singer norm $||\cdot||^{RS}$ on ${\rm det}H(X;\chi)$ is
defined by \cite{Ray}
\begin{equation}
||\cdot||^{RS}\stackrel{def}=|\cdot|^{RS}\prod_{p=0}^{{\rm dim}\,X}
\left[\exp\left(-\frac{d}{ds}
\zeta (s|L_p)|_{s=0}\right)\right]^{(-1)^pp/2}
\mbox{,}
\end{equation}
where the zeta function $\zeta (s|L_p)$ of the Laplacian
acting on the space of
$p$-forms orthogonal to the harmonic forms has been used. For a
closed connected orientable smooth manifold of odd dimension
and for Euler structure
$\eta\in {\rm Eul}(X)$, the Ray-Singer norm of its cohomological
torsion $\tau_{an}(X;\eta)=\tau_{an}(X)\in {\rm det}H(X;\chi)$ is
equal to the positive
square root of the absolute value of the monodromy of $\chi$
along the characteristic class $c(\eta)\in H^1(X)$ \cite{Farber}:
$||\tau_{an}(X)||^{RS}=|{\rm det}_{\chi}c(\eta)|^{1/2}$.
In the special
case where the flat bundle $\chi$ is acyclic, we have
\begin{equation}
\left[\tau_{an}(X)\right]^2
=|{\rm det}_{\chi}c(\eta)|
\prod_{p=0}^{{\rm dim}\,X}\left[\exp\left(-\frac{d}{ds}
\zeta (s|L_p)|_{s=0}\right)\right]^{(-1)^{p+1}p}
\mbox{.}
\label{RS}
\end{equation}
For a  closed oriented hyperbolic three-manifolds of the form
$X = {H}^3/\Gamma$, and for acyclic
$\chi$, the $L^2$-analytic torsion has the form
\cite{Fried,Bytsenko3,Bytsenko4}:
$[\tau_{an}(X)]^2={\mathcal R}(0)$, where
${\mathcal R}(s)$ is the Ruelle function
(it can be continued meromorphically to the entire complex plane
$\mathbb C$).
The function ${\mathcal R}(s)$ is an alternating
product of more complicate factors, each of which is a Selberg zeta function $Z_{\Gamma}(s)$. The relation between the Ruelle and Selberg zeta functions is:
\begin{equation}
{\mathcal R}(s)=\prod_{p=0}^{{\rm dim}\,X-1}Z_{\Gamma}
(p+s)^{(-1)^p}\,.
\label{Ruelle}
\end{equation}
The Ruelle zeta function associated with closed oriented
hyperbolic three-manifold $X$ has the form:
${\mathcal R}(s)=Z_{\Gamma}(s)Z_{\Gamma}(2+s)/Z_{\Gamma}(1+s)$.

\subsection{Spectral functions of hyperbolic geometry}
\label{Dirac}

We recall some results on the eta invariant of a self-adjoint
elliptic differential operator acting on a compact manifold. For
details we refer the reader to \cite{Atiyah1,Atiyah2,Atiyah3}
where the eta invariant was introduced in connection with the
index theorem for a manifold with boundary. One can attach the eta
invariant to any operator of Dirac type on a compact Riemannian
manifold of odd dimension. Dirac operators on even dimensional
manifolds have symmetric spectra and, therefore, trivial eta
invariants. To define a spectral invariant which measures the asymmetry of the spectrum ${\rm Spec}({\mathfrak D})$ of an operator ${\mathfrak D}$, one starts with the following formula: the holomorphic function
\begin{equation}
\eta(s,{\mathfrak D})\stackrel{def}{=}
\sum_{\lambda\in \,{\rm Spec}\,{\mathfrak D}\,-\{0\}}
{\rm sgn}(\lambda)|\lambda|^{-s}={\rm Tr}\left({\mathfrak D}
\left({\mathfrak D}^2\right)^{-(s+1)/2}\right)
\mbox{,}
\end{equation}
is well defined for all ${\rm Re}\, s\gg 0$ and extends to a meromorphic
function on ${\mathbb C}$.
Indeed, from the asymptotic behavior of the heat operator at
$t=0$ \cite{Bismut2},
$
{\rm Tr}\left({\mathfrak D}e^{-t{\mathfrak D}^2}\right)=
{O}(t^{1/2}),
$
and from the identity
\begin{equation}
\eta(s,{\mathfrak D})
=\frac{1}{\Gamma\left((s+1)/2\right)}\int_{{\mathbb R}_{+}}
{\rm Tr}\left({\mathfrak D} e^{-t{\mathfrak D}^2}\right)t^{(s-1)/2}dt
\mbox{,}
\end{equation}
it follows that $\eta (s,{\mathfrak D})$ admits a meromorphic
extension to the
whole $s$-plane, with at most simple poles at
$s=({\rm dim}\,X-k)/({\rm ord}\,{\mathfrak D})$\,\,\,
$(k\in {\mathbb Z}_{+})$
and locally computable residues.
It has been established that the point $s=0$ is not a pole, which makes it
possible to define the eta invariant of ${\mathfrak D}$ by
$\eta(0,{\mathfrak D})$. It also follows directly that
$\eta(0,-{\mathfrak D})=-\eta(0,{\mathfrak D})$ and
$\eta(0,\lambda{\mathfrak D})=\eta(0,{\mathfrak D})$, $\forall \lambda>0$.
As ${\mathfrak D}^{+}$ is isomorphic to ${\mathfrak D}^{-}$, we have
$\eta(s,{\mathfrak D}^{+})=
\eta(s,{\mathfrak D}^{-})= \eta(s,{\mathfrak D})/2$.

An important case of such an operator is the even part of the
tangential signature operator, ${\mathfrak B}$, acting on the even
forms of the given manifold. The eta invariant associated to the
operator ${\mathfrak B}$ is given by
$\eta_X(0)=\eta({0, \mathfrak B})$, and
is called the eta invariant of $X$ \cite{Moscovici1,Moscovici2}.

Let $X$ be a compact oriented $N = (4m-1)$-dimensional Riemannian
manifold of constant negative curvature. A remarkable formula relating
$\eta(s,{\mathfrak B})$,
to the closed geodesics on
$X$ has been derived by Millson \cite{Millson}. More explicitly,
Millson proved the following result for a Selberg type (Shintani) zeta
function.

\begin{definition}
{\rm (J. J. Millson \cite{Millson})}.
Define a zeta function by the
following series, which is absolutely convergent for ${\rm Re}\,s>0$,
\begin{equation}
{\rm log}Z(s,{\mathfrak B})\stackrel{def}{=}\sum_{[\gamma]\neq 1}
\frac{{\rm Tr}\tau^{+}_{\gamma}-{\rm Tr}\tau^{-}_{\gamma}}
{|{\rm det}(I-P_h(\gamma))|^{1/2}}\frac{e^{-s\ell(\gamma)}}{m(\gamma)}
\mbox{,}
\end{equation}
where $[\gamma]$ runs over the nontrivial conjugacy classes in
the fundamental group $\Gamma=\pi_1(X)$, $\ell(\gamma)$ is the
length of the closed geodesic $c_{\gamma}$ {\rm (}with multiplicity $m(\gamma)${\rm )} in the free homotopy class
corresponding to $[\gamma]$, $P_h(\gamma)$ is the restriction
of the linear Poincar\'{e} map $P(\gamma)=d\Phi_1$ at $(c_{\gamma},\dot{c}_{\gamma})\in TX$
to the directions normal to the geodesic flow $\Phi_t$ and
$\tau^{\pm}_{\gamma}$ is the parallel translation around $c_{\gamma}$ on
the $\Lambda^{\pm}_{\gamma}=\pm i$ eigenspace of
$\sigma_{\mathfrak B}(\dot{c}_{\gamma})$ ($\sigma_{\mathfrak B}$
denoting the principal symbol of ${\mathfrak B}$).
Then $Z(s,{\mathfrak B})$ admits a meromorphic continuation to the entire complex plane, which in particular is holomorphic at $0$.
Moreover,
$
{\rm log}Z(0,{\mathfrak B})= \pi i \eta (0,{\mathfrak B})
\mbox{.}
$
\end{definition}
In fact, $Z(s,{\mathfrak B})$ satisfies the functional equation
$
Z(s,{\mathfrak B})Z(-s,{\mathfrak B})=
\exp [2\pi i\eta (s,{\mathfrak B})].
$
Millson's formulae have been extended by Moscovici and Stanton
\cite{Moscovici1} to  operators of Dirac type (acting in
non-positively curved locally symmetric manifolds), even with
additional coefficients in locally flat bundles.

Let ${\mathfrak D}$ denote a generalized Dirac operator associated
to a locally homogeneous Clifford bundle over a compact oriented odd
dimensional locally symmetric space $X$, whose simply
connected cover ${\widetilde X}$ is a symmetric space of
noncompact type. The fixed point set of the geodesic flow, acting
on the unit sphere bundle $S^1X$, is a disjoint union of
submanifolds $X_{\gamma}$. These submanifolds are
parameterized by the nontrivial conjugacy classes $[\gamma] \neq 1$
in $\Gamma$. By ${\cJ}_1(\Gamma)$ we denote the set of
those conjugacy classes $[\gamma]$ for which $X_{\gamma}$
has the property that the Euclidean de Rham factor of ${\widetilde
X}_{\gamma}$ is one-dimensional. A bundle $C{\widetilde
X}_{\gamma}$ over ${\widetilde X}_{\gamma}$, the ``central''
bundle, is determined by the eigenvalues of absolute value 1 of
the linear Poincar\'{e} map $P(\gamma)$. The parallel translation
around $c_{\gamma}$ gives rise to an orthogonal transformation
${\widetilde \tau}_{\gamma}$ of $C{\widetilde X}_{\gamma}$;
$T{\widetilde X}_{\gamma} \subset C{\widetilde X}_{\gamma}$ and we
let $N{\widetilde X}_{\gamma}$ denote the orthogonal component of
$T{\widetilde X}_{\gamma}$ in $C{\widetilde X}_{\gamma}$. The
tangent bundle $T{\widetilde X}_{\gamma}$ corresponds to the
eigenvalue 1 of  ${\widetilde \tau}_{\gamma}$ and $N{\widetilde
X}_{\gamma}$ decomposes as \cite{Moscovici1}
\begin{equation}
N{\widetilde X}_{\gamma}=N{\widetilde X}_{\gamma}(-1)\oplus
\sum_{0<\theta<\pi}N{\widetilde X}_{\gamma}(\theta)
\label{Northogonal}
\end{equation}
according to the other values --1, $\exp(\pm i)$\,
$(0<\theta<\pi)$. The restriction to ${X}_{\gamma}$ of the
exterior bundle can be pushed down to a vector bundle ${\widetilde
\Lambda}_{\gamma}$ over ${\widetilde X}_{\gamma}$ which splits
into a subbundle ${\widetilde {\Lambda}}_{\gamma}^{\pm}$
corresponding to the eigenvalue $\pm i$ of the symbol
${\mathfrak D}$. Thus, we obtain a ${\widetilde
\tau}_{\gamma}-$equivariant complex ${\widetilde
\sigma}_{\gamma}^{\mathfrak D}: {\widetilde \Lambda}_{\gamma}^{+}
\rightarrow  {\widetilde \Lambda}_{\gamma}^{-}$ over $T{\widetilde
X}_{\gamma}$ and a class $[{\widetilde \sigma}_{\gamma}^{\mathfrak
D}] \in K_{{\widetilde \tau}_{\gamma}} (T{\widetilde X})$, the
${\widetilde \tau}_{\gamma}-$equivariant $K-$theory group of
$T{\widetilde X}_{\gamma}$. The cohomology class can be formed as
in \cite{Atiyah} (Section 3), ${\rm ch}({\widetilde
\sigma}_{\gamma}^{\mathfrak D} ({\widetilde \tau}_{\gamma})) \in
H^{+}(T{\widetilde X}_{\gamma}; {\mathbb C})$. We now present the
main results.
\begin{theorem} {\rm (H. Moscovici and R. J. Stanton \cite{Moscovici1},
Theorem 6.3)}. The following function can be defined,
initially for ${\rm Re}(s^2)\gg 0$, by the formula
\begin{equation}
{\rm log}Z(s,{\mathfrak D})\stackrel{def}{=}
\sum_{[\gamma]\in {\cJ}_1(\Gamma)}
(-1)^q\frac{L(\gamma,{\mathfrak D})}
{|{\rm det}(I-P_h(\gamma))|^{1/2}}\frac{e^{-s\ell(\gamma)}}{m(\gamma)}
\mbox{,}
\end{equation}
where $q=(1/2){\rm dim}N{\widetilde X}_{\gamma}$ is an integer
independent of $\gamma$.
The Lefschetz number $L(\gamma,{\mathfrak D})$ is given by
{\rm (}see, for example, {\rm \cite{Hotta}} or {\rm \cite{Moscovici1}}, Eq.~{\rm (5.5))}:
\begin{equation}
L(\gamma,{\mathfrak D}) =
\left\{\frac{{\rm ch}({\widetilde \sigma}_{\gamma}^{\mathfrak D}
({\widetilde \tau}_{\gamma}))
{\mathfrak R}(N{\widetilde X}_{\gamma}(-1))
\prod_{0<\theta<\pi}{\mathfrak S}^{\theta}
(N{\widetilde X}_{\gamma}(\theta)){\rm Td}({\widetilde X}_{\gamma})}
{{\rm det}(I-{\widetilde \tau}_{\gamma}|N{\widetilde X}_{\gamma})}
\right\}
[T{\widetilde X}_{\gamma}]
\mbox{.}
\label{Lef}
\end{equation}
The Lefschetz formula {\rm (\ref{Lef})} is given in terms of the stable
characteristic classes ${\mathfrak R}, {\mathfrak S}^{\theta}$ and
${\rm Td}$ defined in {\rm \cite{Atiyah}} {\rm (}Theorem 3.9{\rm )}.
\end{theorem}
Furthermore ${\rm log}Z(s,{\mathfrak D})$ has a meromorphic
continuation to the plane ${\mathbb C}$ given by the identity
$
{\rm log}Z(s,{\mathfrak D})={\rm log}{\rm det}^{\prime}
[({\mathfrak D}- is)/({\mathfrak D}+ is)],
$
where $s\in i{\rm Spec}^{\prime}({\mathfrak D})$\,\,\,
$({\rm Spec}({\mathfrak D})-\{0\})$, and $Z(s,{\mathfrak D})$
satisfies the functional equation
$
Z(s,{\mathfrak D})Z(-s,{\mathfrak D})=
\exp[2\pi i\eta(s,{\mathfrak D})].
$

Let ${\widetilde X}$ denote a simply connected cover of
$X$, which is a symmetric space of noncompact type, and
let ${\widetilde E}$ denote the pull-back to $\widetilde X$ for
any vector bundle ${E}$ over $X$. We restrict ourselves
to bundles which satisfy a local homogeneity
condition. Namely, a vector bundle ${E}$ over $X$ is
${\mathcal G}$-locally homogeneous, for some Lie group ${\mathcal
G}$, if there is a smooth action of ${\mathcal G}$ on ${E}$ which
is linear on the fibers and covers the action of ${\mathcal  G}$
on $\widetilde X$. Standard constructions from linear algebra
applied to any ${\mathcal G}$-locally homogeneous ${E}$ give, in a
natural way, corresponding ${\mathcal  G}$-locally homogeneous
vector bundles. In particular, all bundles $TX, {\mathbb
C}\ell(X), {\rm End}\,{E}\simeq {E}^{*}\otimes {E}$ are
${\mathcal G}$-locally homogeneous. We will require also that all constructions associated with ${\mathcal G}$-locally homogeneous bundles are ${\mathcal G}$-equivariant \cite{Moscovici1}.

Let ${\mathfrak D}$ denote a generalized Dirac operator
associated to a locally homogeneous bundle $E$ over $X$. We
require ${\mathcal G}$-equivariance for
${\widetilde \nabla}$, the lift of $\nabla$ to
${\widetilde {E}}$, and therefore
the corresponding Dirac operator is then ${\mathcal G}-$invariant.
Suppose now that $\chi: \Gamma\rightarrow U(F)$ is an unitary
representation of $\Gamma$ on $F$. The Hermitian vector bundle
${\mathbb F}={\widetilde X}\times_{\Gamma}F$ over $X$
inherits a flat connection from the trivial connection on
${\widetilde X}\times F$. If ${\mathfrak D}:
C^{\infty}(X,V)\rightarrow C^{\infty}(X,V)$ is a differential
operator acting on the sections of the vector bundle $V$, then
${\mathfrak D}$ extends canonically to a differential operator
\begin{equation}
{\mathfrak D}_{\chi}: C^{\infty}(X,V \otimes{\mathbb
F})\longrightarrow C^{\infty}(X,V\otimes{\mathbb F})\,,
\end{equation}
uniquely characterized by the property that ${\mathfrak D}_{\chi}$ is
locally isomorphic to ${\mathfrak D}\otimes \cdots \otimes {\mathfrak
D}$\,\,\, (${\rm dim}\,F$ times). We specialize to the case of
locally homogeneous Dirac operators ${\mathfrak D}:
C^{\infty}(X,{\mathbb E})\rightarrow C^{\infty}
(X,{E})$ in order to construct a generalized
operator ${\mathfrak D}_{\chi}$, acting on spinors with
coefficients in $\chi$ (see for detail \cite{Moscovici1}). One can
repeat the arguments of the previous discussion to construct a
twisted zeta function $Z(s,{\mathfrak D}_{\chi})$. The main
results can be stated as follows.
\begin{theorem} {\rm (H. Moscovici and J. J. Stanton \cite{Moscovici1},
Section 7)}. There exists a
zeta function  $Z(s,{\mathfrak D}_{\chi})$, meromorphic on ${\mathbb C}$,
given for ${\rm Re}(s^2)\gg 0$ by the formula
\begin{equation}
{\rm log}Z(s,{\mathfrak D}_{\chi})\stackrel{def}{=}
\sum_{[\gamma]\in {\cJ}_1(\Gamma)}
(-1)^q{\rm Tr}\chi(\gamma)\frac{L(\gamma,{\mathfrak D})}
{|{\rm det}(I-P_h(\gamma))|^{1/2}}\frac{e^{-s\ell(\gamma)}}{m(\gamma)}
\mbox{;}
\end{equation}
moreover, one has
$
{\rm log}Z(0,{\mathfrak D}_{\chi})=
\pi i\eta (0,{\mathfrak D}_{\chi}).
$
\end{theorem}

\subsection{Spectral functions of hyperbolic three-geometry}
\label{Three-geometry}

Now let us consider three-geometry with an orbifold description $H^3/\Gamma$. The complex unimodular group $G=SL(2, {\mathbb C})$
acts on the real hyperbolic three-space $H^3$ in a standard way, namely for $(x,y,z)\in H^3$ and $g\in G$, one gets
$g\cdot(x,y,z)= (u,v,w)\in H^3$. Thus for $r=x+iy$,\,
$g= \left[ \begin{array}{cc} a & b \\ c & d \end{array} \right]$,
$
u+iv = [(ar+b)\overline{(cr+d)}+ a\overline{c}z^2]\cdot
[|cr+d|^2 + |c|^2z^2]^{-1},\,
w = z\cdot[
{|cr+d|^2 + |c|^2z^2}]^{-1}\,.
$
Here the bar denotes the complex conjugation. Let $\Gamma \in G$ be the discrete group of $G$
defined as
\begin{eqnarray}
\Gamma & = & \{{\rm diag}(e^{2n\pi ({\rm Im}\,\tau + i{\rm Re}\,\tau)},\,\,  e^{-2n\pi ({\rm Im}\,\tau + i{\rm Re}\,\tau)}):
n\in {\mathbb Z}\}
= \{{\mathfrak g}^n:\, n\in {\mathbb Z}\}\,,
\nonumber \\
{\mathfrak g} & = &
{\rm diag}(e^{2\pi ({\rm Im}\,\tau + i{\rm Re}\,\tau)},\,\,  e^{-2\pi ({\rm Im}\,\tau + i{\rm Re}\,\tau)})\,.
\end{eqnarray}
One can define a Selberg-type zeta function for the group
$\Gamma = \{{\mathfrak g}^n : n \in {\mathbb Z}\}$ generated by a single hyperbolic element of the form ${\mathfrak g} = {\rm diag}(e^z, e^{-z})$, where $z=\alpha+i\beta$ for $\alpha,\beta >0$. In fact, we will take
$\alpha = 2\pi {\rm Im}\,\tau$, $\beta= 2\pi {\rm Re}\,\tau$. For the standard action of $SL(2, {\mathbb C})$ on $H^3$ one has
\begin{equation}
{\mathfrak g}
\left[ \begin{array}{c} x \\ y\\ z \end{array} \right]
=
\left[\begin{array}{ccc} e^{\alpha} & 0 & 0\\ 0 & e^{\alpha} & 0\\ 0
& 0 & \,\,e^{\alpha} \end{array} \right]
\left[\begin{array}{ccc} \cos(\beta) & -\sin (\beta) & 0\\
\sin (\beta) & \,\,\,\,\cos (\beta) & 0
\\ 0 & 0 & 1 \end{array} \right]
\left[\begin{array}{c} x \\ y\\ z \end{array} \right]
\,.
\end{equation}
Therefore, ${\mathfrak g}$ is the composition of a rotation in ${\mathbb R}^2$ with complex eigenvalues $\exp (\pm i\beta)$ and a dilatation $\exp (\alpha)$. There exists the Patterson-Selberg spectral function $Z_\Gamma (s)$,
meromorphic on $\mathbb C$, given for ${\rm Re}\, s> 0$ by
the formula {\rm \cite{Bytsenko07,Bytsenko08}}
\begin{equation}
{\rm log}\, Z_{\Gamma} (s)   =
-\frac{1}{4}\sum_{n=1}^{\infty}\frac{e^{-n\alpha(s-1)}}
{n[\sinh^2\left(\frac{\alpha n}{2}\right)
+\sin^2\left(\frac{\beta n}{2}\right)]}\,.
\label{logZ}
\end{equation}
The Patterson-Selberg function can be attached to ${H}^3/\Gamma$ as follows \cite{Perry}:
\begin{equation}
Z_\Gamma(s) :=\prod_{k_1, k_2 \in \mathbb{Z}_+ \cup \{0\}} [1-(e^{i\beta})^{k_1}(e^{-i\beta})^{k_2}e^{-(k_1+k_2+s)\alpha}]\,.
\label{zeta00}
\end{equation}
Zeros of $Z_\Gamma (s)$ are the complex numbers
$
\zeta_{n,k_{1},k_{2}} = -\left(k_{1}+k_{2}\right)+i\left(k_{1}-
k_{2}\right)\beta/\alpha+ 2\pi  in/\alpha\,\,\,
(n \in {\mathbb Z})
$
(for details, see \cite{Bonora11}).

{\bf Generating and spectral functions.}
Using the equality
$
\sinh^2\left(\alpha n/2\right)
+ \sin^2\left(\beta n/2\right)$ $ = |\sin(n\pi \tau)|^2 =
|1-q^n|^2/(4|q|^{n})
$
and Eq.~(\ref{logZ}), we get
\begin{eqnarray}
{\rm log}\prod_{m=\ell}^{\infty}(1- q^{m+\varepsilon})
& = &
\sum_{m=\ell}^{\infty}
{\rm log}(1 - q^{m+\varepsilon})
=
-\sum_{n=1}^{\infty}
\frac{q^{(\ell + \varepsilon)n}
(1- \overline{q}^{n})|q|^{-n}}
{4n|\sin(n\pi \tau)|^2}
\nonumber \\
&=&
{\rm log}\,\left[\frac{Z_\Gamma(\xi (1-it))}{Z_\Gamma(
\xi (1-it)+1+it)}\right]\,,
\label{modular1}
\\
{\rm log}\prod_{m=\ell}^{\infty}(1- \overline{q}^{m+\varepsilon})
& = &
\sum_{m=\ell}^{\infty}
{\rm log}(1 - {\overline q}^{m+\varepsilon})
=
-\sum_{n=1}^{\infty}
\frac{{\overline q}^{(\ell + \varepsilon)n}
(1- {q}^{n})|q|^{-n}}
{4n|\sin(n\pi \tau)|^2}
\nonumber \\
& = &
{\rm log}\,\left[\frac{Z_\Gamma(\xi (1+it))}{Z_\Gamma(
\xi (1+it)+1-it)}\right]\,,
\label{modular2}
\\
{\rm log}\prod_{m=\ell}^{\infty}(1+ {q}^{m+\varepsilon})
& = &
\sum_{m=\ell}^{\infty}
{\rm log}(1 + {q}^{m+\varepsilon})
=
-\sum_{n=1}^{\infty}
\frac{(-1)^n q^{(\ell + \varepsilon)n}
(1- {\overline q}^{n})|q|^{-n}}
{4n|\sin(n\pi \tau)|^2}
\nonumber \\
& = &
{\rm log}\,\left[\frac{Z_\Gamma(\xi (1-it) +
i\eta(\tau))}{Z_\Gamma(\xi (1+it)+1-it + i\eta(\tau))}\right]\,,
\label{modular3}
\\
{\rm log}\prod_{m=\ell}^{\infty}(1+ {\overline q}^{m+\varepsilon})
& = &
\sum_{m=\ell}^{\infty}
{\rm log}(1 + {\overline q}^{m+\varepsilon})
=
-\sum_{n=1}^{\infty}
\frac{(-1)^nq^{(\ell + \varepsilon)n}
(1- {q}^{n})|q|^{-n}}
{4n|\sin(n\pi \tau)|^2}
\nonumber \\
& = &
{\rm log}\,\left[\frac{Z_\Gamma(\xi(1+it) + i\eta(\tau))}{Z_\Gamma(\xi(1+it)+1-it + i\eta(\tau))}\right]\,,
\label{modular4}
\end{eqnarray}
where $\ell \in {\mathbb Z}_+,\, \varepsilon \in {\mathbb C}$, $t = {\rm Re}\,\tau/{\rm Im}\,\tau$, $\xi = \ell + \varepsilon$ and $\eta (\tau)= \pm(2\tau)^{-1}$.

Let us next introduce some well-known functions and their modular properties under the action of $SL(2, {\mathbb Z})$. The special cases associated with (\ref{modular1}), (\ref{modular2})
are (see  \cite{Kac}):
\begin{eqnarray}
f_1(q) & = & q^{-\frac{1}{48}}\prod_{m\in {\mathbb Z}_+}
(1-q^{m+\frac{1}{2}})\,\, = \,\, \frac{\eta_D(q^{\frac{1}{2}})}{\eta_D(q)}\,,
\\
f_2(q) & = & q^{-\frac{1}{48}}\prod_{m\in {\mathbb Z}_+}
(1+q^{m+\frac{1}{2}})\,\, = \,\, \frac{\eta_D(q)^2}{\eta_D(q^{\frac{1}{2}})\eta_D(q^2)}\,,
\\
f_3(q) & = & \,\, \,\, q^{\frac{1}{24}}\prod_{m\in {\mathbb Z}_+}
(1+q^{m+1})\,\, = \,\, \frac{\eta_D(q^2)}{\eta_D(q)}\,,
\end{eqnarray}
where
$
\eta_D(q) \equiv q^{1/24}\prod_{n\in {\mathbb Z}_+}(1-q^{n})
$
is the Dedekind $\eta$-function. The linear span of $f_1(q), f_2(q)$
and $f_3(q)$ is $SL(2, {\mathbb Z})$-invariant \cite{Kac}
(
$
\!g\in \left[\begin{array}{cc} a & b \\ c & d \end{array}\right],\,
$
$
g \cdot f(\tau) = f\left(\frac{a\tau + b}{c\tau + d}\right)
$
\!)\footnote{
It can be shown that these are modular forms of weight 0 for the principal congruence subgroup $\Gamma^{48}$ (see also \cite{Bonora11} for $H_*(K3; {\mathbb Z})$ in the integral homology of $K3$).
}.

As before for a closed oriented hyperbolic three-manifolds of the form $X= H^3/\Gamma$ (and any acyclic orthogonal representation of $\pi_1(X)$) the analytic torsion takes the form\footnote{
Vanishing theorems for type $(0, q)$ cohomology $H^q(X; E)$
of Hermitian symmetric spaces $X$ have been formulated in \cite{Floyd,Abdalla}.
}:
$[\tau_{\rm an}(X)]^2 = \cR(0)$, where $\cR(s)$ is the Ruelle function
(\ref{Ruelle}).
Let us introduce next the Ruelle functions $\cR(s),\, \cR(\overline{s}),\, \cR(\sigma),\, \cR(\overline{\sigma})$:
\begin{eqnarray}
\prod_{n=\ell}^{\infty}(1- q^{n+\varepsilon})
& = & \prod_{p=0, 1}Z_\Gamma(s+p(1+it))^{(-1)^p} =
\cR(s=\xi(1-it)),
\label{RF1}
\\
\prod_{n=\ell}^{\infty}(1- \overline{q}^{n+\varepsilon})
& = & \prod_{p=0, 1}Z_\Gamma(\overline{s}+p(1-it))^{(-1)^p} =
\cR(\overline{s}=\xi(1+it)),
\\
\prod_{n=\ell}^{\infty}(1+ q^{n+\varepsilon})
& = & \prod_{p=0, 1}Z_\Gamma(\sigma+p(1+it))^{(-1)^p} =
\cR(\sigma =\xi(1-it) + i\eta(\tau)),
\\
\prod_{n=\ell}^{\infty}(1+ \overline{q}^{n+\varepsilon})
& = & \prod_{p=0, 1}Z_\Gamma(\overline{\sigma}+p(1-it))^{(-1)^p} =
\cR(\overline{\sigma}=\xi(1+it) + i\eta(\tau))\,.
\end{eqnarray}
{ As
$
f_1(q)\cdot f_2(q)\cdot f_3(q) = 1,
$
we get
\begin{equation}
\cR(s=3/2-(3/2)it)\cdot \cR(\sigma =3/2-(3/2)it+i\eta(\tau))
\cdot \cR(\sigma =2-2it+i\eta(\tau)) = 1\,.
\end{equation}
}
\emph{}

\section{Elliptic genera of nonlinear sigma models}
\label{Elliptic}

Elliptic genera of nonlinear sigma models have been discussed extensively elsewhere, thus we shall review them only briefly.
In physical applications, an elliptic genus represents the one-loop partition function of a theory with at least (0,2) supersymmetry. In this theory
the right-moving fermions are all in the {\it R sector}
(in addition we have the partition function of a half-twisted theory),
and possibly the left-moving states are also twisted in some way.
To be more precise, we will consider elliptic genera which are of the form
\begin{equation}
\mbox{Tr } (-)^{F_R} \exp(i \gamma J_L) q^{L_0} \overline{q}^{
\overline{L}_0}\,,
\label{Trace}
\end{equation}
where $q$ is the modular parameter, and the current $J_L$ is a left-moving
$U(1)$ current, which is implicitly assumed to exist.
Computations of such genera have been initiated in \cite{Witten87} and repeatedly discussed in the physical literature. Following the lines of \cite{Ando09}, we shall consider  nonlinear sigma models with (0,2) supersymmetry, defined on a complex K\"ahler manifold $X$ of dimension $n$ with a gauge bundle ${\mathcal E}$ of rank $r$ satisfying
$
\Lambda^{top} {\cE} = K_X
$
$
{\rm ch}_2(TX) = {\rm ch}_2({\cE})
$
We shall assume $X$ to be a Calabi-Yau space (though we shall note special cases in which sensible results can be obtained more generally). It has been shown that the elliptic genus mentioned above is an index \cite{Witten87,Witten88}. Thus, this index is invariant under smooth deformations of the theory, and one can consistently deform the theory to the large-radius limit, where the computation of the index becomes a free-field computation. Since the right-movers are all in the R sector, it is clear that the nonzero modes of the right-moving fermions and bosons cancel out, leaving only the left-movers and right-moving zero modes to contribute. The right-moving zero modes are defined by a Fock vacuum transforming as a spinor lift of $TX$ \cite{Ando09}.
As a result, all of the states appearing in the elliptic genera have spinor indices. It should be stressed again that the elliptic genus is the index of the Dirac operator (as we have seen in Sections \ref{Dirac} and \ref{Lefschetz}) coupled to various bundles defined by the nonzero modes of the fields. To make this more specific, below we list bosonic oscillators at a few mass levels and corresponding bundles, for a nonlinear sigma model on $X$:
\begin{center}
\begin{tabular}{ccc}
Mass level \,\,\, & Oscillator \,\,\,& Bundle
\\
\\
\hline
\\
\\
1 & $\alpha^{\mu}_{-1}$ & $TX$ \\
2 & $\alpha^{\mu}_{-2}$, $\alpha^{\mu}_{-1} \alpha^{\mu}_{-1}$ &
$TX \oplus \mbox{Sym}^2 (TX)$ \\
3 & $\alpha^{\mu}_{-3}$, $\alpha^{\mu}_{-2} \alpha^{\nu}_{-1}$,
$\alpha^{\mu}_{-1} \alpha^{\nu}_{-1} \alpha^{\rho}_{-1}$ &
$TX \oplus \left( TX \otimes TX \right) \oplus
\mbox{Sym}^3(TX)$
\\
\\
\hline
\end{tabular}
\end{center}
At mass level $n$, it is straightforward to check that the bundle
obtained above is the coefficient of $q^n$ in the following
element of the Grothendieck group of vector bundles:
$
\bigotimes_{n\in {\mathbb Z}_+} S_{q^n}(TX)
$.
Each factor of $S_{q^n}(TX)$ corresponds to a set of states
of the form \,
$
\left\{1, \alpha^{\mu}_{-n}, \alpha^{\mu_1}_{-n} \alpha^{\mu_2}_{-n},
\alpha^{\mu_1}_{-n} \alpha^{\mu_2}_{-n} \alpha^{\mu_3}_{-n}, \cdots
\right\}
$
and so the tensor product encodes all products of all nonzero
oscillator creation operators.
Thus, for example, the final result for the elliptic genus will involve
computing the index of a bundle which has, among other things, a factor
of the tensor product. Furthermore, because we have been implicitly working with complex manifolds, holomorphic bundles, and we distinguish $\alpha_{-1}^i$
from $\alpha_{-1}^{\overline{i}}$, we have
$
\bigotimes_{n\in {\mathbb Z}_+} S_{q^n}\left((TX)^{\mathbb C} \equiv
TX \oplus \overline{TX}\right).
$

{\bf (0, 2) supersymmetric nonlinear sigma models.}
For later use let us introduce the following notations:
\begin{eqnarray}
&& B{\left[ \begin{array}{c} \xi \\ \sigma q\end{array} \right]}
(\cP)^{\mathbb C}  :=  \bigotimes_{n\in {\mathbb Z}_+}
S_{\sigma q^n}\left((\xi{\cP})^{\mathbb C}\right)\,, \,\,\,\,\,\,
{\widehat B}{\left[ \begin{array}{c} \xi \\ \sigma q \end{array} \right]}
(\cP)^{\mathbb C}  :=  \bigotimes_{n\in {\mathbb Z}_+/2}
S_{\sigma q^n}\left((\xi{\cP})^{\mathbb C}\right)\,,
\\
&& F{\left[ \begin{array}{c} \zeta  \\ \lambda q \end{array} \right]}
(\cQ)^{\mathbb C}
:=
\bigotimes_{n\in {\mathbb Z}_+}
\Lambda_{\lambda q^n}\left((\zeta{\cQ})^{\mathbb C}\right)\,, \,\,\,\,\,\,
{\widehat F}{\left[ \begin{array}{c} \zeta \\ \lambda q \end{array} \right]}
(\cQ)^{\mathbb C} :=  \bigotimes_{n\in {\mathbb Z}_+/2}
\Lambda_{\lambda q^n}\left((\zeta{\cQ})^{\mathbb C}\right)\,,
\\
&& BF{\left[ \begin{array}{cc} \xi \,\, \zeta\\ \sigma q \,\, \lambda q \end{array} \right]}
(\cP, \cQ)^{\mathbb C}
:= \,\,\,\,
\bigotimes_{n \in {\mathbb Z}_+}S_{\sigma q^n}((\xi\cP)^{\mathbb C})
\,\,\,\,\, \bigotimes_{n\in {\mathbb Z}_+}
\Lambda_{\lambda q^n}\left((\zeta{\cQ})^{\mathbb C}\right)\,,
\\
&& B{\widehat F}{\left[ \begin{array}{cc} \xi \,\, \zeta\\ \sigma q\,\, \lambda q\end{array} \right]}
(\cP, \cQ)^{\mathbb C}
:= \,\,\,\,\,
\bigotimes_{n \in {\mathbb Z}_+}S_{\sigma q^n}((\xi\cP)^{\mathbb C})
\bigotimes_{n\in {\mathbb Z}_+/2}
\Lambda_{\lambda q^n}\left((\zeta{\cQ})^{\mathbb C}\right)\,,
\\
&& {\widehat B}F{\left[ \begin{array}{cc} \xi \,\, \zeta\\ \sigma q\,\, \lambda q\end{array} \right]}(\cP, \cQ)^{\mathbb C}
:=
\bigotimes_{n\in {\mathbb Z}_+/2}S_{\sigma q^n}((\xi\cP)^{\mathbb C})
\,\, \bigotimes_{n\in {\mathbb Z}_+}
\Lambda_{\lambda q^n}\left((\zeta{\cQ})^{\mathbb C}\right)\,,
\\
&& {\widehat B}{\widehat F}{\left[ \begin{array}{cc} \xi \,\, \zeta\\ \sigma q\,\, \lambda q\end{array} \right]}
(\cP, \cQ)^{\mathbb C}
:=
\bigotimes_{n\in {\mathbb Z}_+/2}S_{\sigma q^n}((\xi\cP)^{\mathbb C})
\!\! \bigotimes_{n\in {\mathbb Z}_+/2}
\Lambda_{\lambda q^n}\left((\zeta{\cQ})^{\mathbb C}\right)\,.
\end{eqnarray}
In the (0,2) supersymmetric nonlinear sigma models
we consider, the current $J_L$ exists by virtue of the condition
$\Lambda^{top} {\mathcal E} \cong {\mathcal O}_X$ on ${\mathcal E}$
(and becomes the left R-current in the special case of (2,2) supersymmetry).

{\bf NS-sector genera.}
If the left-moving fermions are in an NS sector, then
the trace (\ref{Trace}) is given by \cite{Witten87,Witten88}
\begin{equation}
q^{-(1/24)(2n + r)}
\int_X {\rm Td}(TX)
\wedge {\rm ch}\left(
B{\widehat F}{\left[ \begin{array}{cc} e^{i\gamma} \,\, 1 \\ q \,\,\,\,\,\, q \end{array} \right]}
(TX, \cE)^{\mathbb C}\right)\,,
\end{equation}
where $S_q(TX)$ is as above,
$\Lambda_q({\mathcal E})$ denotes an element of the Grothendieck
group of vector bundles on $X$, defined as the linear combinations
$\Lambda_q = 1 + \sum_j q^j{\rm Alt}^j(\cE)$
(it arises physically from the left-moving fermion oscillator modes,
just as the factor $B{\left[ \begin{array}{c} 1 \\ q \end{array} \right]}(TX)$ arose from bosonic oscillator
modes), and the ${\mathbb C}$ symbol indicates complexification:
$
(TX)^{\mathbb C} =  T^{1,0}X \oplus \overline{ T^{1,0}X},\,
(z{\mathcal E})^{\mathbb C} =  z{\mathcal E} \oplus \overline{z} \overline{{\mathcal E}}.
$
The prefactor of $q$ is due to the zero energy of the vacuum:
each periodic complex boson contributes $-1/12$, and each
antiperiodic complex fermion contributes $-1/24$.
The fact that the $S_{q^n}$'s are tensored together for integer $n$
reflects the fact that the bosonic oscillators are integrally moded;
the fact that the $\Lambda_{q^n}$'s are tensored together for
half-integer $n$'s reflects the fact that the fermionic oscillators
are half-integrally moded.

{\bf R-sector genera.}
If the left-moving fermions are in a R sector rather than a NS sector,
then the elliptic genus
$
\mbox{Tr}_{{\rm RR}} (-)^{F_R} \exp\left( i \gamma J_L \right)
q^{L_0} \overline{q}^{\overline{L}_0}
$
is given by
\begin{equation} \label{eq:32}
q^{+(1/12)(r - n)} \int_X {\widehat A} (TX)
\wedge {\rm ch}\left(z^{-r/2} \left( \det {\mathcal E} \right)^{+1/2}
\Lambda_1\left( z {\mathcal E}^{\vee}\right)
BF{\left[ \begin{array}{cc} z^{-1} \, 1\\ q \,\,\,\,\,\, q \end{array} \right]}
(TX, \cE)^{\mathbb C}
\right)
\end{equation}
where $z = \exp(-i \gamma)$ (cf. \cite{Witten88}).
In the case $X$ is Calabi-Yau, $\det {\mathcal E}$ is trivial, since
$\Lambda^{\rm top} {\mathcal E} \cong K_X$, so the expression above is
well-defined.
Readers familiar with elliptic genera computations elsewhere
should note that the ``Witten genus'' can be obtained as a special
case of the R sector genus above.  Specifically, for $z=-1$,
the R sector genus above is proportional to
\begin{eqnarray} \label{eq:31}
{\rm Tr}_{RR} (-)^{F_R} (-)^{F_L}
q^{L_0} \overline{q}^{\overline{L}_0}
& = & q^{+(1/12)(r - n)} \int_X {\widehat A} (TX)
\nonumber \\
& \bigwedge&
\!\!\!\!{\rm ch}\left(
\left( \det {\mathcal E} \right)^{+1/2}
\Lambda_{-1}\left(  {\mathcal E}^{\vee} \right)
BF{\left[ \begin{array}{cc} 1 \,\,\,\,\,\, 1 \\ q  -q \end{array}
\right]}
(TX, \cE)^{\mathbb C}
\right)
\end{eqnarray}
which is precisely the Witten genus \cite{Witten87}.
This genus has been shown to play a fundamental role in elliptic cohomology \cite{Ando01}.  When
$z=1$, both the genus \eqref{eq:32} and \eqref{eq:31} are modular
provided the two conditions $\Lambda^{top}{\cE} = K_X,\, {\rm ch}_2(TX)= {\rm ch}_2(\cE)$ hold \cite{Ando09}. However it is not necessary to require that $X$ be Calabi-Yau.

As an example, suppose that Chern polynomial has the form
$c (TX) = \prod_{i} (1+x_{i})$. For $\bigotimes _{n\in {\mathbb Z}_+}S_{q^n}((TX)^{\mathbb C})$ the resulting Chern character is
\begin{eqnarray}
{\rm ch}(\bigotimes_{n\in {\mathbb Z}_+} S_{q^{n}}
((TX)^{\mathbb C})) & = &
{\rm ch}(B{\left[ \begin{array}{c} 1 \\ q\end{array} \right]}
(TX)^{\mathbb C}) =
\prod_i\prod_{n\in {\mathbb Z}_+}[(1-q^n e^{x_i})(1-q^n e^{-x_i})]^{-1}
\nonumber \\
& = &
\prod_i[\cR(s= x_i(1-it))\cdot \cR(s= -x_i(1-it))]^{-1}\,.
\end{eqnarray}

The factor of
\begin{equation}   \label{fock-vacua-reln}
z^{-r/2} \left( \det {\mathcal E} \right)^{+1/2} \Lambda_1( z {\mathcal E}^{\vee} )
\: = \:
z^{+r/2} \left( \det {\mathcal E} \right)^{-1/2} \Lambda_1( z^{-1} {\mathcal E} )
\end{equation}
arises above from the zero modes of the left-moving fermions.
It reflects the ambiguity in the Fock vacuum:
if we define $|0\rangle$ by $\lambda_-^a |0\rangle = 0$\,\,
(or $|0 \rangle$ by $\lambda_-^{\overline{a}} |0\rangle = 0$)
then we have a set of vacua
\begin{equation}
| 0 \rangle, \:
\lambda_-^{\overline{a}} | 0 \rangle, \:
\cdots,
\lambda_-^{\overline{a}_1} \cdots \lambda_-^{\overline{a}_r}
| 0 \rangle\,\,\,\,\,\,\,\,
({\rm or}\,\,\,\,| 0 \rangle, \:
\lambda_-^{a} | 0 \rangle, \:
\cdots,
\lambda_-^{a_1} \cdots \lambda_-^{a_r}
| 0 \rangle)\,.
\end{equation}
The existence of these two equivalent characterizations of the Fock
vacua corresponds to the two sides of equation~(\ref{fock-vacua-reln}).
Furthermore, these vacua correspond to spinor lifts of ${\mathcal E}$:
note that we can write
$
\Lambda_1( z {\mathcal E}^{\vee} ) =
{\mathcal S}_+(z {\mathcal E}^{\vee}) \oplus {\mathcal S}_-(z{\mathcal E}^{\vee}),
$
where
${\mathcal S}_{\pm}$ denote the two chiral Spin$^c$ lifts
of ${\mathcal E}^{\vee}$, {\it i.e.}
\begin{equation}
{\mathcal S}_+({\mathcal E}^{\vee}) \equiv  \bigoplus_{n \: {\rm even}}
\Lambda^n {\mathcal E}^{\vee}\,,\,\,\,\,\,\,\,\,
{\mathcal S}_-({\mathcal E}^{\vee})  \equiv  \bigoplus_{n \: {\rm odd}}
\Lambda^n {\mathcal E}^{\vee}
\end{equation}
which are made into honest spinors via the $\sqrt{ \det {\mathcal E} }$
factors.
(Physically, every vector bundle comes with a hermitian fiber metric,
so we will often fail to distinguish ${\mathcal E}^{\vee}$ from
$\overline{ {\mathcal E} }$.)
The prefactor of $q$
is due to the vacuum zero energy:  each periodic complex boson contributes
$-1/12$, and each periodic complex fermion contributes $+1/12$.

Note that in the spinor lifts of ${\mathcal E}$, ${\mathcal E}$ is not
complexified, unlike the nonzero modes.  This is because, for the
ambiguity in the Fock vacuum, we use the relation $\{ \psi_0^i,
\psi_{0 j} \} \propto \delta^i_j$, so we take one of either
$\psi_0^i$, $\psi_{0 j}$ to be creation operators and the other to
be annihilation operators -- the choice does not matter, as the resulting
collection of states are the same.

\subsection{Elliptic genera of Landau-Ginzburg models over
vector spaces}
\label{Landau}

{\bf R-sector genera.} We have already reviewed
elliptic genus computations
in nonlinear sigma models; next, let us review the computation of
elliptic genera in Landau-Ginzburg models on topologically trivial
spaces, with quasi-homogeneous superpotentials,
as first discussed in \cite{Witten94}.  In particular, we will focus
on the special case of a Landau-Ginzburg model over the complex
line, with a monomial superpotential.
In \cite{Witten94} a Landau-Ginzburg model over the complex
line ${\mathbb C}$ was considered, with superpotential $W = \Phi^{k+2}$.
The elliptic genus was defined there as the trace
\begin{equation}
\mbox{Tr } (-)^{F_R} q^{L_0} \overline{q}^{\overline{L}_0}
\exp(i \gamma J_L)
\end{equation}
over states. The boundary conditions along timelike directions require more explanation. Let us work out the left R-charges of the fields, so as to understand the $\exp(i \gamma J_L)$ factor in the trace.
Because of the superpotential interactions, the left R-symmetry no longer merely rotates the $\psi_-$'s by a phase, leaving other fields invariant, but rather rotates all of the fields by some phase.
It is straightforward to check that the
left R-charges are as follows:
\begin{center}
\begin{tabular}{ccc}
Field \,\,\,\,\,\,\,\, & R-charge \,\,\,\,\,\,\,\, & R boundary conditions \\
\\
\hline
\\
$\phi$ \,\,\,\,\,\,\,\, & $1$ \,\,\,\,\,\,\,\, & $\phi(x_1+1,x_2) =  \phi(x_1,x_2)$\\
$\psi_+$ \,\,\,\,\,\,\,\, & $1$  \,\,\,\,\,\,\,\,& $\psi_+(x_1+1,x_2) =  \psi_+(x_1,x_2)$\\
$\psi_-$  \,\,\,\,\,\,\,\,& $-(k+1)$  \,\,\,\,\,\,\,\,
& $\psi_-(x_1+1,x_2) = \psi_-(x_1,x_2)$
\\
\\
\hline
\end{tabular}
\end{center}
Furthermore, also because of the superpotential interactions,
$(-)^{F_R}$ no longer merely corresponds to a sign on $\psi_+$'s;
rather, it generates a sign on both $\psi_+$ and $\psi_-$ simultaneously.

We do not list here the timelike boundary conditions, but the
attentive reader should recall, for example, that
fields with Ramond boundary conditions along
timelike directions correspond to traces with $(-)^F$ factors.
The zero modes of $\psi_-$ contribute a factor of
$
\exp(- i \gamma(k+1)/2 ) - \exp( +i \gamma (k+1)/2 ),
$
the zero modes of $\psi_+$ contribute a factor of
$
\exp(i \gamma/2) - \exp(- i \gamma/2).
$

{\it Fermion contribution.} The nonzero modes of the fermions contribute
\begin{eqnarray}
&&
\prod_{n\in {\mathbb Z}_+}[1-z^{k+1}q^n][1-z^{-(k+1)}q^n]
[1-z^{-1}\overline{q}^n][1-z\overline{q}^n]
\nonumber \\
\!\!\!\!\!\!\!\!\!\!\!\!\!\!\!\!
&&
= \cR(s= -i\gamma(k+1)(1-it))\cdot \cR(s= i\gamma(k+1)(1-it))
\nonumber \\
\!\!\!\!\!\!\!\!\!\!\!\!\!\!\!\!
&&
\times \,\,
\cR(\overline{s}= -i\gamma(1-it))\cdot
\cR(\overline{s}= i\gamma(1-it))\,,
\end{eqnarray}
where as before $z = \exp(-i \gamma)$ and
where the minus signs are due to the $(-)^{F_R}$ factor in the trace
(and the fact that because of the superpotential interactions,
$(-)^{F_R}$ multiplies both $\psi_-$ and $\psi_+$ simultaneously by
a sign). We can rewrite this in the form of the index of a Dirac operator. If we let $L$ denote the tangent bundle of ${\mathbb C}$ restricted to the origin, then the expression above for the contribution from the nonzero modes of the fermions is of the form
\begin{equation}
\mbox{ch}\left(
F{\left[ \begin{array}{c} z^{k+1} \\ -q \end{array} \right]}
(L)^{\mathbb C}
F{\left[ \begin{array}{c} z^{-1} \\ -q \end{array} \right]}
(L)^{\mathbb C}\right)
\end{equation}

{\it Boson contribution.} The nonzero modes of the bosons contribute
\begin{eqnarray}
\!\!\!\!\!\!\!\!\!\!\!\!\!\!\!\!
&&
\prod_{n\in {\mathbb Z}_+}\left([1-zq^n][1-z^{- 1}q^n]
[1-z\overline{q}^n][1-z^{- 1}\overline{q}^n]\right)^{-1}
\nonumber \\
\!\!\!\!\!\!\!\!\!\!\!\!\!\!\!\!
&&
= \left[\cR(s= -i\gamma(1-it))\cdot\cR(s= i\gamma(1-it)) \right]^{-1}
\nonumber \\
\!\!\!\!\!\!\!\!\!\!\!\!\!\!\!\!
&&
\times \, \left[\cR(\overline{s}= -i\gamma(1-it))\cdot
\cR(\overline{s}= i\gamma(1-it))\right]^{-1}\,.
\end{eqnarray}
which we can rewrite as
\begin{equation}
\mbox{ch}\left(
B{\left[ \begin{array}{c} z^{-1} \\ q \end{array} \right]}
(L)^{\mathbb C}
B{\left[ \begin{array}{c} z^{-1} \\ -\overline{q} \end{array} \right]}
(L)^{\mathbb C}\right)
\end{equation}
Note that the $\overline{q}$ contributions from the bosons and fermions
cancel out; this can be seen also directly from the product formula above.

{\it Case of zero modes.}  Finally, in this special case, the zero modes\footnote{
Taking into account the timelike boundary conditions, there are no
zero modes from the point of view of path integral quantization.
The $\phi_0$, $\overline{\phi}_0$ referred to here are solely an artifact of the periodic moding in canonical quantization.
}
$\phi_0$, $\overline{\phi}_0$ also contribute
a factor of $[(1- z)(1 - z^{-1})]^{-1}$ as discussed in \cite{Witten94}. Putting this together, we get the genus
\begin{eqnarray}\label{eq:9}
\!\!\!\!\!\!
f (z,q) & = & \frac{ z^{-1/2} \left( z^{(k+1)/2}  \: - \: z^{-(k+1)/2}
\right)}{\left( 1 \: - \: z^{-1} \right)}
\prod_{n\in {\mathbb Z}_{+}} \frac{
\left( 1 \: - \: z^{k+1} q^n   \right)
\left( 1 \: - \: z^{-(k+1)} q^n  \right)
}{
\left( 1 \: - \: z^{-1} q^n  \right)
\left( 1 \: - \: z q^n \right)
}
\nonumber \\
& = &
\frac{\sin \frac{\gamma}{2}(k+1)}{\sin \frac{\gamma}{2}}
\left[
\frac{\cR(s= -i\gamma(k+1)(1-it))\cdot \cR(s= (i\gamma(k+1)(1-it))}
{\cR(\overline{s}= -i\gamma(1-it))\cdot
\cR(\overline{s}= i\gamma (1-it))}\right]
\end{eqnarray}
We can interpret this as index theory over a point, the fixed-point locus of a $U(1)$ action, {\it i.e.} the space
of bosonic zero modes satisfying the boundary condition
$
\phi(x_1,x_2+1)= \exp(i \gamma) \phi(x_1,x_2)
$
for generic $\gamma$, which is to say, $\phi_0 = \{ 0 \}$.
Note that the expression for the nonzero
modes is given by
\begin{equation}
\mbox{ch} \left(
B{\left[ \begin{array}{c} 1 \\ z^{-1}q \end{array} \right]}
(L)^{\mathbb C}
B{\left[ \begin{array}{c} 1 \\ zq \end{array} \right]}
({\overline L})^{\mathbb C}
F{\left[ \begin{array}{c} 1 \\ -z^{k+1}q \end{array} \right]}
(L)^{\mathbb C}
F{\left[ \begin{array}{c} 1 \\ -z^{-k-1}q \end{array} \right]}
({\overline L})^{\mathbb C} \right)
\end{equation}
Instead of working with fields with R boundary conditions along spacelike directions, one could instead try to
compute elliptic genera in which the $\psi_-$
have NS boundary conditions in spacelike directions.  Note, however,
that because of the $\psi_+ \psi_- \phi^k$ Yukawa coupling in the theory, if the $\psi_-$ have NS boundary conditions, then so too must the $\psi_+$, and then the right-moving contributions would no longer cancel out.

{\bf NS-sector genera.}
In the last subsection we reviewed the results of \cite{Witten94} on computing elliptic genera of Landau-Ginzburg models over vector spaces,
in the R sector.  In this subsection we will extend the results of
\cite{Witten94} to the NS sector.

From the table of left R-charges in the last subsection,
we see that fields in the NS sector have spacelike boundary conditions
\,\,
$
\phi(x_1+1,x_2) = - \phi(x_1,x_2),\,\,
\psi_+(x_1+1,x_2)= - \psi_+(x_1,x_2),\,\,
\psi_-(x_1+1,x_2)= (-)^{k+1} \psi_-(x_1,x_2).
$
From the spacelike boundary conditions above, we see that we must
consider the cases of $k$ even and odd separately.

{\it The case $k$ is even.}
In the case $k$ is even, there are no zero modes at all.
The fermions contribute
\begin{eqnarray}
&&
\prod_{n\in {\mathbb Z}_+/2} \left[ 1 - z^{k+1} q^n \right]
\left[ 1 - z^{-(k+1)} q^n \right]
\left[ 1 - z^{-1} \overline{q}^n \right]
\left[ 1 - z \overline{q}^n \right]
\nonumber \\
&&
= \,\,\cR(s= -i\gamma(k+1)(1-it))\cdot \cR(s= i\gamma(k+1)(1-it))
\nonumber \\
&&
\times
\,\,\,\,\cR(\overline{s}= -i\gamma(1-it))\cdot
\cR(\overline{s}= i\gamma(1-it))
\end{eqnarray}
and the bosons contribute
\begin{eqnarray}
&&
\prod_{n \in {\mathbb Z}_+/2}
\left[ 1 - z^{-1} q^n \right]^{-1}
\left[ 1 - z q^n \right]^{-1}
\left[ 1 - z^{-1} \overline{q}^n \right]^{-1}
\left[ 1 - z \overline{q}^n \right]^{-1}
\nonumber \\
&&
= \,\,\,\, \left[\cR(s= -i\gamma(1-it))\cdot \cR(s= i\gamma(1-it)) \right]^{-1}
\nonumber \\
&&
\times \,\,\,\, \left[\cR(\overline{s}= -i\gamma(1-it))\cdot
\cR(\overline{s}= i\gamma(1-it))\right]^{-1}\,.
\end{eqnarray}
Putting this together, we see that for $k$ even, the elliptic genus is
given by
\begin{eqnarray}
&&
\prod_{n\in {\mathbb Z}_+/2}
\left[ 1 - z^{k+1} q^n \right]
\left[ 1 - z^{-(k+1)} q^n \right]
\left[ 1 - z^{-1} q^n \right]^{-1}
\left[ 1 - z q^n \right]^{-1}
\nonumber \\
&&
= \left[\frac{\cR(s= -i\gamma(1-it))\cdot \cR(s= i\gamma(1-it))}
{\cR(s = -i\gamma(1-it))\cdot
\cR(s= i\gamma(1-it))}\right]\,.
\end{eqnarray}

{\it The case $k$ is odd.}
In the case $k$ is odd, there are $\psi_-$ zero modes.
In this case, the total contribution from the fermions is
\begin{eqnarray}
\!\!\!\!\!\!\!
&&
\left( z^{(k+1)/2} - z^{-(k+1)/2} \right)
\prod_{n\in {\mathbb Z}_+}
\left[ 1 - z^{k+1} q^n \right]
\left[ 1 - z^{-(k+1)} q^n \right]
\prod_{n \in {\mathbb Z}_+/2}
\left[ 1 - z^{-1} \overline{q}^n \right]
\left[ 1 - z \overline{q}^n \right]
\nonumber \\
&&
= -2i \sin \frac{\gamma}{2}(k+1)
\cR(s= -i\gamma(k+1)(1-it))\cdot \cR(s= i\gamma(k+1)(1-it))
\nonumber \\
&{}&
\times \,\, \cR(s = -i\gamma(1-it))\cdot
\cR(s = i\gamma (1-it))
\end{eqnarray}
and the bosons contribute
\begin{displaymath}
\prod_{n\in {\mathbb Z}_+/2}
\left[ 1 - z^{-1} q^n \right]^{-1}
\left[ 1 - z q^n \right]^{-1}
\left[ 1 - z^{-1} \overline{q}^n \right]^{-1}
\left[ 1 - z \overline{q}^n \right]^{-1}
\end{displaymath}
Putting this together, we see that for $k$ odd, the elliptic genus is
given by
\begin{eqnarray}
\!\!\!\!\!\!\!\!\!\!
&&
\left( z^{(k+1)/2} - z^{-(k+1)/2} \right)
\prod_{n\in {\mathbb Z}_+}
\left[ 1 - z^{k+1} q^n \right]
\left[ 1 - z^{-(k+1)} q^n \right]
\prod_{n\in {\mathbb Z}_+-1/2}
\left[ 1 - z^{-1} q^n \right]^{-1}
\left[ 1 - z q^n \right]^{-1}
\nonumber \\
\!\!\!\!\!\!\!\!\!\!
&&
= -2i \sin \frac{\gamma}{2}(k+1) \left[
\frac{\cR(s= -i\gamma(k+1)(1-it))\cdot \cR(s= (i\gamma(k+1)(1-it))}
{\cR({s}= -i\gamma(1-it))\cdot
\cR({s}= i\gamma (1-it))}\right]
\end{eqnarray}

\section{Chiral primary states}
\label{Chiral}

In this section we will study the $CY_3$ black hole component of chiral primaries. A very important notion in string theory is that of a ${\cN}=2$ superconformal field theory. The primary chiral fields of a ${\cN}=2$ superconformal field theory form an algebra \cite{Lerche,Warner}.
Superconformal quantum mechanics describes $D0$ branes in the $AdS_{2}\times S^2\times CY_3$ attractor geometry of a Calabi-Yau black hole with $D4$ brane charges \cite{Gaiotto}. This superconformal theory contains a large degeneracy of chiral primary bound states. The degeneracy arises from $D0$ branes in the lowests Landau level, which tile the
$CY_3\times S^2$ horizon, and apparently such a multi-$D0$ brane $CFT_1$ is holographically dual to IIA string theory on $AdS_2\times S^2\times CY_3$. It is important to note the appearance of a dual $CFT_1$, potentially relevant for non-BPS states (for details, see \cite{Gaiotto}).

We summarize the results for the case when a $D2$ brane wraps the horizon $S^2$ and carries $N$ units of magnetic flux thereby including $N$ units of $D0$ charge. The $D2$ brane can be considered as a point particle in $CY_3$ with a two-form magnetic field $F_{CY}$.
We are interested in the $CY_3$ component of chiral primaries.
The magnetic field divides the $CY_3$ into
${\Phi} = (1/6)\int F_{CY}\wedge F_{CY}\wedge F_{CY}$ cells, corresponding to the lowest Landau levels. The chiral primary conditions can be written as
$
{\overline{D}\omega =\overline{D}^*\omega=0, \label{eq:chp}}
$
where $\omega$ is a $(p,q)$ form
and $\overline D$ is the holomorphic covariant derivative with connection. Solutions of this equation are in one-to-one correspondence with the elements of $H^q(CY_3, \Omega^p\otimes \mathfrak{L})$, where ${\mathfrak{L}}$ is the line bundle
for which $c_1({\mathfrak{L}})=[F_{CY}]$; \,$c_1({\mathfrak{L}})$ is the first Chern class of ${\mathfrak{L}}$. $H^q(CY_3,
\Omega^p\otimes \mathfrak{L})$ vanishes for $q>0$ and large
$c_1(\mathfrak{L})$. Thus we need compute the cohomology
$H^q (X, \Omega^p\otimes \mathfrak{L})$ for the moduli space of a brane on $X$; the black hole entropy counting can be reduced to a cohomology problem. In some cases one can
compute the dimension of $H^q(X, \Omega^p \otimes \mathfrak{L})$
using mirror symmetry\footnote{Let $X$ be mirror to $Y$, then because of homological mirror symmetry, a 0-brane on $X$ is mirror to a three-torus $T^3$ on $Y$.
In addition, the moduli space of a $D3$-brane wrapping such a $T^3$ on $Y$ is $X$. One can compute
$H^q(X, \Omega^p \otimes \mathfrak{L})$ and, therefore, the black hole entropy, using mirror symmetry \cite{Aspinwall}.} while $\dim H^0(CY_3, \Omega^p\otimes \mathfrak{L})$ can be obtained using the Riemann-Roch formula, the result being (see also \cite{Gaiotto})
\begin{eqnarray}
h_0 & = & {\rm dim}\, H^0(CY_3, {\mathfrak{L}})\,\,\,\,\,\,\,\,\,\,\,\,\,\, =
\int \left[ \frac{F_{CY}^3}{6}+\frac{c_2\wedge F_{CY}}{12} \right]
\,\,\,\,\,\,\,\,\,\,\,\,\,\,\,\,\, = {\Phi} + \frac{1}{12}c_2\cdot {C},
\nonumber \\
h_1 & = & {\rm dim}\, H^0(CY_3, \Omega^1\otimes {\mathfrak{L}}) =
\int \left[
\frac{F_{CY}^3}{2}-\frac{3c_2\wedge F_{CY}}{4}+\frac{c_3}{2}\right]
= 3{\Phi} -\frac{3}{4}c_2\cdot {C}-\frac{\chi}{2},
\nonumber \\h_2 & = & {\rm dim}\, H^0(CY_3, \Omega^2\otimes {\mathfrak{L}}) =
\int \left[
\frac{F_{CY}^3}{2}-\frac{3c_2\wedge F_{CY}}{4}-\frac{c_3}{2}
\right] = 3{\Phi} -\frac{3}{4}c_2\cdot {C} +\frac{\chi}{2},
\nonumber \\
h_3 & = & {\rm dim}\, H^0(CY_3, \Omega^3\otimes {\mathfrak{L}}) = h_0
\mbox{.}
\end{eqnarray}
Here $c_k$ is the $k$-th Chern class, $\chi$ the Euler
characteristic, and $C$ is a four-cycle of a certain manifold (for more
notations and details see, for example, \cite{Maldacena1,Gaiotto}).

{\bf The partition function.}
We now have to count the multiparticle primaries, choosing a basis of states. We can consider partitioning the $D0$ branes into $k$ clusters of $n_\ell$ $D0$ branes each, such that $\sum_{\ell = 1}^kn_\ell = N$.
Each cluster forms a wrapped $D2$ brane (we ignore the possibility of multiwrapped $D2$ bound states) with $n_\ell$ units of magnetic flux. Then, each one of the such $D2$ branes (there are $k$ of them) can sit in one of the $\sum h_j$ chiral primary states.
Actually the counting of configurations is in one to one
correspondence with the counting of states for conformal field theory
with $\sum_{(j\,\, {\rm even})}h_j$ bosons and $\sum_{(j\,\,
{\rm odd})}h_j$ fermions and total momentum $N$.
The partition function can be calculated by taking the trace over chiral primaries, with the result
\begin{equation}
Z_{\rm CFT}(\tau) =
\prod_n \frac{(1+q^n)^{\sum_{(j\,\,{\rm odd})}h_j}}
{(1-q^n)^{\sum_{(j\,\,{\rm even})}h_j}}
= \frac{\left[\cR(\sigma = 1-it + i\eta(\tau))
\right]^{\sum_{(j\,\,{\rm odd})}h_j}}
{\left[\cR(s= 1-it)\right]^{\sum_{(j\,\,{\rm even})}h_j}}\,.
\label{generating}
\end{equation}

\section{Hilbert schemes of points}
\label{HH}

In this section we will study the moduli space
parameterizing 0-dimensional subschemes of length $n$ in a nonsingular quasi-projective surface $X$ over $\mathbb C$. It is called {\it the Hilbert scheme of points}, and denoted by $X^{[n]}$.
A simple example of a 0-dimensional subscheme is a collection of distinct points. The length is equal to the number of points. In the case when some points collide, more complicated subschemes appear. (When two points collide, we get infinitely many
nearby points, which makes a pair of a point x and and a 1-dimensional subspace of the tangent space $T_xX$.) This marks the difference between $X^{[n]}$ and the $n$-th symmetric product $S^nX$ in which the information of the one-dimensional subspace is lost.

When ${\rm dim}\,X = 1$ the Hilbert scheme $X^{[n]}$ is isomorphic
to $S^nX$, while for ${\rm dim}\,X = 2$, $X^{[n]}$ is smooth and
there is a morphism $\pi : X^{[n]}\rightarrow S^nX$ which is a
resolution of the singularities, by a result of \cite{Fogarty}.
This case is in a contrast with Hilbert schemes for ${\rm dim}\,X
> 2$. For a projective $X$ the scheme is also projective, and it
follows from Grothendieck's construction of Hilbert schemes. Note
a nontrivial example which is a result of \cite{Beauville}:
$X^{[n]}$ has a holomorphic symplectic form when $X$ has one.

Interesting noncompact examples are: $X = {\mathbb C}^2$ or
$X = T^{*}\Sigma$ where $\Sigma$ is a Riemann surface. For these examples there exists a ${\mathbb C}^{*}$-action, which naturally induces an action on $X^{[n]}$ \cite{Nakajima99}.
One can construct a representation of products of Heisenberg and Clifford algebras on the direct sum of homology groups of all
components $\bigoplus_n H_{*}(X^{[n]})$. Thus the generating function can be interpreted as a character of the Heisenberg algebra (see also Sect. \ref{Super}).

The relation between $X^{[n]}$ and
$S^nX$ have many similarities with the relation between string and field theory. Note, for example, the $S$ duality (or Montonen-Olive duality) conjecture implies that the generating function of Euler numbers of moduli spaces of instantons has modular
properties  \cite{Vafa94}. For the base $K3$ surface, the Euler numbers of moduli spaces of instantons are the same as those of Hilbert schemes of points (strictly speaking, we must consider moduli spaces of stable sheaves instead of moduli spaces of instantons, which are usually noncompact). Then G\"{o}ttsche's formula (see Eq.~(\ref{GFP}) below) gives the desired answer.
Moreover, one can see that homology groups of moduli spaces of sheaves on an ALE space (the minimal resolution of a simple singularity) form an integrable representation of an affine Lie algebra \cite{Nakajima94}.
The modular properties of the characters of the representation (or the
string functions) were explained in the language of the conformal field theory \cite{Kac}. Perhaps Heisenberg algebras and affine Lie algebras could be understood in the framework of the heterotic-type IIA duality \cite{Vafa96,Harvey98}.

\subsection{Hilbert schemes of points on surfaces}
\label{Surface}

As preliminaries to the subject of symmetric products and their connection with spectral functions, we explain briefly the relation between the Heisenberg algebra and its representations, and the Hilbert scheme of points, mostly following the lines of \cite{Nakajima99}. To be more specific:
\begin{itemize}
\item{}
Recall that the infinite dimensional Heisenberg algebra (or, simply, the Heisenberg algebra) plays a fundamental role in the representation theory of the affine Lie algebras. As it has been discussed in Sect. \ref{Super}, an important representation of the Heisenberg algebra is the Fock space representation on the polynomial ring of infinitely many variables. The degrees of polynomials (with different degree variables) give a direct sum decomposition of the representation, which is called weight space decomposition.
\item{}
The Hilbert scheme of points on a complex surface appears in the algebraic geometry. The Hilbert scheme of points decomposes into infinitely many connected components according to the number of points. Betti numbers of the Hilbert scheme have been computed in \cite{Gottsche}. The sum of the Betti numbers of the Hilbert scheme of $N$-points is equal to the dimension of the subspaces of the Fock space representation of degree $N$.
\end{itemize}
\begin{remark}
If we consider the generating function of the Poincar\'{e} polynomials associated with set of points we get the character of the Fock space representation of the Heisenberg algebra.
The character of the Fock space representation of the Heisenberg algebra {\rm (}in general the integrable highest weight representations of affine Lie algebras{\rm )} are known to have modular invariance as has been proved in {\rm \cite{Kac2}}. This occurrence is naturally explained through the relation to partition functions of conformal field theory on a torus. In this connection
the affine Lie algebra has close relation to conformal field theory.
\end{remark}

{\bf Algebraic preliminaries.}
Let ${R} = {\mathbb Q}[p_1, p_2, ...]$
be the polynomial ring of infinite many variables $\{p_j\}_{j=1}^{\infty}$. Define $P[j]$ as $j\partial/\partial p_j$
and $P[- j]$ as a multiplication of $p_j$ for each positive $j$. Then the commutation relation holds:
$
[\,P[i],\, P[j]\,] = i\delta_{i+j, 0}\,{\rm Id}_{R},\,
$
$i, j \in {\mathbb Z}/\{0\}.$ We define the infinite dimensional Heisenberg algebra as a Lie algebra generated by $P[j]$ and $K$ with defining relation
\begin{equation}
[\,P[i],\, P[j]\,] = i\delta_{i+j, 0} K_{R},\,\,\,\,\,
[\,P[i],\, K\,] = 0,\,\,\,\,\, i, j \in {\mathbb Z}/\{0\}.
\end{equation}
The above ${R}$ labels the representation.
If $1\in {R}$ is the constant polynomial, then $P[i]1 = 0,\, i\in {\mathbb Z_+}$ and
\begin{equation}
{R} =  {\rm Span}\{ P[-j_1] \cdots  P[-j_k]\,1\mid
k\in {\mathbb Z}_+\cup \{0\}, \,\,\, j_1, \dots, j_k \in
{\mathbb Z}_+\}\,.
\end{equation}
1 is a highest weight vector.
This is known in physics as the {\it bosonic Fock space}.
The operators $P[j]\, (j< 0)$ ($P[j]\, (j>0)$) are the {\it creation {\rm (}annihilation{\rm )} operators}, while 1 is  the {\it vacuum vector}.

Define the degree operator
${\cD}: {R}\rightarrow {R}$ by \,
$
{\cD}(p_1^{m_1} p_2^{m_2}\cdot\cdot\cdot \,)\stackrel{def}{=}
(\sum_i i m_i) p_1^{m_1} p_2^{m_2}\ldots
$
The representation ${R}$ has $\cD$ eigenspace decomposition; the eigenspace with eigenvalue $N$ has a basis
$
p_1^{m_1} p_2^{m_2}\cdot\cdot\cdot
(\sum_i i m_i) = N.
$
Recall that partitions of $N$ are defined by a non-increasing sequence of nonnegative integers $\nu_1\geq \nu_2 \geq \ldots$ such that $\sum_\ell \nu_\ell = N$. One can represent $\nu$ as $(1^{m_1}, 2^{m_2}, \cdot\cdot\cdot)$ (where 1 appears $m_1$-times, 2 appears $m_2$-times, \ldots in the sequence). Therefore, elements of the basis corresponds bijectively to a partition $\nu$. The generating function of eigenspace dimensions, or the {\it character}
in the terminology of the representation theory, is well--known to have the form
\begin{equation}
{\rm Tr}_{R}\, q^{\cD}\, \stackrel{def}{=}
\sum_{N\in {\mathbb Z}_+\cup \{0\}}
q^N {\rm dim}\,\{ r\in {R}\, \mid \, {\cD}r = Nr\,\}
= \prod_{n\in {\mathbb Z}_+} (1 - q^n)^{-1}\,.
\end{equation}

Let us define now the Heisenberg algebra associated with a finite dimensional $\mathbb Q$-vector space $V$ with non-degenerate symmetric bilinear form $(\, ,\, )$. Let $W = (V\otimes t\,{\mathbb Q}[t])\oplus (V\otimes t^{-1}\,{\mathbb Q}[t^{-1}])$, then define a skew-symmetric bilinear form on $W$ by
$(r\otimes t^i,\,s\otimes t^j) = i \delta_{i+j, 0} (r, s)$.
The Heisenberg algebra associated with $V$ can be defined as follows: we take the quotient of the free algebra $L(W)$ divided
by the ideal $\mathcal I$ generated by $[r,\, s] - (r,\,s)1\,\, (r, s\in W)$. It is clear that when $V = {\mathbb Q}$ we have the above Heisenberg algebra. For an orthogonal basis
$\{ r_j\}_{j=1}^p$ the Heisenberg algebra associated with $V$ is isomorphic to the tensor product of $p$-copies of the above Heisenberg algebra.

Let us consider next the {\it super}-version of the Heisenberg algebra, the super-Heisenberg algebra. The initial data are a vector space
$V$ with a decomposition $V = V_{\rm even}\oplus V_{\rm odd}$
and a non-degenerate bilinear form satisfying
$(r, s) = (-1)^{|r||s|}(r, s)$. In this formula $r, s$ are either elements of $V_{\rm even}$ or $V_{\rm odd}$, while $|r| = 0$ if
$r \in V_{\rm even}$ and $|r|=1$ if $r\in V_{\rm odd}$.
As above we can define $W$, the bilinear form on $W$, and $L(W)/{\mathcal I}$, where now we replace the Lie bracket
$[\, ,\,]$ by the super-Lie bracket. In addition, to construct the free-super Lie algebra in the tensor algebra we set
$
(r\otimes t^i,\,s\otimes t^j) = (r \otimes t^i)(s \otimes t^j)
+ (s \otimes t^j)(r \otimes t^i)
$
for $r, s \in V_{\rm odd}$.
By generalizing the representation on the space of polynomials of infinite many variables one can get a representation of the super-Heisenberg algebra on the symmetric algebra ${R}= S^*(V \otimes t\, {\mathbb Q} [t])$ of the positive degree part $V\otimes t\, {\mathbb Q}\,[t]$. As above we can define the degree operator $\cD$. The following character formula holds:
\begin{equation}
{\rm Tr}_{R}\, q^{\cD} = \prod_{n\in {\mathbb Z}_+}
\frac{(1 + q^n)^{{\rm dim}\, V_{\rm odd}}}
{(1 - q^n)^{{\rm dim}\, V_{\rm even}}}
=
\left[\frac{\cR(\sigma = 1-it+ i\eta(\tau))^{{\rm dim}\, V_{\rm odd}}}
{\cR(s = 1-it)^{{\rm dim}\, V_{\rm even}}}\right]\,.
\label{trace-usual}
\end{equation}
Counting the odd degree part by $-1$ we can replace the usual trace by the super-trace (denoted by ${\rm STr}$) and get the result\footnote{In the case when $V$ has one-dimensional odd degree part only
(the bilinear form is $(r, r) =1$ for a nonzero vector $r\in V$) the above condition is not satisfied. We can modify the definition of the corresponding super-Heisenberg algebra by changing the bilinear form on $W$ as $(r\otimes t^i, r\otimes t^j) = \delta_{i+j, 0}$. The resulting algebra is called {\it infinite dimensional Clifford algebra}. The above representation $R$ can be modified as follows and it is the {\it fermionic Fock space} in physics. The representation of the even degree part was realized as the space of polynomials of infinity many variables; the Clifford algebra is realized on the exterior algebra
${R} = \wedge^*(\bigoplus_j{\mathbb Q} dp_j)$ of a vector space with a basis of infinity many vectors. For $j>0$ we define $r\otimes t^{- j}$ as an exterior product of $dp_j$, $r\otimes t^j$ as an interior product of $\partial/\partial p_j$.
}
\begin{equation}
{\rm STr}_{R}\, q^{\cD} = \prod_{n\in {\mathbb Z}_+}
(1 - q^n)^{{\rm dim}\, V_{\rm odd} - {\rm dim}\, V_{\rm even}}
= \cR(s = 1-it)^{{\rm dim}\, V_{\rm odd} - {\rm dim}\, V_{\rm even}}\,.
\end{equation}

\subsection{G\"{o}ttsche's formula}
\label{Gottsche}

For a one-dimensional higher variety (i.e. for a surface) the following results hold:
\begin{itemize}
\item{}
For a Riemann surface (${\rm dim}\,X = 1$) $\cS^NX$ and $X^{N}$
are isomorphic under the Hilbert-Chow morphism.
\item{}
If $X$ is a nonsingular quasi-projective surface, the Hilbert-Chow morphism $\pi: \, X^{[N]}\rightarrow \cS^NX$ gives a resolution of singularities of the symmetric product $\cS^NX$ \cite{Fogarty}. In particular $X^{[N]}$ is a nonsingular quasi-projective variety of dimension $2N$.
\item{}
If $X$ has a symplectic form, $X^{[N]}$ also has a symplectic form.
For $N= 2$ it has been proved in \cite{Fujiki}, for $N$ general in \cite{Beauville}.
\item{}
The generating function of the Poincar\'{e} polynomials $P_t(X^{[N]})$ of $X^{[N]}$ is given by
\begin{eqnarray}
\!\!\!\!\!\!\!\!\!\!\!
\sum_{N=0}^{\infty}q^N\, P_t(X^{[N]})
& = &
\prod_{n\in {\mathbb Z}_+} \frac{(1+t^{2n-1}q^n)^{b_1(X)}(1+t^{2n+1}q^n)^{b_3(X)}}
{(1-t^{2n-2}q^n)^{b_0(X)}(1-t^{2n}q^n)^{b_2(X)}
(1-t^{2n+2}q^n)^{b_4(X)}}
\nonumber \\
& = &
\frac{\prod_{j=0, 2}\,\,\,\,[\cR(\sigma = (1+(j-1){\rm log}\,t)(1-it)+
i\eta (\widetilde{\tau}))]^{b_{j+1}(X)}}
{\!\!\!\! \!\!\!\!\prod_{j=0, 2, 4}[\cR(s = (1+(j-2){\rm log}\,t)(1-it)(1-it))]^{b_j(X)}}\,.
\label{GFP}
\end{eqnarray}
In the last line of Eq.~(\ref{GFP}) we put
${\rm Im}\,\widetilde{\tau} = {\rm Im}\,\tau - (2\pi)^{-1}
{\rm log}\,t$.
\end{itemize}

\section{String partition functions and elliptic genera of symmetric products}
\label{Symmetric}

In this section we study the string partition functions of some local Calabi-Yau geometries, in particular, the Gopakumar-Vafa conjecture for them \cite{Gopakumar}. The identification of the Gromov-Witten partiton functions as equivariant genera suggests an approach to prove the Gopakumar-Vafa conjecture. Gromov-Witten invariants are in general rational numbers, but as has been conjectured by Gopakumar and Vafa
\cite{Gopakumar} using M-theory, the generating series of these
invariants in all degrees and all genera has a particular form, determined by some integers. There have been various proposals for the proof of this conjecture (see, for instance, \cite{Katz,Hosono});
for example, the computation of the Gromov-Witten invariants by diagrammatic methods as sum over partitions and Gopakumar-Vafa type infinite products by series manipulations \cite{Li-Liu04}.

\subsection{Partition functions}
\label{GV}

The computation of the equivariant elliptic genera for symmetric products is naturally reduced to some infinite product expansion. To prove the Gopakumar-Vafa conjecture, one needs to rewrite the sums over partition as infinite products. We propose a spectral function formulation for it.
Denote by $F$ the generating series of Gromov-Witten invariants of a Calabi-Yau three-fold $X$. Intuitively, one counts the number of stable maps with {\em connected} domain curves to $X$ in any given nonzero homology classes. However, because of the existence of automorphisms, one has to perform the weighted count by dividing by the order of the automorphism groups (hence Gromov-Witten invariants are in general rational numbers). Based on $M$-theory considerations, Gopakumar and Vafa \cite{Gopakumar} made a remarkable conjecture on the structure of $F$, in particular, on its integral properties. More precisely, integers $n^g_{\cC}$ are conjectured to exists such that
\begin{equation}
F =  \sum_{\cC \in H_2(X)-\{0\}} \sum_{g \geq 0} \sum_{k\in {\mathbb Z}_+} {k}^{-1}
n^g_{\cC} (2\sin (k\lambda/2))^{2g-2}Q^{k\cC}\,.
\label{Gopakumar}
\end{equation}
For given $\cC$, there are only finitely many nonzero $n^g_{\cC}$,
$Q^{\cC} = \exp(-\int_{\cC}\omega)$, the holomorphic curve $\cC\in H_2(X, {\mathbb Z}):= H_2(X)$ is given by $\int_{\cC}\omega$, where $\omega$ is the K\"{a}hlerian form on $X$.
Let us regard $q = \exp (i\lambda)$, for some real $\lambda$, as an element of $SU(2)$ represented by the diagonal matrix
${\rm diag}\,(q , q^{-1})$. The generating series of {\it disconnected} Gromov-Witten invariants is given by the string partition function:
$Z = \exp F$. The Gopakumar-Vafa conjecture can be reformulated as follows (see \cite{Hollowood}):
\begin{equation} \label{eqn:GV}
Z = \prod_{\cC \in H_2(X)}\prod_j \prod_{k=-j}^j \,\,\prod_{m \in {\mathbb Z}_+ \cup\, \{0\}}
(1- q^{2k+m+1}Q^{\cC})^{(-1)^{2j+1}(m+1)N_{\cC}^g}.
\end{equation}
where $j = g/2$,\, $k=-j, -j+1, \dots, j-1, j$.
Along the same lines of \cite{Li-Liu04}, we have
\begin{eqnarray}
F & = &
\sum_{\cC \in H_2(X)} \sum_{g \geq 0} \sum_{k\in {\mathbb Z}_+} \frac{(-1)^{g-1}}{k (q^{\frac{k}{2}} - q^{-\frac{k}{2}})^2 }
N^g_{\cC} \sum_{a=0}^{g} q^{k(g-2a)}Q^{k\cC}
\nonumber \\
& = & \sum_{\cC \in H_2(X)} \sum_{g \geq 0} \sum_{k\in {\mathbb Z}_+} \frac{(-1)^{g -1}}{k (1 - q^k)^2}
N^g_{\cC} \sum_{a=0}^{g} q^{k(g-2a+1)} Q^{k\cC}
\nonumber \\
& = & \sum_{\cC \in H_2(X)} \sum_{g \geq 0} \sum_{k\in {\mathbb Z}_+} \frac{(-1)^{g-1}}{k}
N^g_{\cC} \sum_{a=0}^{g} q^{k(g-2a+1)}
Q^{k\cC} \sum_{m \in {\mathbb Z}_+ \cup\, \{0\}} (m+1) q^{km}
\nonumber \\
& = & -\sum_{\cC \in H_2(X)} \sum_{g \geq 0} (-1)^g N^g_{\cC} \sum_{a=0}^{g} \sum_{m\in {\mathbb Z}_+ \cup\, \{0\}} (m+1)
\sum_{k \geq 1} \frac{1}{k} (q^{g-2a+m+1} Q^{\cC})^k \nonumber \\
& = & \sum_{\cC \in H_2(X)} \sum_{g \geq 0} (-1)^g N^g_{\cC} \sum_{a=0}^{g} \sum_{m\in {\mathbb Z}_+ \cup\, \{0\}} (m+1)
\log (1- q^{g-2a+m+1} Q^{\cC})\,.
\label{G-V}
\end{eqnarray}
In calculating Eq.~(\ref{G-V}) we can use the Ruelle function (\ref{RF1}). Indeed, for $\varepsilon \equiv g-2a+1- i(\cC/\lambda){\rm log}\,Q$, we have
\begin{eqnarray}
\sum_{m\in {\mathbb Z}_+}m\, {\rm log}\,(1-q^{m+\varepsilon}) & = &
\sum_{m\in {\mathbb Z}_+}m \sum_{k=m}^{\infty}{\rm log}\,(1-q^{k+\varepsilon}) -
\sum_{m\in {\mathbb Z}_+}\sum_{k=m}^{\infty}{\rm log}\,
(1-q^{k+\varepsilon +1})
\nonumber \\
& = &
\sum_{m\in {\mathbb Z}_+}m \,{\rm log}\left[\frac{\cR(s=(m+\varepsilon)(1-it))}{\cR(s= (m+\varepsilon +1)(1-it))}\right]
\nonumber \\
& = &
\sum_{m\in {\mathbb Z}_+}{\rm log} \cR(s=(m+\varepsilon)(1-it))\,,
\end{eqnarray}
and by Eq.~(\ref{modular1}),
\begin{equation}
\sum_{m\in {\mathbb Z}_+ \cup \,\{0\}}{\rm log}\,(1-q^{m+\varepsilon})
= {\rm log}\,\cR(s= \varepsilon(1-it))\,.
\end{equation}
Therefore
\begin{eqnarray}
\sum_{m\in {\mathbb Z}_+ \cup \,\{0\}}(m+1){\rm log}\,(1-q^{m+\varepsilon})
& = & \sum_{m\in {\mathbb Z}_+ \cup \,\{0\}} {\rm log}\,
\cR(s=(m+\varepsilon)(1-it))\,,
\label{GV2}
\end{eqnarray}
and finally we have
\begin{equation}
F =
-\sum_{\cC \in H_2(X)} \sum_{g \geq 0} (-1)^g N^g_{\cC} \sum_{a=0}^{g} \sum_{m\in {\mathbb Z}_+ \cup \,\{0\}} {\rm log}\,
\cR(s=(m+\varepsilon)(1-it))\,.
\label{GV3}
\end{equation}

\subsection{Orbifold elliptic genera}
\label{Orbifold}

Infinite product expressions, as those discussed above, arise naturally when
one considers genera of symmetric products. This motivates the proposal of using symmetric products to prove the Gopakumar-Vafa Conjecture (for references see e.g. \cite{Zhou99, Zhou991,Borisov}). Recall that, for $g \in G$, the Lefschetz number is defined by (see Eq.~(\ref{R-R})):
$
L(X, E)(g) = \sum_{p=0}^{\dim X} (-1)^p {\rm Tr} (g\,|\,H^p(X; \cO(E)))\,.
$
It can be computed by means of the holomorphic Lefschetz formula \cite{Atiyah}:
\begin{equation}
L(X, E)(g) = \int_{X^g} \frac{{\rm ch} E^g}{{\rm ch} \Lambda_{-1} (N_{X^g/X})}\,,
\end{equation}
where $X^g$ denotes the set of points fixed by $g$ and
$N_{X^g/X}$ denotes the normal bundle of $X^g$ in $X$.
Here and in the following it is understood that one considers each connected component of $X^g$ separately. There is a natural decomposition
$TX|_{X^g} = \oplus_j N^g_{\lambda_j},$
where each $N^g_{\lambda_j}$ is a holomorphic subbundle on which $g$ acts as $\exp \,(2\pi i \lambda_j)$, where $0 \leq \lambda_j < 1$.
Define the {\em fermionic shift} by \cite{Zaslow}:
$
F(X^g) = \sum_j (\rank N_{\lambda_j}) \lambda_j.
$
For ${\frak X} := TX^g$ let us follow \cite{Borisov} and define
\begin{eqnarray}
\!\!\!\!\!\!\!\!\!\!\!\!\!\!\!\!\!\!\!\!\!\!\!\!
&&
E(X;q,y)^g  =  y^{- \frac{d}{2} + F(X^g_i)}
\bigotimes_j \left(
\bigotimes_{n \in {\mathbb Z}_+}
\left( S_{q^n} {\frak X}^*\otimes S_{q^n} {\frak X} \right) 
\bigotimes_{n \in {\mathbb Z}_+} \left(
\Lambda_{-yq^{n - 1}}{\frak X}^*\otimes
\Lambda_{-y^{-1} q^{n}} {\frak X}
\right)\right)
\nonumber \\
\!\!\!\!\!\!\!\!\!\!\!\!\!\!\!\!\!\!\!\!\!\!\!\!
&&
\bigotimes_{\lambda_j \neq 0}\left(
\bigotimes_{n \in {\mathbb Z}_+}
( S_{q^{n -1 + \lambda_j}} N_{\lambda_j}^*\otimes
S_{q^{n - \lambda_j}} N_{\lambda_j}) 
\bigotimes_{n \in {\mathbb Z}_+} \left(
\Lambda_{-yq^{n - 1+\lambda_j}}N^*_{\lambda_j}\otimes
\Lambda_{-y^{-1} q^{n-\lambda_j}} N_{\lambda_j}
\right) \right).
\end{eqnarray}
For a compact complex $d$-manifold $X$,
one is often interested in its Hirzebruch $\chi_y$ genus \cite{Li-Liu04}, which is defined by
\begin{equation}
\chi_y(X) = \sum_{p=0}^d (-y)^p \sum_{q=0}^d (-1)^q \dim H^q(X, \Lambda^pT^*X).
\end{equation}
From the Hirzebruch-Riemann-Roch theorem, it can be computed as
\begin{equation}
\chi_y(X) =  \int_X \prod_{j=1}^d \frac{x_j(1- ye^{-x_j})}{(1- e^{-x_j})}\,.
\end{equation}
The $\chi_y$-genus reduces to other invariants for special values of $y$,
e.g., $\chi_1(X)$ is the Euler number,
$
\chi_0(X) = \sum_{q=0}^d (-1)^q \dim H^q(X, \cO_X)
$
is the geometric genus of $X$.
An important generalization of the $\chi_y$-genus is the elliptic genus $\chi(X, y, q)$,
which can be defined as the generating series of dimensions of some cohomology
groups of series of vector bundles.
It can be computed as
\begin{eqnarray}
\chi(X, y, q) & = & y^{-d/2} \int_X \prod_{j=1}^d x_j
\prod_{n\in {\mathbb Z}_+}\frac{(1-yq^{n-1}e^{-x_j})(1- y^{-1}q^ne^{x_j})}
{(1-q^{n-1}e^{-x_j})(1-q^ne^{x_j})}
\nonumber \\
& = & y^{-d/2}\int_X \prod_{j=1}^d x_j
\left[\frac{\cR(s= \xi_{1j}(1-it))\cdot \cR(s= \xi_{2j}(1-it))}
{\cR(s= \xi_{3j}(1-it))\cdot \cR(s= \xi_{4j}(1-it))}\right]\,,
\end{eqnarray}
where
\begin{eqnarray}
&&
\xi_{1j} =  1-x_j + {\rm log}\,y\,, \,\,\,\,\,\,\,
\xi_{2j} = x_j - {\rm log}\,y\,,\,\,\,\,\,\,\,
\xi_{3j} =  -1 -x_j\,,\,\,\,\,\,\,\,
\xi_{4j} = x_j\,,
\nonumber \\
&&
(\xi_{1j} + \xi_{2j}) + (\xi_{3j} + \xi_{4j}) = 1+ (-1) = 0\,.
\end{eqnarray}
It is easy to see that $\chi_y(X) = y^{d/2} \chi(X, y, 0).$
\begin{definition}
The orbifold elliptic genus of $X/G$ is defined by
$$\chi(X, G; q, y)
=\sum_{[g] \in G_*} \frac{1}{|Z(g)|}
\sum_{h \in Z(g)} L(X^g, E(X; q, y)^g)(h),$$
where $G_*$ denotes the set of conjugacy classes of $G$,
and $Z(g)$ denotes the centralizer of $g$.
\end{definition}

{\bf Equivariant orbifold elliptic genera.}
For any $h \in Z(g)$,
let $X^{(g, h)}$ be the set of points of $X$
fixed by both $g$ and $h$. By the Lefschetz formula, we have
\begin{equation}
L(X^g, E(X; q, y)^g)(h)
= \int_{X^{g, h}} \frac{{\rm ch}(E(X; q, y)^g)}
{{\rm ch}(\Lambda_{-1}(N^*_{X^{g,h}/X^g}))}
{\rm Td}(TX^{g, h}).
\end{equation}
Assume now that the $G$-manifold, $X$,
admits an $S^1$-action which commutes with the $G$-action.
The equivariant orbifold elliptic genus is given by
\begin{equation}
\chi(X, G; q, y)(s) = \sum_{[g] \in G_*}
\frac{1}{|Z(g)|} \sum_{h \in Z(g)}
L(X^g, E(X; q, y)^g)(h, s)\,,
\end{equation}
where $s \in S^1$. This defines a character for $S^1$.
Denote by $M^{(g, h, s)}$ the points on $M$ which are fixed by
$g$, $h$, and $s$. For our applications,
it suffices to assume that $X^{g, h, s}$ consists of an isolated point.
Then, using the Lefschetz formula, we have
\begin{equation}
L(X^g, E(X; q, y)^g)(h, s)
=  \frac{{\rm ch}(E(X; q, y)^g)}
{{\rm ch}(\Lambda_{-1}(N^*_{X^{g,h,s}/X^g}))}.
\end{equation}

{\bf Symmetric products.}
Assume that $X$ is a nonsingular projective variety admitting an action by a torus group $T$.
The diagonal action by $T$ and the natural action by the permutation group $S_N$ on the $N$-fold cartesian product $X^N$ commute with each other.
\begin{theorem} \label{thm:Symmetric}
Let the equivariant elliptic genera $\chi(X; q, y)(t_1, \dots, t_r)$ of $X$ be written as
$
\chi(X; q, y)(t_1, \dots, t_r)
= \sum_{m \geq 0, l, k} c(m, l, k) q^my^lt_1^{k_1} \cdots t_r^{k_r},
$
then, one has
\begin{eqnarray} \label{eqn:SymmEll}
\sum_{N\in {\mathbb Z}_+\cup \,\{0\}}\!\!\!  Q^N\chi(X^N, S_N; q, y)(t_1, \dots, t_r) \!\!
& = & \!\!
\exp \!\!\sum_{N, n \in {\mathbb Z}_+}\!\! \frac{Q^{Nn}}{Nn}
\sum_{i=0}^{n-1} \chi(X)((\omega_n^iq^{1/n})^{N}, y^{N})(t_1^{N}, \cdots, t_r^{N})
\nonumber \\
& = & \!\!\!\!\! \!\!\!\!\!\prod_{n > 0, m \geq 0, l, k_1, \dots, t_r}
\!\!\!(1-Q^nq^my^lt_1^{k_1} \cdots t_r^{k_r})^{- c(nm, l, k_1, \dots, k_r)},
\label{Prod1}
\end{eqnarray}
where $\omega_n = \exp (2\pi i/n)$.
\end{theorem}
Also, we have
\begin{equation}
\sum_{N \in {\mathbb Z}_+\cup\, \{0\}}  Q^N\chi_y(X^N, S_N)(t_1, \dots, t_r)
= \sum_{N \in {\mathbb Z}_+\cup\, \{0\}} (yQ)^N\chi(X^N, S_N; 0, y) (t_1, \dots, t_r),
\end{equation}
and, hence, by taking $q=0$ in (\ref{eqn:SymmEll}) we get
\begin{eqnarray} \label{eqn:SymmChiy}
\sum_{N\in {\mathbb Z}_+\cup\, \{0\}}  Q^N\chi_y(X^N, S_N)(t_1, \dots, t_r)
& = & \exp \sum_{m\in {\mathbb Z}_+} \frac{Q^m\chi_{y^m}(X)(t_1^m, \dots, t_r^m)}{m(1-y^mQ^m)}
\nonumber \\
& = & \prod_{n > 0,  l, k_1, \dots, k_r}
(1-(yQ)^ny^lt^k)^{- c(0, l, k_1, \dots, k_r)}\,.
\label{Prod2}
\end{eqnarray}
Finally, taking $y=0$ in (\ref{eqn:SymmChiy}),
\begin{eqnarray} \label{eqn:SymmChi0}
&& \sum_{N\in {\mathbb Z}_+\cup \{0\}}  Q^N\chi_0(X^N, S_N)(t_1, \dots, t_r)
= \exp (\sum_{m\in {\mathbb Z}_+} \frac{Q^m}{m}  \chi_{0}(X)(t_1^m, \dots, t_r^m)).
\end{eqnarray}
For the nonequivariant version of (\ref{eqn:SymmEll})-(\ref{eqn:SymmChi0}), see \cite{Zhou99,Zhou991, Borisov}. From (\ref{eqn:GV}), one sees that in order to prove the Gopakumar-Vafa conjecture one needs an infinite product expression for the string partition function. These partition functions are usually given by a sum of expressions (in the diagrammatic method, for example). We will next show  how one can convert this sum of expressions into an infinite product.

\section{Homological aspects of combinatorial identities}
\label{Identities}

We would like now to spend some time on what we believe is the
mathematical origin of the combinatorial identities that are at
the basis of the relation between partition functions and formal
power series and homologies of Lie algebras. The following is
meant to be a brief introduction to homological aspects of
differential complexes, and in particular we would like to show how
combinatorial identities could be derived from initial complex of
(graded) Lie algebras. In recalling the main results we will
follow the book \cite{Fuks}. Our interest is the
Euler-Poincar\'{e} formula associated with a complex consisting of
finite-dimensional linear spaces. The relationship between Lie
algebras and combinatorial identities was first discovered by
Macdonald \cite{Macdonald,Macdonald1} and the Euler-Poincar\'{e}
formula is useful for combinatorial identities known as {\it
Macdonald identities}. Macdonald identities are related to Lie
algebras in one way or other, and can be associated with
generating functions (in particular, elliptic genera) in quantum theory.

Let ${\mathfrak g}$ be an {\it infinite-dimensional} Lie algebra, and assume that it has a grading, i.e. ${\mathfrak g}$ is a direct sum of homogeneous components
${\mathfrak g}_{(\lambda)}$, where the $\lambda$'s are elements of an abelian group, $[{\mathfrak g}_{(\lambda)}, {\mathfrak g}_{(\mu)}]
\subset {\mathfrak g}_{(\lambda + \mu)}$ (recall, e.g., the Virasoro algebra). Let us consider a module $\bf k$ over ${\mathfrak g}$, or ${\mathfrak g}$-module. ${\bf k}$ is a vector with the property that there exists a bilinear map $\mu : {\mathfrak g}\times {\bf k} \rightarrow {\bf k}$ such that
$[{\mathfrak g}_1, {\mathfrak g}_2] ={\mathfrak g}_1({\mathfrak g}_2k) - {\mathfrak g}_2 ({\mathfrak g}_1k)$, for all $k\in {\bf k},\, {\mathfrak g}_1, {\mathfrak g}_2 \in {\mathfrak g}$.
In other words, our ${\mathfrak g}$-module is a left module over the
universal enveloping algebra $U({\mathfrak g})$ of ${\mathfrak g}$.
Let $C^n({\mathfrak g}; {\bf k})$ be the space of all cochains, an
$n$-dimensional cochain of the algebra ${\mathfrak g}$, with coefficients in $\bf k$, being a skew-symmetric $n$-linear functional on ${\mathfrak g}$, with values in $\bf k$.
Since $C^n({\mathfrak g}; {\bf k}) = {\rm Hom} (\Lambda^n, {\bf k})$,
the cochain space $C^n({\mathfrak g}; {\bf k})$ becomes a
${\mathfrak g}$-module.
The differential $d= d_n : C^n({\mathfrak g}; {\bf k})\rightarrow
C^{n+1}({\mathfrak g}; {\bf k})$ can be defined as follows
\begin{eqnarray}
dc({\mathfrak g}_1, \ldots ,{\mathfrak g}_{n+1})
& = &
\sum_{1\leq s\leq t\leq n+1}(-1)^{s+t-1}c
([{\mathfrak g}_s, {\mathfrak g}_t], {\mathfrak g}_1, \ldots , {\widehat {\mathfrak g}}_s, \ldots
{\widehat {\mathfrak g}}_t, \ldots,{\mathfrak g}_{n+1})
\nonumber \\
& + &
\sum_{1\leq s\leq n+1}(-1)^{s}{\mathfrak g}_sc
({\mathfrak g}_1, \ldots,{\widehat{\mathfrak g}}_s,...,{\mathfrak g}_{n+1})
\mbox{,}
\label{diff}
\end{eqnarray}
where $c\in C^n({\mathfrak g}; {\bf k}),\, {\mathfrak g}_1, \ldots,
{\mathfrak g}_{n+1}\in {\mathfrak g}$, and
$C^n({\mathfrak g}; {\bf k}) =0, \, d_n= 0$ for $n<0$.
As $d_{n+1}\circ d_n =0$, for all $n$, the set
$C^{\bullet}({\mathfrak g}; {\bf k}) \equiv \{C^n({\mathfrak g}; {\bf k}), d_n\}$ is an algebraic complex, while the corresponding cohomology
$H^n({\mathfrak g}; {\bf k})$ is referred to as the cohomology of the
algebra ${\mathfrak g}$ with coefficients in $\bf k$.
Let $C_n({\mathfrak g}; {\bf k})$ be the space of $n$-dimensional chains of the Lie algebra ${\mathfrak g}$. It can be defined as
${\bf k}\otimes \Lambda^n{\mathfrak g}$. The differential $\delta= \delta_n:
C_n({\mathfrak g}; {\bf k})\rightarrow C_{n-1}({\mathfrak g}; {\bf k})$ is given by the expression
\begin{eqnarray}
\delta (a\otimes ({\mathfrak g}_1 \wedge \ldots \wedge
{\mathfrak g}_{n}))
& = &
\sum_{1\leq s\leq t\leq n}(-1)^{s+t-1}a\otimes
([{\mathfrak g}_s, {\mathfrak g}_t]\wedge {\mathfrak g}_1\wedge \ldots
{\widehat {\mathfrak g}}_s \ldots {\widehat {\mathfrak g}}_t \ldots \wedge
{\mathfrak g}_{n})
\nonumber \\
& + &
\sum_{1\leq s\leq n}(-1)^{s}{\mathfrak g}_sa\otimes
({\mathfrak g}_1\wedge \ldots {\widehat{\mathfrak g}}_s \ldots \wedge
{\mathfrak g}_{n})
\mbox{.}
\label{diff1}
\end{eqnarray}
The homology $H_n({\mathfrak g}; {\bf k})$ of the complex $\{C_n({\mathfrak g}; {\bf k}), \delta_n\}$ is referred to as the homology of the algebra ${\mathfrak g}$.
Suppose now that the $\mathfrak g$-module $\bf k$ can be graded by homogeneous components ${\bf k}_{(\mu)}$ in such a way that ${\mathfrak g}_{(\lambda)}{\bf k}_{(\mu)}\subset {\bf k}_{(\lambda+\mu)}$. In the case when the module $\bf k$ is trivial, we assume that
${\bf k}= {\bf k}_{(0)}$.
The grading of our Lie algebra endows with a grading both the chain and the cochain spaces,
for we can define
\begin{eqnarray}
C^n_{(\lambda)}({\mathfrak g}; {\bf k}) & = &
\{c\in C^n({\mathfrak g}; {\bf k})|
c({\mathfrak g}_1, \ldots,{\mathfrak g}_n)\in
{\bf k}_{(\lambda_1+ \ldots +\lambda_n-\lambda)} \,\,\,\,\,
{\rm for}\,\,\,\,\,{\mathfrak g}_i\in {\mathfrak g}_{(\lambda_i)}\};
\nonumber \\
C_n^{(\lambda)}({\mathfrak g}; {\bf k}) & {\rm is} & \, {\rm
generated}\,\,\, {\rm by}\,\,\, {\rm the}\,\,\, {\rm
chains}\,\,\, a\otimes({\mathfrak g}_1\wedge \ldots \wedge
{\mathfrak g}_n),\, a \in {\bf k}_{(\mu)},
\nonumber \\
& {\mathfrak g}_i & \!\!\!\!\in {\mathfrak g}_{(\lambda_i)}, \,\,\,
\lambda_1+ \ldots +\lambda_n+\mu = \lambda \,.
\label{chain}
\end{eqnarray}
We get $d(C^n_{(\lambda)}({\mathfrak g}; {\bf k})) \subset
C^{n+1}_{(\lambda)}({\mathfrak g}; {\bf k})$ and
$\delta (C_n^{(\lambda)}({\mathfrak g}; {\bf k})) \subset
C_{n-1}^{(\lambda)}({\mathfrak g}; {\bf k})$ and both spaces acquire gradings\footnote{
The cohomological (and homological) multiplicative structures
are compatible with these gradings, for example,
$H^m_{(\lambda)}({\mathfrak g})\otimes H^n_{(\mu)}({\mathfrak g})\subset
H^{m+n}_{(\lambda + \mu)}({\mathfrak g})$.}.
The chain complex
$C_{\bullet}({\mathfrak g})$,\,
${\mathfrak g}= \oplus_{\lambda=1}^{\infty}{\mathfrak g}_{(\lambda)},\,
{\rm dim}\,{\mathfrak g}_{(\lambda)}< \infty$,
can be decomposed as
\begin{equation}
0\longleftarrow C_{0}^{(\lambda)}({\mathfrak g}) \longleftarrow
C_{1}^{(\lambda)}({\mathfrak g}) \ldots\longleftarrow C_N^{(\lambda)}({\mathfrak g}) \longleftarrow 0\,,
\end{equation}
and the Euler-Poincar\'{e} formula reads
\begin{equation}
\sum_m(-1)^m {\rm dim}\,C_m^{(\lambda)}({\mathfrak g}) =
\sum_m(-1)^m {\rm dim}\,H_m^{(\lambda)}({\mathfrak g})\,.
\label{E-P}
\end{equation}
As a consequence, we can
introduce the $q$ variable and rewrite the identity (\ref{E-P}) as a formal power
series\footnote{If $\bf k$ is the main field, the notation $C_n({\mathfrak g}; {\bf k}),\, H_n({\mathfrak g}; {\bf k})$ is abbreviated as $C_n({\mathfrak g}),\, H_n({\mathfrak g})$.
For a finite-dimensional algebra $\mathfrak g$ we obviously have $C^n({\mathfrak g})=[C_n({\mathfrak g})]^{*}$ and $H^n({\mathfrak g})=[H_n({\mathfrak g})]^{*}$, where the symbol $*$ denotes dual space. It is clear that in the case of cohomologies we have a power series similar to (\ref{EP}). For the finite-dimensional case, as well as for the infinite-dimensional one, an element of the space $H^n({\mathfrak g}; {\bf k})$ defines a linear map $H_n({\mathfrak g})\rightarrow {\bf k}$; if the algebra $\mathfrak g$ and the module $\bf k$ are finite dimensional, then $H^n({\mathfrak g}; {\bf k}^{*})= [H_n({\mathfrak g}; {\bf k})]^{*}$.
}:
\begin{equation}
\sum_{m,\lambda}(-1)^m q^\lambda {\rm dim}\,C_m^{(\lambda)}({\mathfrak g}) = \sum_{m,\lambda}(-1)^m q^\lambda {\rm dim}\,H_m^{(\lambda)}({\mathfrak g})
=\prod_n(1-q^n)^{{\rm dim}\,{\mathfrak g}_{n}}\,.
\label{EP}
\end{equation}
In order to get the identity in its final form the
homology $H_m^{(\lambda)}({\mathfrak g})$ has to be computed.
Let ${\mathfrak g}$ be a (poly)graded Lie algebra
$
{\mathfrak g} = \bigoplus_{\scriptstyle \lambda_1\geq 0, ..., \lambda_k\geq 0
\atop\scriptstyle \lambda_1+...+\lambda_k>0}
{\mathfrak g}_{(\lambda_1, ..., \lambda_k)},
$
satisfying the condition
$
{\rm dim}\, {\mathfrak g}_{(\lambda_1, ..., \lambda_k)} < \infty.
$
For formal power series in $q_1,...,q_k$, we have the following identity \cite{Fuks}:
\begin{equation}
\sum_{m,\lambda_1,...,\lambda_k}(-1)^m
q_1^{\lambda_1}...q_k^{\lambda_k} H_m^{(\lambda_1,...,\lambda_k)}
=
\prod_{n_1,...,n_k}\left(1-q_1^{n_1}\cdots q_k^{n_k}\right)^
{{\rm dim}\, {\mathfrak g}_{n_1, ..., n_k}}\,.
\label{poly}
\end{equation}
We would like to stress that partition functions can indeed be converted into product expressions. As has been noted in \cite{Hollowood}, the expression on the right-hand side of (\ref{eqn:GV}) look like {\it counting} the states in the Hilbert space of a second quantized theory.
This is achieved by relating the string partition functions to equivariant genera \cite{Dijkgraaf}, first of Hilbert schemes and then of symmetric products (see Section \ref{Orbifold}). In general, formulas for Poincar\'{e} polynomials might be associated with dimensions of homologies of topological spaces and linked with generating functions and elliptic genera. Infinite products are the heritage of appropriate polygraded Lie algebras (Eq.~(\ref{poly})). The reader can find in \cite{Fuks} computations of the homologies $H_m^{(\lambda)}$ and, therefore, corresponding partition functions in their final form. Some algebras can be constructed from exterior automorphisms of simple algebras possessing a finite center. In this case, we have to find the dimensions of the space $H_{*}^{(\lambda_1,...,\lambda_n)}(\mathfrak g)$, which is not too difficult to do, but the answer turns out to be rather cumbersome.\footnote{Interesting combinatorial identities can be obtained by applying (\ref{poly}) to the subalgebras of Kac-Moody algebras (see Section \ref{Represent}), $N^{+}({\mathfrak g}^{\bf k})$, where $N^{+}({\mathfrak g})$ is the subalgebra of the algebra $\mathfrak g$, consisting of upper triangular matrices with zero diagonal. For further details, we refer the reader to the book \cite{Fuks}.
}

\subsection*{Acknowledgements}

AAB would like to acknowledge the Conselho Nacional
de Desenvolvimento Cient\'ifico e Tecnol\'ogico (CNPq, Brazil), Funda\c cao Araucaria (Parana, Brazil) and the ESF Research Network CASIMIR for financial support. Research of AAB was performed in part while on leave at the High Energy Sector of SISSA, Trieste, Italy. EE's research was performed in part at Dartmouth College, NH, USA, and his investigation has been partly supported by MICINN (Spain), contract PR2011-0128 and projects FIS2006-02842 and FIS2010-15640, by the CPAN Consolider Ingenio Project, and by AGAUR (Generalitat de Ca\-ta\-lu\-nya), contract 2009SGR-994.

\end{document}